\documentclass{raa}           
\usepackage{graphicx,times}  
\usepackage{natbib}
\usepackage{amssymb,amsmath}
\usepackage{xcolor}
\bibpunct{(}{)}{;}{a}{}{,}
\usepackage{threeparttable}
\usepackage{textcomp}
\usepackage[a4paper=true, pagebackref=true]{hyperref}
\hypersetup{pdftitle = The title of my PDF, pdfauthor = My name, pdfsubject= The subject, pdfkeywords = keyword1 keyword2 keyword3} 
\hypersetup{colorlinks = true, linkcolor = green, anchorcolor = red, citecolor = blue, filecolor = red, pagecolor = red, urlcolor = red}

\begin{document}
\title{ Symmetry properties and widths of the filamentary structures in the Orion A giant molecular cloud}
\volnopage{ {\bf 20XX} Vol.\ {\bf X} No. {\bf XX}, 000--000}
\setcounter{page}{1}
\author{Yuqing Zheng \inst{1, 2} \and Hongchi Wang \inst{1, 2} \and Yuehui Ma \inst{1, 3} \and Chong Li \inst{1}} 
\institute{Purple Mountain Observatory and Key Laboratory of Radio Astronomy, Chinese Academy of Sciences, 10 Yuanhua Road, Nanjing 210033, P. R. China; {\it zhengyq@pmo.ac.cn}\\
	\and 
	    School of Astronomy and Space Science, University of Science and Technology of China, 96 JinZhai Road, Hefei 230026, P. R. China\\
    \and
		University of Chinese Academy of Sciences, 19A Yuquan Road, Beijing 100049, P. R. China}
\vs \no
{\small Received 20XX Month Day; accepted 20XX Month Day}

\abstract{We identify 225 filaments from an $\rm{H_2}$ column density map constructed using simultaneous $^{12}$CO, $^{13}$CO, and C$^{18}$O (J=1-0) observations carried out as a part of the MWISP project.  We select 46 long filaments with lengths above 1.2 pc to analyze the filament column density profiles. We divide the selected filaments into 397 segments and calculate the column density profiles for each segment. The symmetries of the profiles are investigated. The proportion of intrinsically asymmetrical segments is 65.3\%, and that of intrinsically symmetrical ones is 21.4\%. The typical full width at half maximum (FWHM) of the intrinsically symmetrical filament segments is $\sim$ 0.67 \mbox{pc} with the Plummer-like fitting, and $\sim$ 0.50 pc with the Gaussian fitting, respectively. The median FWHM widths derived from the second-moment method for intrinsically symmetrical and asymmetrical profiles are $\sim$ 0.44 and 0.46 \mbox{pc}, respectively. Close association exists between the filamentary structures and the YSOs in the region.
}

\authorrunning{Y.-Q. Zheng et al.}            
\titlerunning{Filamentary structures in the Orion A GMC}  
\maketitle
 
\section{Introduction} \label{sec1}
Filamentary structures are commonly seen both in observations of the interstellar medium (ISM) \citep[e.g.,][]{Schneider1979, Joncas1992, Abergel1994, Johnstone1999, Falgarone2001, Myers2009, Andre2010, Molinari2010, Men'shchikov2010, Hill2011, Arzoumanian2011, Wang2015, Arzoumanian2019} and in numerical simulations of gravity \citep[e.g.,][]{Lin1965, Zel'Dovich1970, Burkert2004, Gomez2014} and interstellar turbulence \citep[e.g.,][]{Padoan2001, Pudritz2013, Inoue2013, Chen2014, Inutsuka2015}. Interstellar filaments are usually defined as elongated over-dense ISM structures with large aspect ratios ($>$3). They are detected initially through dust extinction by \citet{Schneider1979} and their existence in star-forming regions has been confirmed through observations with other tracers, such as dust emission \citep[e.g.,][]{Abergel1994} and CO line emission \citep[e.g.,][]{Falgarone2001}. The ubiquitous presence of filamentary structures either in diffuse or in dense star-forming molecular clouds has been revealed through Herschel observations \citep[e.g.,][]{Andre2010, Molinari2010, Arzoumanian2011, Arzoumanian2019}. These observations suggested that there is a ``universal width'', $\sim$0.1 pc, of the observed filaments regardless of their central column densities. \citet{Arzoumanian2011}  propose that a uniform filament width should  be the result of the dissipation of large-scale turbulence. Some observations toward molecular filaments in active star-forming regions reveal accretion signatures along the axis of filaments \citep[e.g.,][]{Kirk2013, Palmeirim2013}.  \citet{Pon2011}  found that local collapases occur most favorably under the filamentary geometry compared with the spheres and disks. A possible scenario is that molecular gas is firstly compressed into filamentary structures, and then break up into molecular cores that eventually form stars \citep[e.g.,][]{Kainulainen2013, Takahashi2013, Andre2014, Teixeira2016, Kainulainen2017}. \citet{Li2016}  produced a Galaxy-wide catalogue of dense filaments and have shown that these filaments are correlated with the spiral arms and make a significant contribution to star formation in the Galaxy.

The definition of ``width'' for filaments varies in different studies,  e.g., there are two commonly used forms of functions for the column density profiles of filaments, the Gaussian function and the Plummer-like function  \citep[e.g.,][]{Arzoumanian2011}. The radius of the inner flat core and the exponent of the outside power-law wing of the Plummer-like column density profile are related to the dynamical state of the filaments \citep[e.g.,][]{Ostriker1964, Heitsch2013}. Besides, the typical width of filaments usually varies with different tracers of molecular gas and different observational resolutions. For example, the typical width of molecular filaments traced by \element[][13]{CO} line emission in the Taurus molecular cloud is found to be $\sim$0.4 pc \citep{Panopoulou2017}, whereas the filament width derived through \element[][]{C}\element[][18]{O} line emission in the Orion giant molecular cloud is around 0.1 pc \citep{Suri2019}. The typical width of the dense filaments in the Orion integral-shaped filament region, the so called molecular fibers observed with ALMA and traced by N$_2$H$^+$ (1$-$0) line emission \citep{Hacar2018}, is found to be 0.035 pc. \citet{Smith2014} and \citet{Suri2019}  suggested that the reported widths of filaments are influenced by the fitting ranges used in the analysis of the column density profiles. Moreover, the morphology of the column density profiles of molecular filaments can also provide us information about the interaction between the filaments and their environment \citep[e.g.,][]{Peretto2012}. Therefore, detailed investigations of the radial profiles of molecular filaments are still needed.

The Orion A giant molecular cloud (GMC) is one of the most studied star-forming regions in the Galaxy ($d = 414\ \mbox{pc}$; \citealp{Menten2007}). The GMC itself is a giant molecular filament \citep[e.g.,][]{Ragan2014} composed of a dense and hot (above $\sim$ 50 K) integral-shaped filament (ISF) in the northern part and a relatively extended and cold (below $\sim$20 K) tail in the southern part \citep[e.g.,][]{Bally1987, Stutz2015, Kong2018}. The northern ISF, which contains the star-forming regions OMC 1-4, hosts thousands of protostars and Class II sources as revealed by the Spitzer catalog of young stellar objects \citep{Megeath2012, Megeath2016}, while the southern tail of the Orion A GMC, which contains the L 1641 S region, is relatively inactive in star formation. The northern part of the Orion A GMC is exposed to the UV radiation from the Trapezium stars, which are $\sim$1 pc in front of the GMC \citep[e.g.,][]{vanderWerf2013, Suri2019}. \citet{Stutz2015} and \citet{Ma2020}  investigated the column density structure of molecular hydrogen in the Orion A GMC using the probability distribution function (PDF) method. They found that the Orion A GMC exhibits an evolutionary trend in terms of star formation along the main ridge of the GMC from north to south, and the whole GMC is composed of two giant filaments. The filamentary constituents of the GMC have been studied using observations of high angular resolutions, such as the ALMA \citep[e.g.,][]{Hacar2018} and the CARMA-NRO surveys \citep[e.g.,][]{Suri2019}. However, the former investigation is restricted to the ISF region and is concentrated on the small scale fibers, while the latter is focused on the properties of filaments traced by \element[][]{C}\element[][18]{O} (J = 1--0) emission. The tracers of column density used in these studies can not probe the regions of medium column densities, and neither of the above two studies observed the southernmost part of the Orion A GMC.

In this study, we use the \element[][12]{CO} and \element[][13]{CO} (J = 1--0) emission line data from \citet{Ma2020} to investigate the properties of radial profiles of the filaments in the Orion A GMC. The spatial coverage of the survey is $\sim$3.5 deg$^2$, from $\delta\sim-$4.5$^{\circ}$ to $\sim-$8.7$^{\circ}$. The paper is organized as follows. The observation is described in Section \ref{sec2}, and the methods and calculations of the H$_2$ column density are presented in Section \ref{sec3}. The identification of molecular filaments and the analysis of their column density profiles are presented in Sections \ref{sec4} and \ref{sec5}, respectively. We discuss the results in Section \ref{sec6} and make a summary in Section \ref{sec7}. 

\section{Observations and data reduction} \label{sec2}
The data we used in this work are the same as those used by \citet{Ma2020}. We give a brief review of the observations in this section. The observations were made in June 2011 using the PMO-13.7 m millimeter-wavelength telescope which is equipped with a nine-beam Superconducting Spectroscopic Array Receiver (SSAR) \citep{Shan2012}. The \element[][12]{CO}, \element[][13]{CO} and \element[][]{C}\element[][18]{O} (J = 1--0) emission lines were observed simultaneously. The front end of the receiver is a two-sideband Superconductor-Insulator-Superconductor (SIS) mixer. The \element[][12]{CO} (J = 1--0) line emission is contained in the upper sideband, while the \element[][13]{CO} and \element[][]{C}\element[][18]{O} J = 1--0 line emission is contained in the lower sideband. A Fast Fourier Transform Spectrometer (FFTS) with a total bandwidth of 1 \mbox{GHz} and 16\,384 frequency channels  worked as the back end of the receiver, which provides a velocity resolution of 0.17 km $\rm{s^{-1}}$ at 110 \mbox{GHz}. The half-power beam width (HPBW) of the PMO-13.7 telescope is about 52\arcsec at 110 \mbox{GHz}, and 50\arcsec at 115 \mbox{GHz}, respectively, which corresponds to $\sim$0.10 \mbox{pc} at the distance of 414 \mbox{pc} of the Orion A GMC.

Along the direction of right ascension and declination, the observations were made in position-switch on-the-fly (OTF) mode toward twelve 30\arcmin $\times$ 30\arcmin cells to cover the Orion A GMC, with a scanning rate of 50\arcsec per second and a dump time of 0.3 s. The spectra are re-gridded into pixels of size of 30$\arcsec\times30\arcsec$ ($\sim$0.06$\times$0.06 pc$^2$ at the distance of 414 \mbox{pc}) in the final datacube. During the data reduction processes, we calibrated the antenna temperature according to $T_{\rm{MB}} = T^{\rm{*}}_{A}/\eta_{\rm{MB}}$. At the \element[][12]{CO} and \element[][13]{CO} (J = 1--0) wavelengths, the main beam efficiencies are 44\% and 48\%, respectively. We used the GILDAS/CLASS package to reduce the data, including the subtraction of a second-order baseline from each spectrum and re-griding of the raw data. The spatial coverage of the reduced data is about 4.4 $\rm{deg^2}$. The spectra at the edges of the surveyed area have higher noise levels and we trim the surveyed area to 3.5 $\mathrm{deg^2}$. In the reserved area, the median RMS noise level at 115 \mbox{GHz} is 0.61 K per channel, and the median RMS noise level at 110 \mbox{GHz} is 0.37 K per channel. 
\begin{figure*}[t]
	\centering
	\begin{minipage}[t]{0.45\linewidth}
		\centering
		\includegraphics[trim = 10cm 2cm 7cm 3cm, height = 1\linewidth]{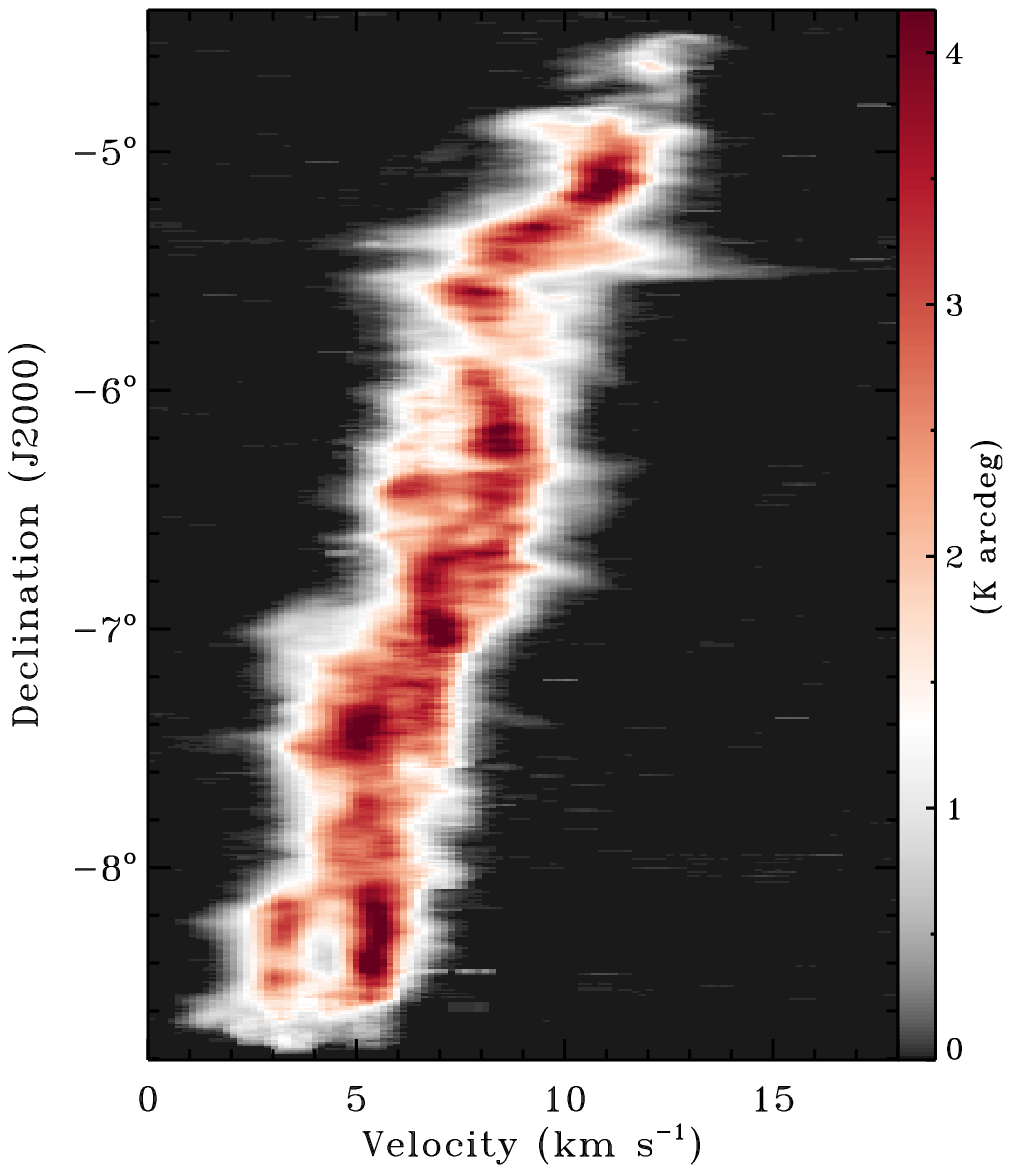}
		\\(a)
	\end{minipage}
	\begin{minipage}[t]{0.45\textwidth}
		\centering
		\includegraphics[trim = 4cm 0cm 3cm 0cm, width = 1.1\linewidth]{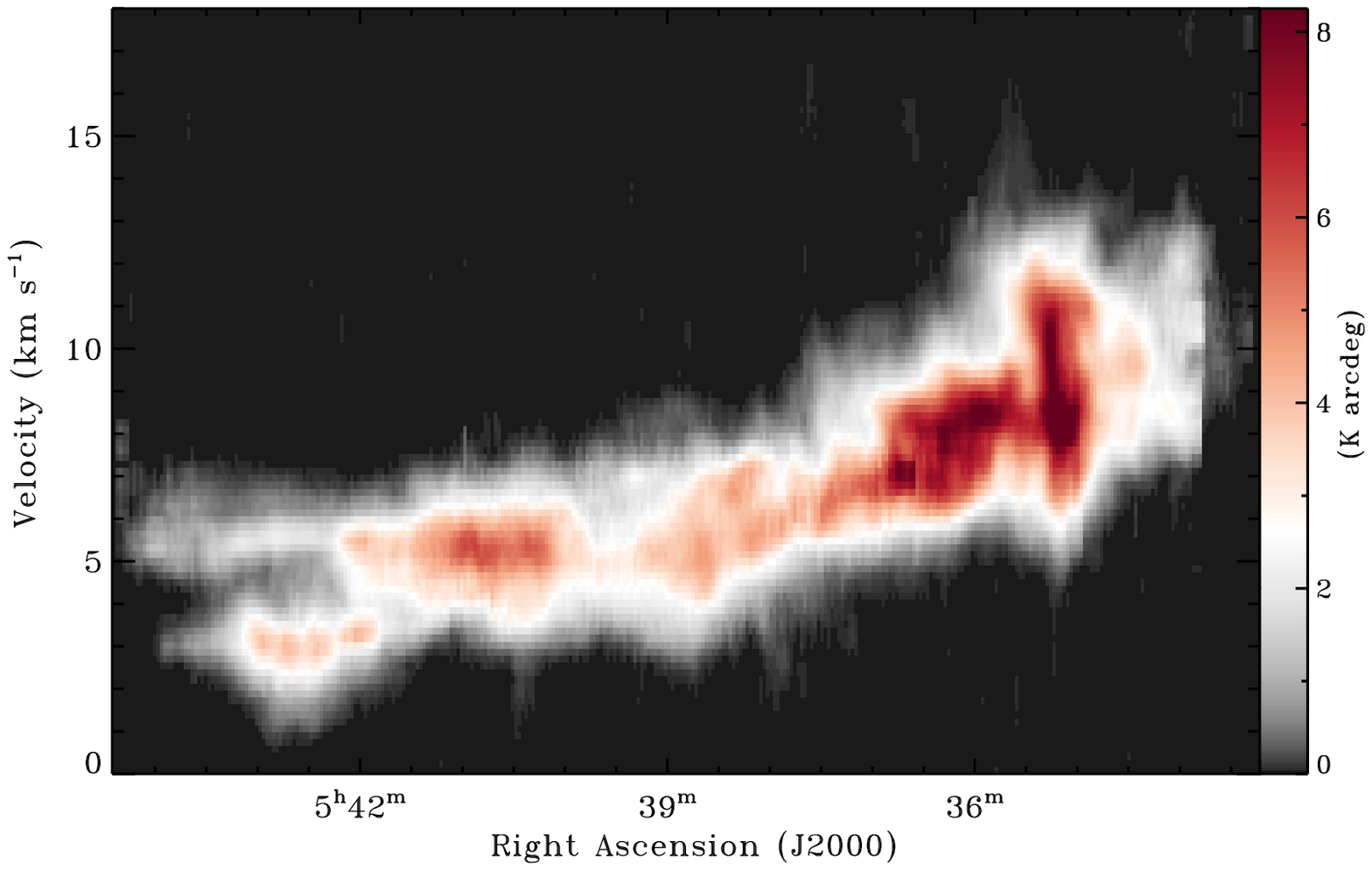}
		\\(b)
	\end{minipage}
	\caption{(a) Declination-velocity map of the \element[][13]{CO} J = 1--0 line emission of the Orion A GMC integrated within the entire observed RA range. (b) RA-velocity map of the \element[][13]{CO} J = 1--0 line emission of the Orion A GMC integrated within the entire observed Declination range.}
	\label{fig1}
\end{figure*}
\section{Methods and calculation} \label{sec3}
\subsection{Velocity Distribution of the Molecular Cloud}
Figure \ref{fig1} gives the integrated position-velocity diagrams of the \element[][13]{CO} (J = 1–0) line emission of the Orion A GMC along the directions of RA and Dec, respectively. The integration covers all the observed ranges of RA or DEC for panels a and b, respectively. We only use the spectra that have at least five contiguous velocity channels with intensities above 1.5 times the RMS noise level for the integration. The most prominent feature in Figure \ref{fig1} is that the Orion A GMC has a systematic velocity gradient, $\sim0.3$ km s$^{-1}$ pc$^{-1}$, from north to south over its entire extent, which has been observed by previous studies with CO or other molecular or atomic tracers \citep[e.g.,][]{Bally1987,Nagahama1998,Ikeda2002,Nishimura2015}. The \element[][13]{CO} (J = 1--0) emission line shows significant broadening of $\sim$10 km s$^{-1}$ at the location of the Orion KL region, which corresponds to the horizontal spike at $\delta\sim-5.5^{\circ}$ in Figure \ref{fig1}(a) and the vertical spike at $\alpha\sim5^h35^m$ in Figure \ref{fig1}(b). At the southern end of the $\delta-v$ diagram, we can see a bifurcation structure with one velocity component located at $\sim$3 km s$^{-1}$ and the other at $\sim$5.5 km s$^{-1}$. The velocity splitting is also clear in Figure \ref{fig1}(b) at the eastern end of the Orion A GMC. We have checked the spatial location of this bifurcation and found that it corresponds to the merging position of the ``fish-tail'' structures discovered by \citet{Fukui1991}. Their observations reveal that the southern end of the Orion A GMC is composed of two filamentary structures that are nearly perpendicular to each other and are twisting together into a ``rope'' toward the main ridge of the Orion A GMC. Although the internal velocity of the Orion A GMC is complex, the overall distributions of the \element[][12]{CO} and \element[][13]{CO} emission are continuous in the position-position-velocity (PPV) space in our data, which indicates the GMC is a coherent structure. The spatial resolution of the observations in this work is not high enough to allow to investigate the fine internal filamentary structures in the PPV space. Therefore,  it is feasible for us to perform filament identification on the $\mathrm{H_2}$ column density map of the Orion A GMC. In the following  two subsections, we introduce how we calculate the column density and its measurement error from the  \element[][13]{CO} data.

\subsection{Calculation of H$_2$ column density} \label{subsec3.1}
In this work, the calculation process for the H$_2$ column density is the same as \citet{Ma2020}. For convenience and clarity, we describe the method and the used formulae as follows. Assuming the local thermal equilibrium for the molecular gas and that the \element[][12]{CO} (J = 1--0) line emission is optically thick, we can obtain the $\rm{H_2}$ column density from the \element[][12]{CO} and \element[][13]{CO} (J = 1--0) data.

Firstly, we calculate the excitation temperature $T_{\mathrm{ex}}$ from the peak brightness temperature of the \element[][12]{CO} line \citep[e.g.,][]{Pineda2010,Li2018,Ma2020},  
\begin{equation}
	T_{\mathrm{ex}} = 5.53[\ln(1 + \frac{5.53}{T_{\mathrm{peak}}(\element[][12]{CO}) + 0.819})]^{-1},
	\label{equation1}
\end{equation}
where $T_{\mathrm{peak}}$ is the peak brightness temperature of the \element[][12]{CO} emission line. Secondly, the optical depth of the \element[][13]{CO} emission can be derived from the excitation temperature $T_{\mathrm{ex}}$ and the brightness temperature, $T_{\mathrm{MB}}(\element[][13]{CO})$, of the \element[][13]{CO} emission line  \citep[e.g.,][]{Pineda2010, Li2018, Ma2020} according to the following fomula, 
\begin{equation}
	\tau_{\mathrm{v}}^{\mathrm{13}} = -\ln \left\{ 1 - \frac{T_{\mathrm{MB}}(\element[][13]{CO})}{5.29[J(T_{\mathrm{ex}})-0.164]} \right\},
	\label{equation2}
\end{equation}
where $J(T_{\mathrm{ex}}) = [\mathrm{e}^{5.29/T_{\mathrm{ex}}} - 1]^{-1}$. Then, the \element[][13]{CO} column density can be calculated through \citep[e.g.,][]{Pineda2010, Li2018, Ma2020}
\begin{equation}
	N_{\mathrm{\element[][13]{CO}}} = 2.42\times10^{14}\frac{T_{\mathrm{ex}}+0.88}{1-\mathrm{e}^{-5.29/T_{\mathrm{ex}}}}\int \tau_{\mathrm{v}}^{13}\, \mathrm{d}v.
	\label{equation3}
\end{equation}
The item $T_{\mathrm{ex}}\int \tau_{\mathrm{v}}^{13}\, \mathrm{d}v$ can be approximated to $\frac{\tau_0}{1-\mathrm{e}^{-\tau_0}}\int T_{\mathrm{MB}}(\element[][13]{CO})\, \mathrm{d}v$, where $\tau_0$ is the optical depth at the brightness peak of the \element[][13]{CO} spectra \citep{Pineda2010}. We assume that the $\mathrm{H_2}$-to-\element[][13]{CO} ratio is $7\times10^5$ \citep[e.g.,][]{Solomon1972,Herbst1973,Wilson1999,Ma2020}. Finally, the $\mathrm{H_2}$ column density can be written as \citep{Li2018, Ma2020}:
\begin{equation}
	N_{\mathrm{H_2}} = 1.694\times10^{20}\frac{\tau_0}{1-\mathrm{e}^{-\tau_0}}\frac{1+0.88/T_{\mathrm{ex}}}{1- \mathrm{e}^{-5.29/T_{\mathrm{ex}}}}\int T_{\mathrm{MB}}(\element[][13]{CO})\,\mathrm{d}v.
	\label{equation4}
\end{equation}

\begin{figure}[!htb]
		\centering
		\includegraphics[trim = 0cm 5cm 0cm 5cm, width = 0.7\linewidth, clip]{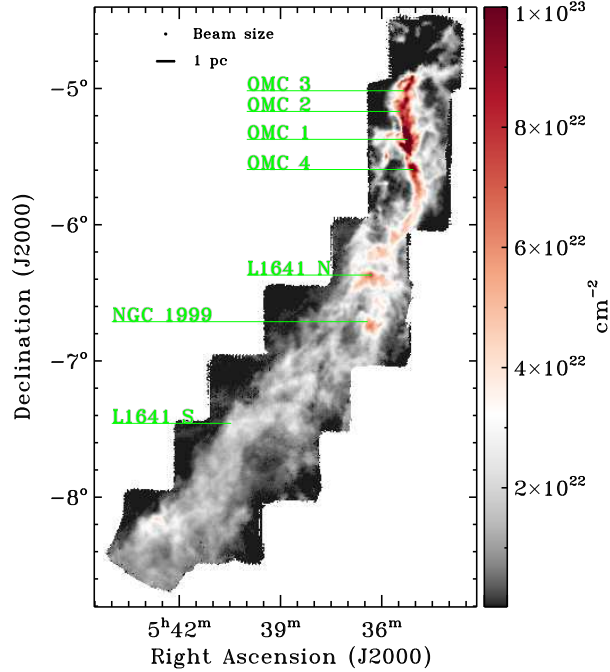}
	\caption{Spatial distribution of the $\mathrm{H_2}$ column density of Orion A GMC. Active star-forming regions in the GMC are indicated with green letters. The 1-pc scale and beam size are indicated at the upper-left corner of this figure.}
	\label{fig2}
\end{figure}

We only use the spectra that have at least five contiguous velocity channels with intensities above 1.5 times the RMS noise level for the calculation. The velocity range used to calculate the peak brightness temperature of the \element[][12]{CO} emission, $T_{\mathrm{peak}}$, optical depth, and the integrated intensity of \element[][13]{CO} emission, $\int T_{\mathrm{MB}}(\element[][13]{CO})\,\mathrm{d}v$, is from 0 to 18 km $\mathrm{s^{-1}}$. The spatial distribution of the $\mathrm{H_2}$ column density of the Orion A GMC is displayed in Figure \ref{fig2}, which is similar to Figure 3 in \citet{Ma2020}. 

\begin{figure*}[!htb]
	\centering
	\includegraphics[trim = 1cm 1.5cm 5cm 5cm, width = 0.6 \linewidth, clip]{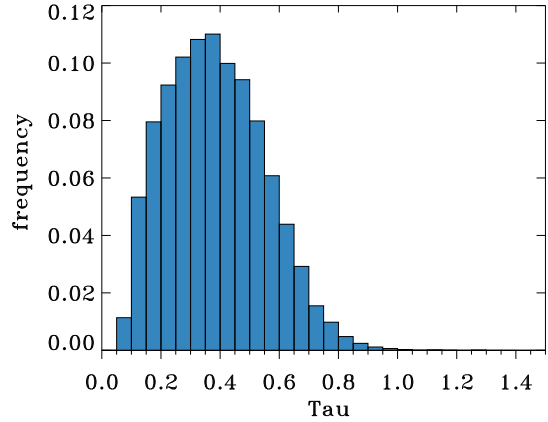}
	\caption{Histogram of the optical depth, $\tau_0(^{13}\mathrm{CO})$, at the peak of the $^{13}\mathrm{CO}$ J=1-0 emission line. The bin size is 0.05.}
	\label{figure3}
\end{figure*}

The optical depth of $^{13}$CO emission may introduce some uncertainties in the estimation of H$_2$ column density. For example, if the $^{13}$CO emission is optically thick, the observed $^{13}$CO brightness may saturate in regions with high column densities. We present the histogram of the optical depth of $^{13}$CO emission in Figure \ref{figure3}. We can see from Figure \ref{figure3} that the median $\tau_0$ in this region is $\sim$0.37, and only 0.1\% (23 pixels) of the \element[][13]{CO} spectra have a $\tau_0$ higher than unity. Moreover, the effect of $^{13}$CO emission optical depth on column density calculation has been corrected for through the use of the factor $\frac{\tau_0}{1-\mathrm{e}^{-\tau_0}}$ in equation \ref{equation4} \citep{Pineda2010}. 

\begin{figure}[!htb]
		\centering
		\includegraphics[trim = 0cm 5cm 0cm 5cm, width = 0.7\linewidth, clip]{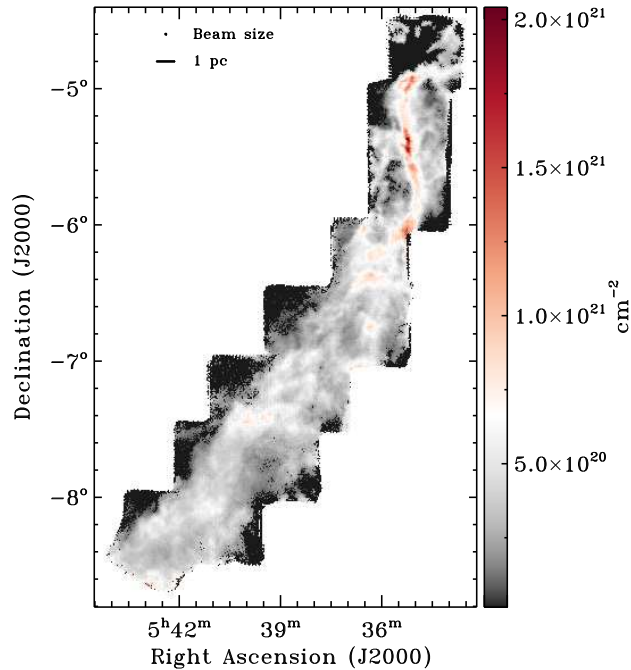}
	\caption{Spatial distribution of estimated standard deviation of calculated $\mathrm{H_2}$ column densities. The 1-pc scale and beam size are indicated at the upper-left corner of this figure.}
	\label{figg2}
\end{figure}

\subsection{Monte Carlo estimation for the measurement errors of the H$_2$ column density} \label{subsec3.2}
\begin{figure}[h]
	\begin{minipage}[t]{0.5\linewidth}
		\centering
		\includegraphics[trim = 0.2cm 0cm 0cm 0cm, width = \linewidth, clip]{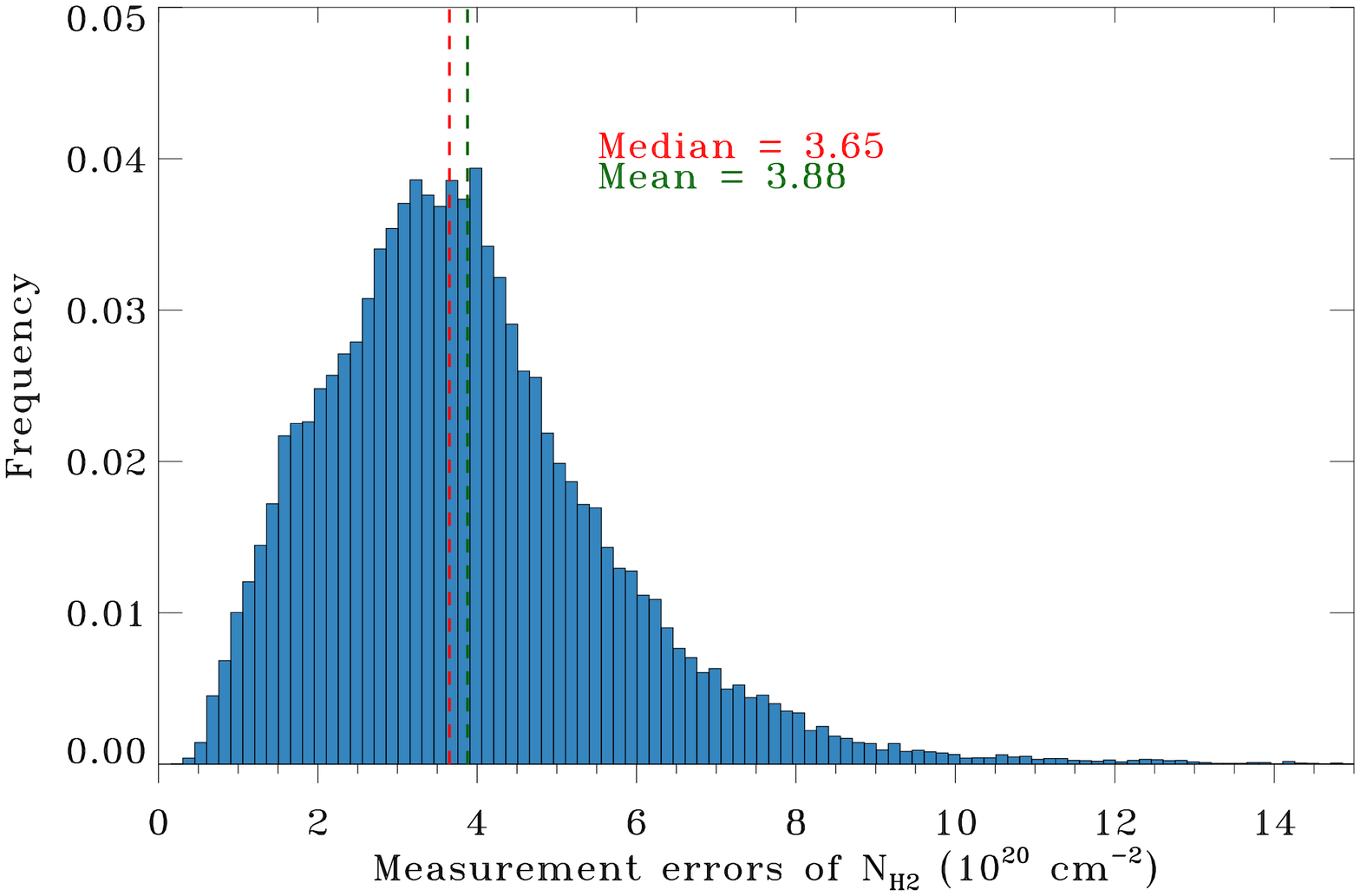}
		\\(a)
	\end{minipage}
	\begin{minipage}[t]{0.5\linewidth}
		\centering
		\includegraphics[trim = 0.2cm 0cm 0cm 0cm, width = \linewidth, clip]{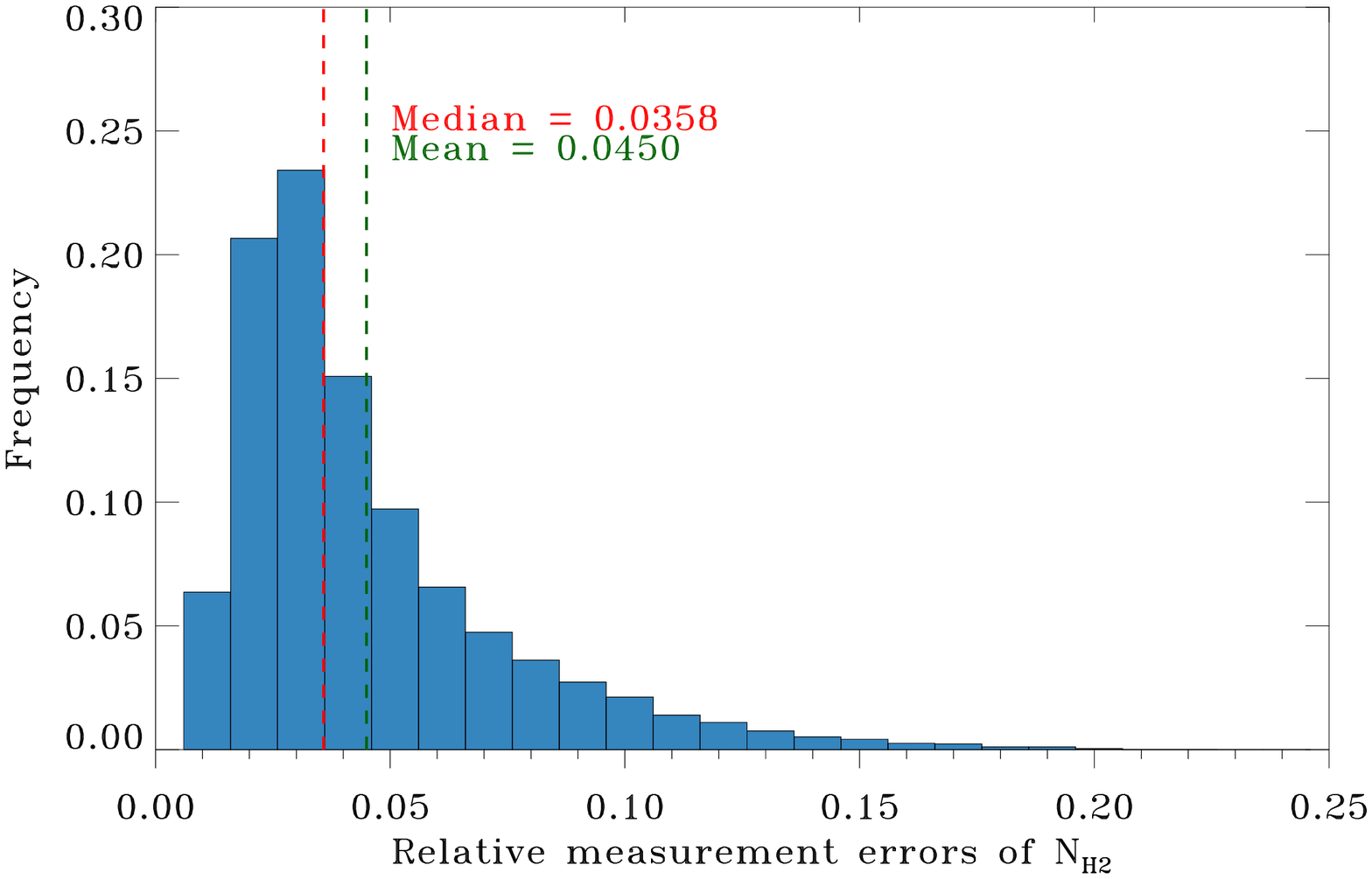}
		\\(b)
	\end{minipage}
	\caption{(a) Histogram of the measurement errors of calculated $\mathrm{H_2}$ column densities. The bin size is $1.5 \times 10^{19} \ \mbox{cm}^{-2}$. (b) Histogram of the relative measurement errors of calculated $\mathrm{H_2}$ column densities. The bin size is 0.01. The red dashed lines mark the median values, and the green dashed lines mark the mean values.}
	\label{fig3}
\end{figure}

Since we intend to derive the width of molecular filaments in the Orion A GMC, the radial column density profiles of the filaments are analyzed. One significant step is to average the radial profiles of the filament over a selected length (a few pixels in this work, see Section \ref{subsec5.1}) to improve the signal-to-noise ratio. Therefore, the measurement errors of the $\mathrm{H_2}$ column densities are needed for the calculation. We adopt a Monte Carlo method to estimate the error in the measurement of the $\mathrm{H_2}$ column density for each pixel. For each pixel, a set of values of the \element[][12]{CO} and \element[][13]{CO} peak brightness temperatures, and the \element[][13]{CO} integrated intensity are extracted randomly from the Gaussian distributions centered on the corresponding measured values of each quantity, and then the extraction is repeated for 2000 times. The dispersions of the Gaussian distributions from which we extracted the quantities are set to be the \element[][12]{CO} and \element[][13]{CO} RMS noise levels, and $\sigma(\element[][13]{CO})\sqrt{\Delta v\delta v}$, where $\sigma(\element[][13]{CO})$ is the \element[][13]{CO} RMS noise level, $\Delta v$ is the integrated velocity range, and $\delta v$ is the velocity resolution \citep{Ripple2013}. Two thousand values of N$_\mathrm{H{_2}}$ can be generated for each pixel. Then, for a given pixel, the dispersion of the two thousand N$_\mathrm{H{_2}}$ is considered as the measurement error for the N$_\mathrm{H{_2}}$ derived using the method in Section \ref{subsec3.1}. Figure \ref{figg2} shows the spatial distribution of the estimated measurement errors, and Figure \ref{fig3} shows the histogram of the measurement errors and the histogram of the relative errors, which are defined as the ratios between the measurement errors and the N$_\mathrm{H{_2}}$. From Figure \ref{fig3}, we can see the median measurement error is $3.65 \times 10^{20}\ \mathrm{cm^{-2}}$, and the mean is $3.88 \times 10^{20}\ \mathrm{cm^{-2}}$, while the median and mean of the relative errors are  3.58\% and 4.50\%, respectively. 

\subsection{Column density maps from different tracers}
\begin{figure}[h]
	\begin{minipage}[t]{0.3\linewidth}
		\centering
		\includegraphics[trim = 2.3cm 5cm 3.1cm 5cm, width = \linewidth, clip]{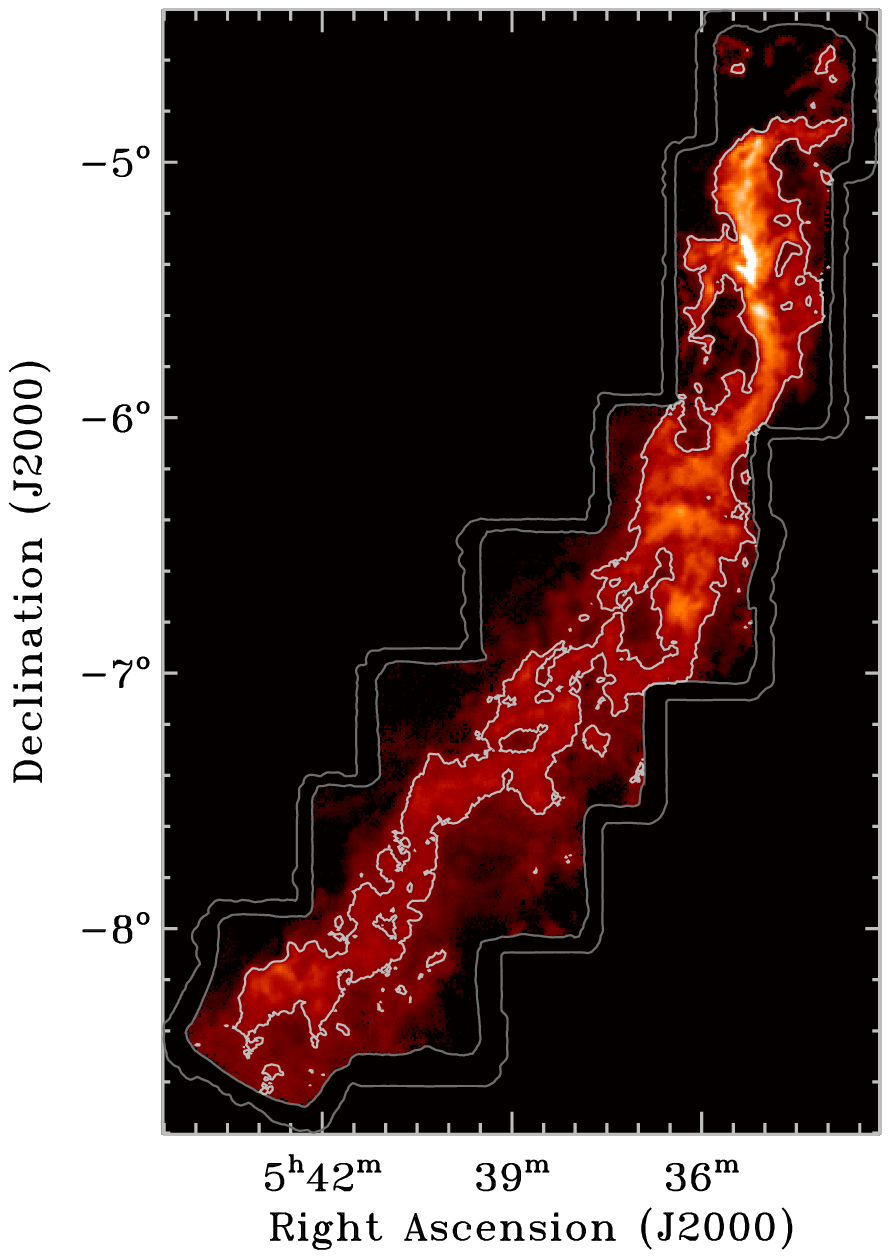}
		\\(a)
	\end{minipage}
	\begin{minipage}[t]{0.3\linewidth}
		\centering
		\includegraphics[trim = 2.3cm 5cm 3.2cm 5cm, width = \linewidth, clip]{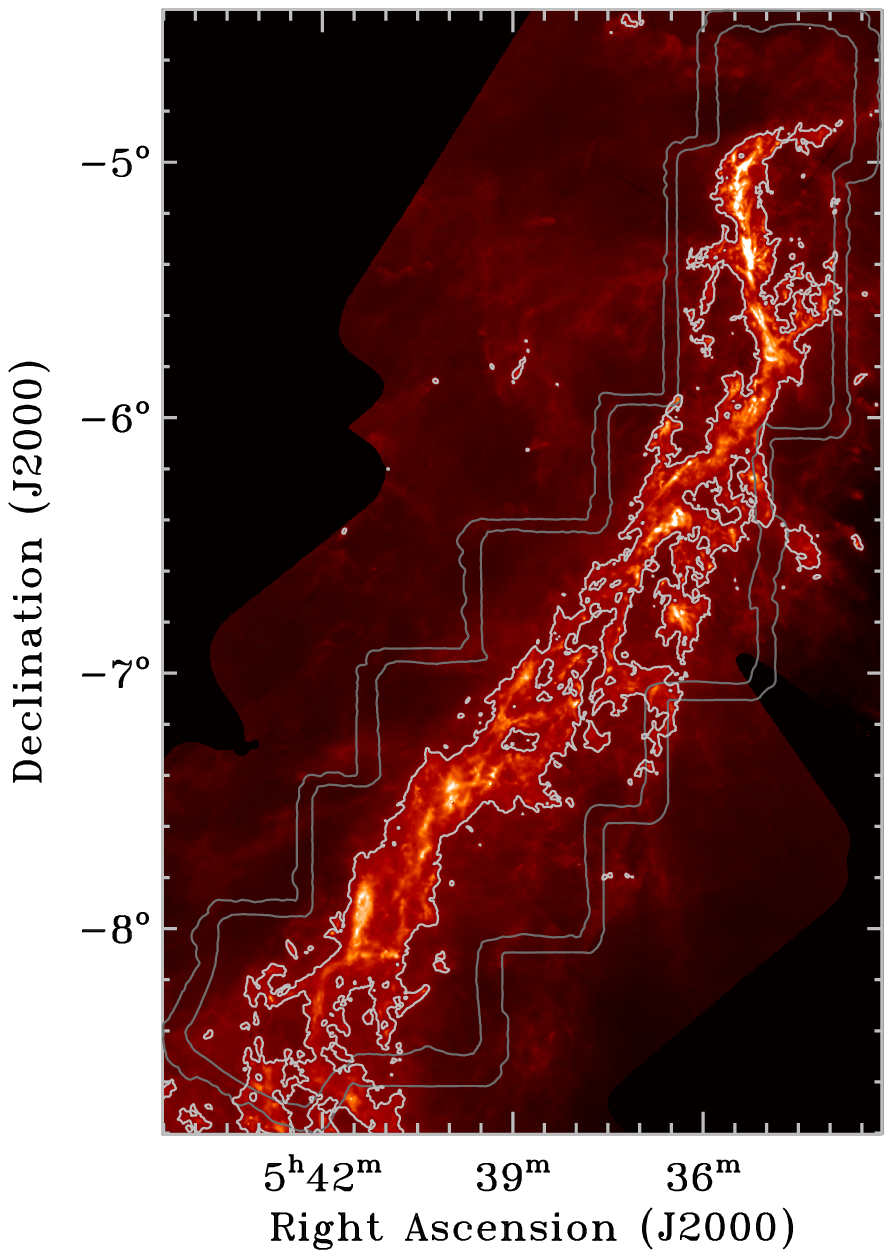}
		\\(b)
	\end{minipage}
	\begin{minipage}[t]{0.37\linewidth}
		\centering
		\includegraphics[trim = 2.3cm 5cm 1cm 5cm, width = \linewidth, clip]{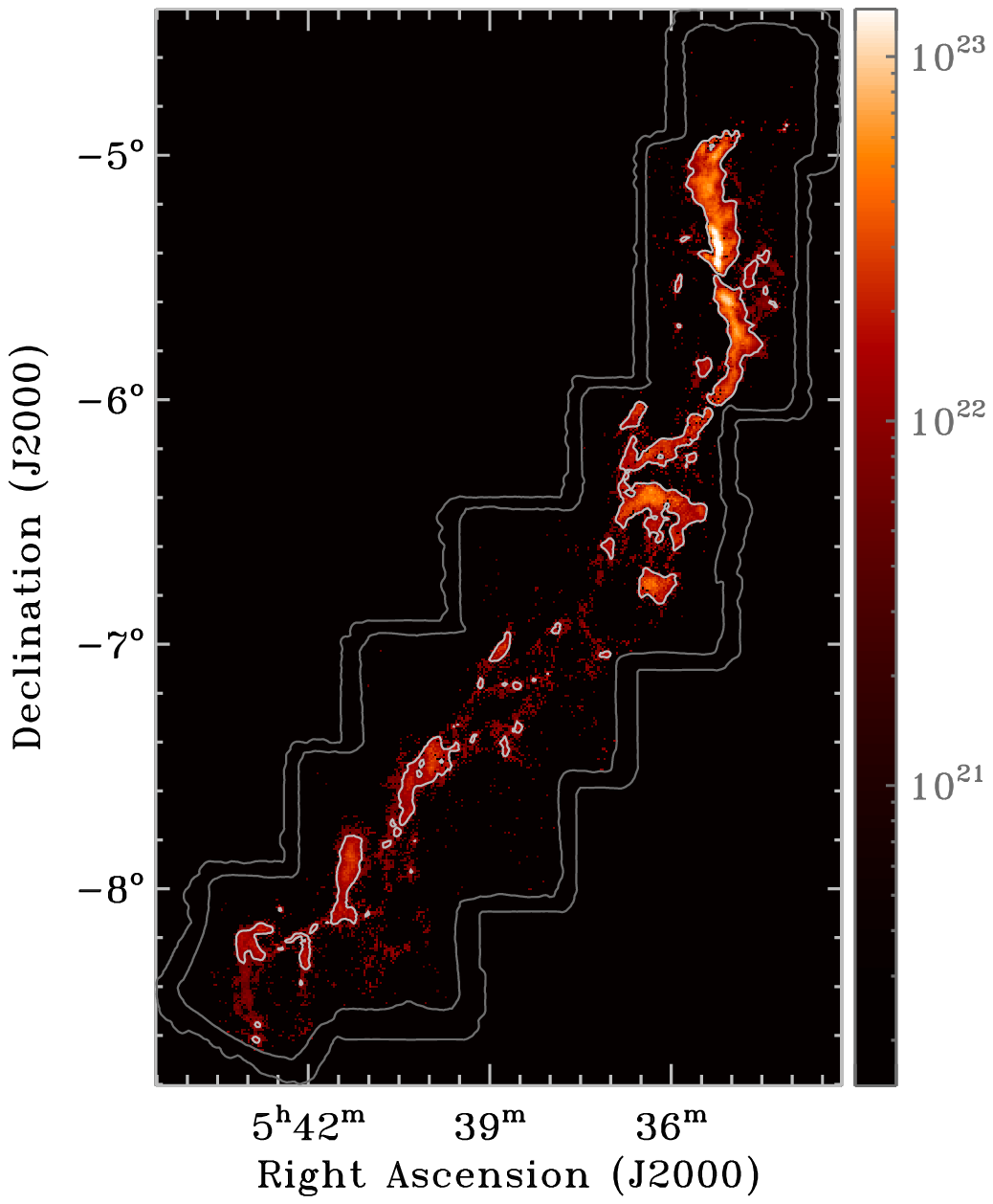}
		\\(c)
	\end{minipage}
	\caption{(a) $\mathrm{H_2}$ column density map derived from $^{13}$CO J=1-0 emission. (b) $\mathrm{H_2}$ column density map derived from Herschel observation (multiplied by three). (c) $\mathrm{H_2}$ column density map derived from C$^{18}$O J=1-0 emission. The color scales in the three panels are identical and range from $1.5\times10^{20}$ cm$^{-2}$ to $1.35\times10^{23}$ cm$^{-2}$. The grey lines outline the noisy edge of the surveyed area in the MWISP observation that has been trimmed in the analysis. The white contours correspond to N(H$_2$)$=1.25\times10^{22}$ cm$^{-2}$.}
	\label{col}
\end{figure}

\begin{figure}[!htb]
	\centering
	\includegraphics[trim = 0cm 0cm 0cm 0cm, width = 0.7\linewidth, clip]{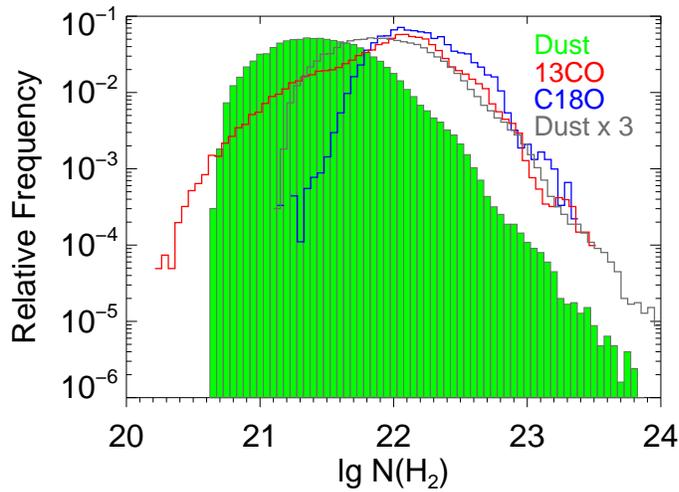}
	\caption{Histogram of column densities. The green columns represent the H$_2$ column densities from dust emission. The grey line represents the H$_2$ column densities from dust emission multiplied by three. The red and the blue lines represent the column densities from the $^{13}$CO and C$^{18}$O J=1-0 emissions, respectively. The bin size of these histograms is 0.05.}
	\label{col_hist}
\end{figure}

Before implementing filament identification, we made a brief comparison between the H$_2$ column density maps derived using the \element[][13]{CO} J=1-0 line emission, far-infrared dust emission, and C\element[][18]{O} J=1-0 line emission as the tracers respectively. These column density maps are presented in Figure \ref{col} with the same color scale.  The column density map in Figure \ref{col}(b) is converted from the N(H) map from Herschel observations which is obtained by private contact with \citet{Stutz2018}. We converted N(H) to N(H$_2$) by dividing a constant of two. For Figure \ref{col}(b) to have the same scale as Figure \ref{col}(a) and \ref{col}(c), the N(H$_2$) map in Figure \ref{col}(b) has been multiplied by a factor of three. The histograms of column densities  from the three tracers within the same area as indicated by the grey contours in Figure \ref{col} are given in Figure \ref{col_hist}. The H$_2$ column density from C$^{18}$O data is calculated according to formula 8 in \citet{Li2018}. The N(H$_2$) distributions from the $^{13}$CO and C$^{18}$O tracers have similar shapes above their peaks, while the distribution of N(H$_2$) traced by dust emission is systematically shifted toward the lower column density when compared with the other two tracers. After multiplying the N(H$_2$) from dust emission by three, the N(H$_2$) distribution match well with those from $^{13}$CO and C$^{18}$O in the range from $\sim10^{22}$ cm$^{-2}$ to 3$\times10^{23}$ cm$^{-2}$. \citet{Stutz2013} and \citet{Launhardt2013} suggested that a systematic uncertainty could exist when using different dust opacity models in the calculation of N(H). Furthermore, the adopted gas-to-dust ratios used in \citet{Stutz2015} and \citet{Stutz2018} and the H$_{2}$-to-$^{13}$CO ratio used in this work may also introduce a systematic difference between the column densities derived from dust emission and those from molecular line emission.   The white contours in Figure \ref{col} corresponds to N(H$_2$)$=1.25\times10^{22}$ cm$^{-2}$, which is the average column density in the southern sub-regions \citep{Ma2020} and is also approximate to the peaks of the blue, red, and grey histograms in Figure \ref{col_hist}. In Figure \ref{col}, the Herschel N(H$_2$) map shows the cloud structures in more detail than the $^{13}$CO and C$^{18}$O N(H$_2$) maps because of its higher spatial resolution. However, the main bright structures, such as the ISF, the L1641 regions, and the southern filaments in the Orion A GMC, are quite consistent in the three maps.

Filaments from different tracers may exhibit different properties. The $^{13}$CO line emission is a good tracer of gas column density. The main features in Figure \ref{col}(a) are consistent with those in Figure \ref{col}(b) and \ref{col}(c).  With less abundance in $^{13}$CO than in C$^{18}$O, the N(H$_2$) map from C$^{18}$O emission exhibits much fewer filaments than the N(H$_2$) map from $^{13}$CO. Furthermore, the C$^{18}$O filaments in Figure \ref{col}(c) usually do not have enough pixels across the filaments which are essential for profile symmetry analysis and filament width calculation.  Therefore, in this work, we perform filament identification and property analysis on the H$_2$ column density map derived from $^{13}$CO data.


\section{Filament identification} \label{sec4}
\subsection{DisPerSE algorithm} \label{subsec4.1}
We use the DisPerSE algorithm \citep{Sousbie2011} to identify filamentary structures on the $\mathrm{H_2}$ column density map. The DisPerSE  can be applied to identify persistent topological structures, such as voids, walls, and filamentary structures \citep{Sousbie2011}. Its first step of filament identification is to find all critical points in the $\mathrm{H_2}$ column density map. Critical points are the set of points where the gradient of the $\mathrm{H_2}$ column density is null. In a column density map, there are three types of critical points, the maxima, the saddle points, and the minima \citep{Sousbie2011}. The second step is to connect the saddle points and the maxima along the integral lines, which are the curves tangent to the gradient field in each point in the map \citep{Sousbie2011}. The part of integral lines that connect the saddle points and maxima are filament candidates in the $\mathrm{H_2}$ column density map.  
\begin{figure*}[h]
	\begin{minipage}[t]{0.56\linewidth}
		\centering
		\includegraphics[trim = 2cm 5cm 0cm 5cm, width = \linewidth, clip]{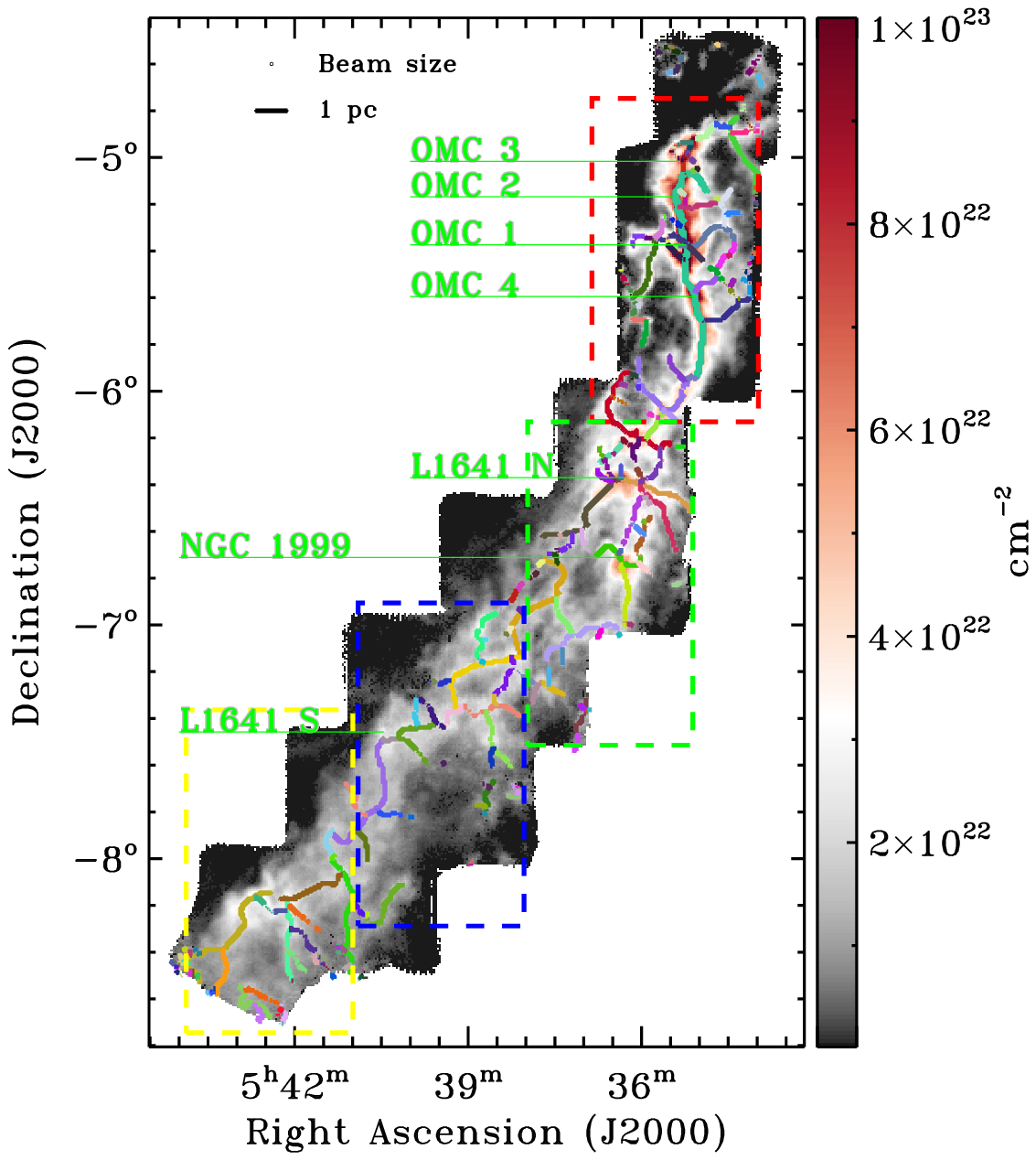}
		\\(a)
	\end{minipage}
	\begin{minipage}[t]{0.32\linewidth}
		\centering
		\includegraphics[trim = 8.5cm 5.5cm 7.9cm 5cm, width = \linewidth, clip]{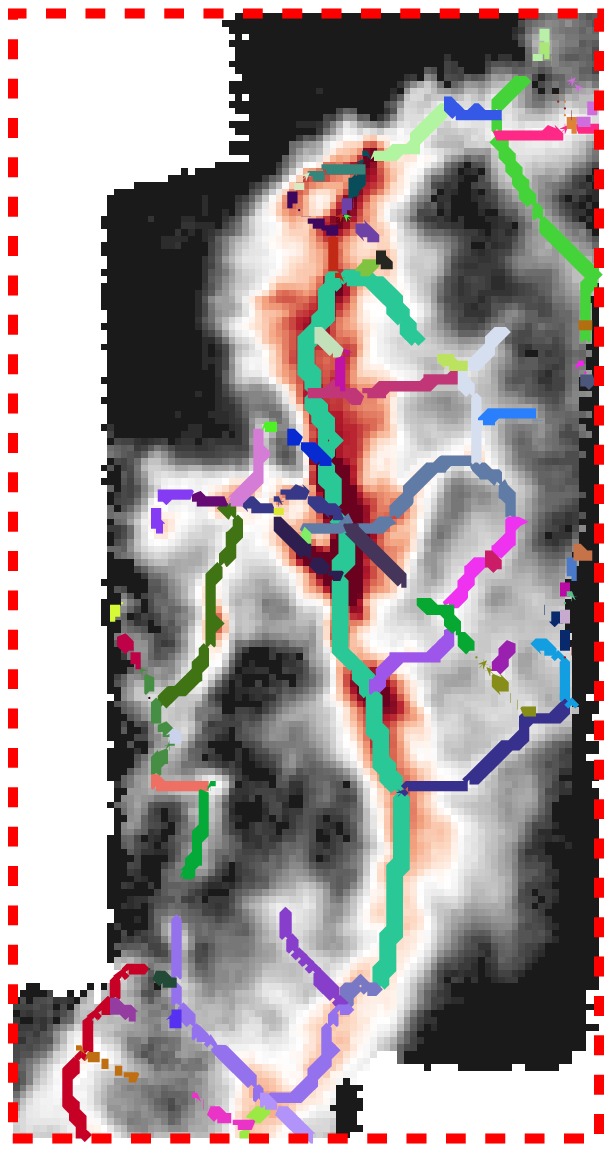}
		\\(b)
	\end{minipage}
	\begin{minipage}[t]{0.33\linewidth}
		\centering
		\includegraphics[trim = 8.5cm 6cm 7.9cm 5cm, width = \linewidth, clip]{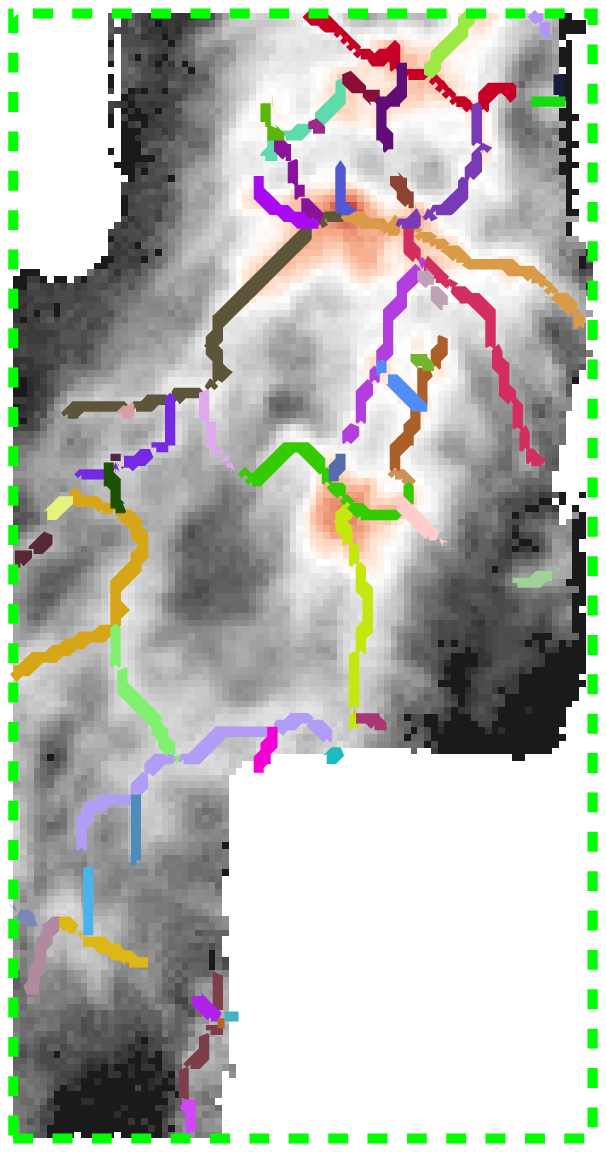}
		\\(c)
	\end{minipage}
	\begin{minipage}[t]{0.33\linewidth}
		\centering
		\includegraphics[trim = 8.5cm 6cm 7.9cm 5cm, width = \linewidth, clip]{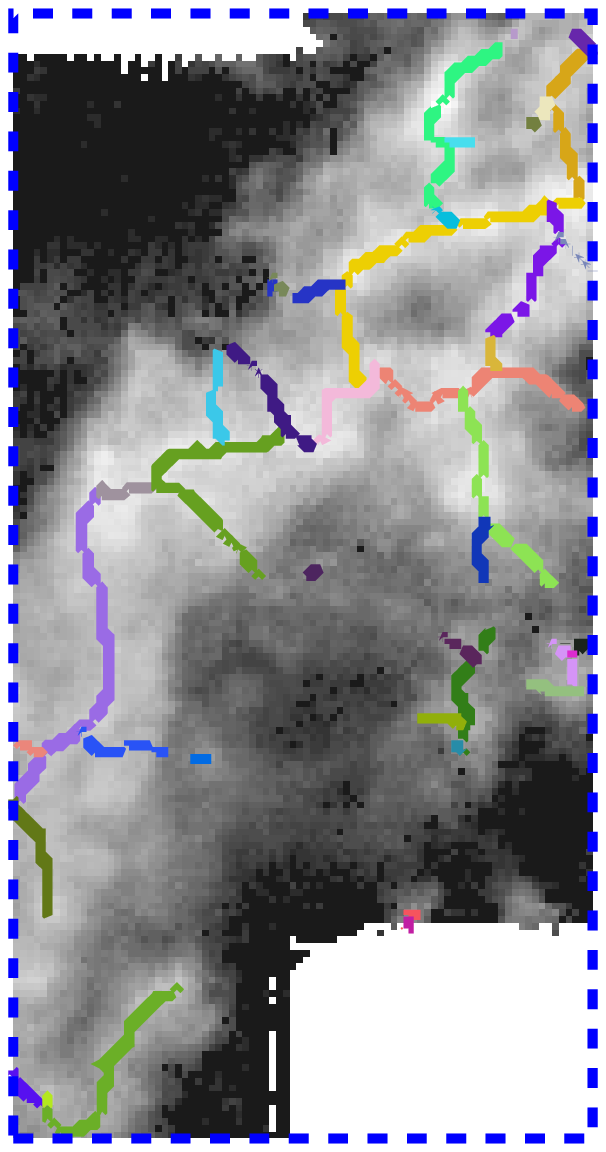}
		\\(d)
	\end{minipage}
	\begin{minipage}[t]{0.33\linewidth}
		\centering
		\includegraphics[trim = 8.5cm 6cm 7.9cm 5cm, width = \linewidth, clip]{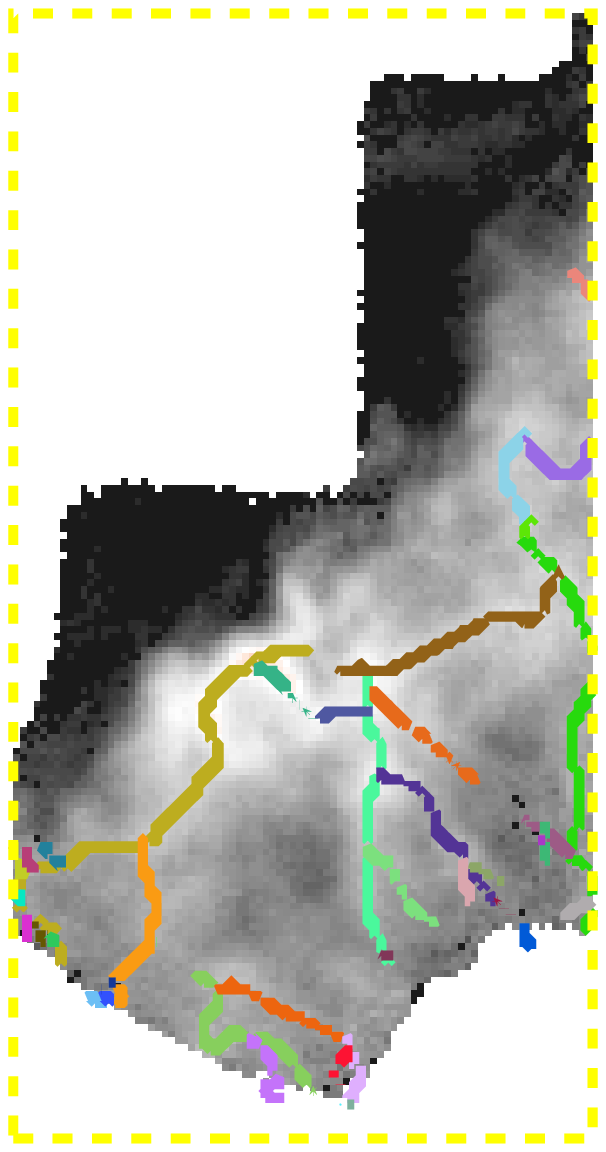}
		\\(e)
	\end{minipage}
	\caption{(a) Spatial distribution of the 225 filament skeletons identified in Orion A GMC. Different colors  represent different filament skeletons. The background is the  map of the $\mathrm{H_2}$ column density of Orion A GMC. (b)-(e) are enlarged views of the red, green, blue and yellow boxes in panel a, respectively.}
	\label{fig4}
\end{figure*}

Two threshold parameters, the persistence and the robustness, need to be set for the selection of filaments. Persistence \citep{Sousbie2011} is the absolute  value difference of a pair of the saddle point and maximum in the filament candidate. The persistence threshold can be used to eliminate noise and non-physical structures. Robustness \citep{Weinkauf2009,Sousbie2011} can be understood as the contrast between the column density of the filament and that of the environment surrounding the filament. The persistence threshold we set is $3.5 \times 10^{21}\ \mathrm{cm^{-2}}$, which corresponds to about nine times the noise level of H$_2$ column density.  The robustness threshold is set to be $8 \times 10^{21}\ \mathrm{cm^{-2}}$, which corresponds to approximately twenty times the noise level of H$_2$ column density. We choose such high threshold settings to make sure that the resulting filaments are real internal structures of the Orion A GMC.

\begin{figure}[h]
	\centering
    \includegraphics[trim = 0.5cm 0cm 1.5cm 0cm, width = 0.6\linewidth, clip]{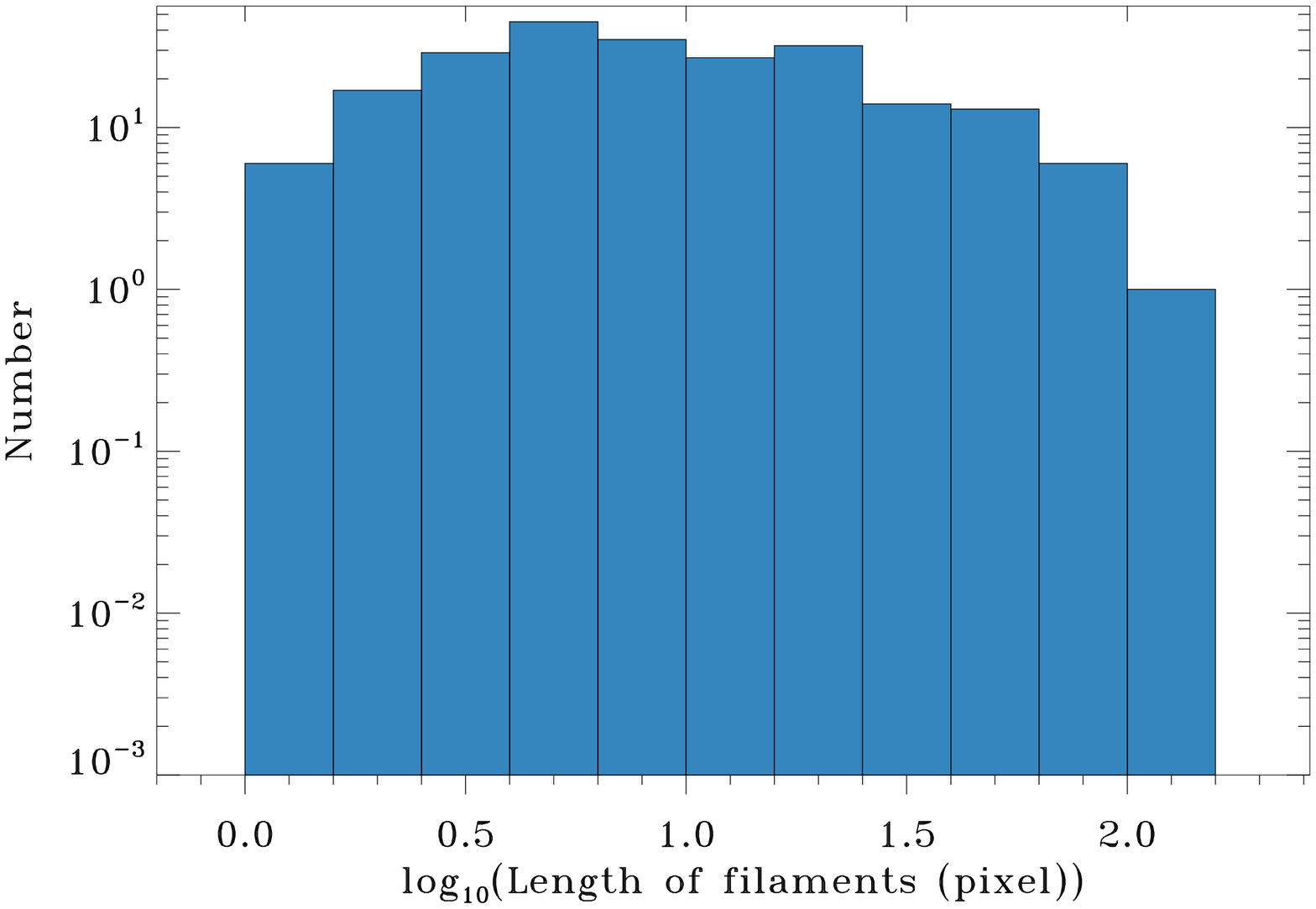}
	\caption{Histogram of the lengths of the 225 filament skeletons identified in Orion A GMC in logarithmic scale. The bin size is  0.2.}
	\label{fig5}
\end{figure}

\subsection{Selection and partition of filaments for analysis of filament density profiles} \label{subsec4.2}
\begin{figure}[h]
	\begin{minipage}[t]{0.56\linewidth}
		\centering
		\includegraphics[trim = 2cm 5cm 0cm 4cm, width = \linewidth, clip]{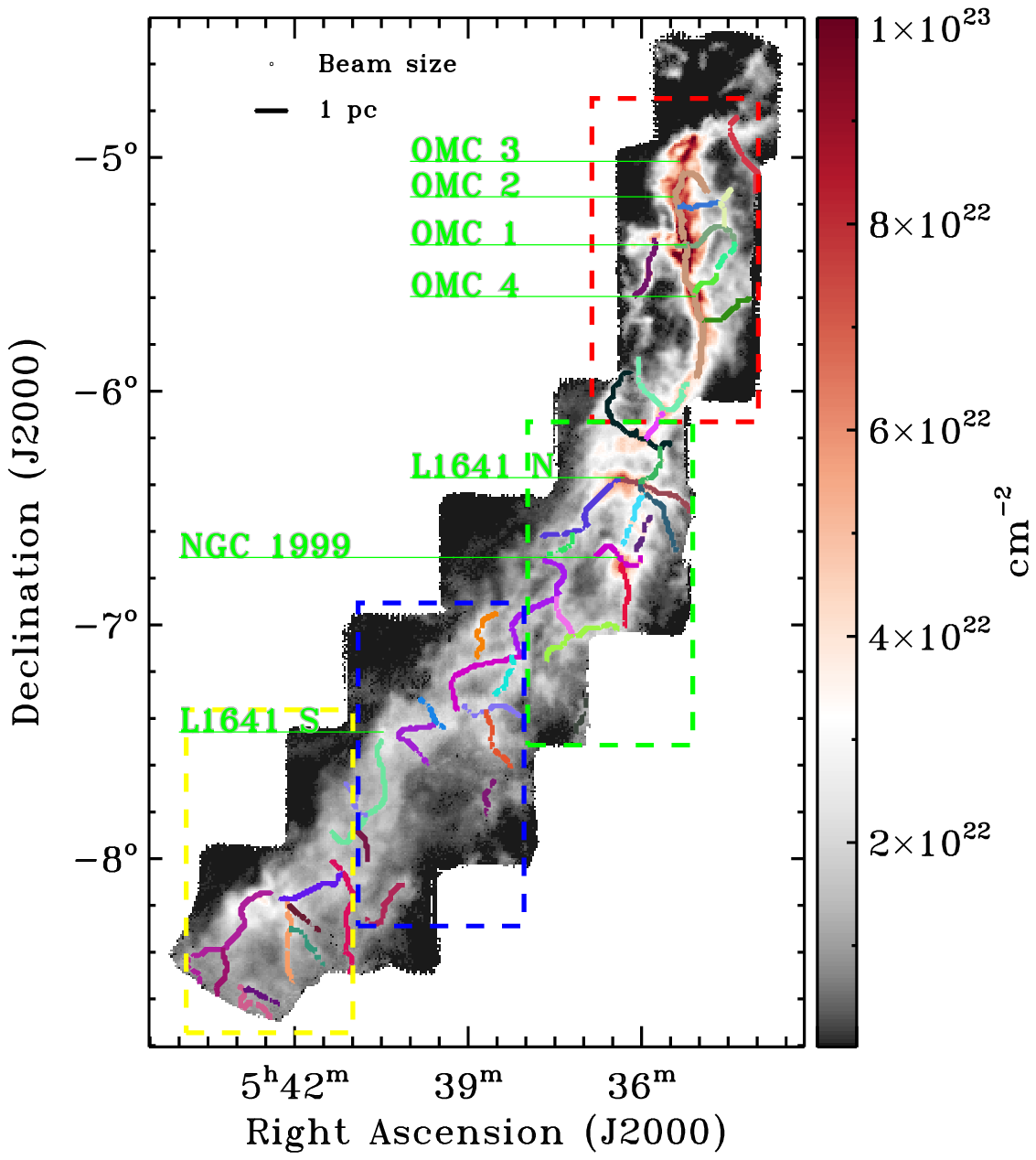}
		\\(a)
	\end{minipage}
	\begin{minipage}[t]{0.32\linewidth}
		\centering
		\includegraphics[trim = 8.5cm 5.5cm 7.9cm 5cm, width = \linewidth, clip]{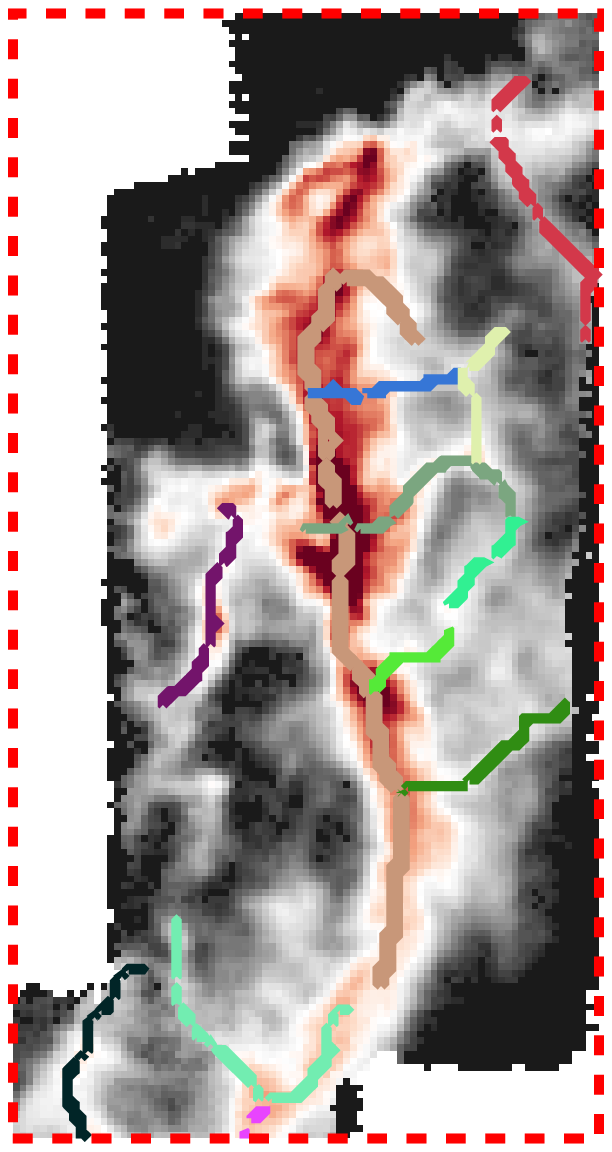}
		\\(b)
	\end{minipage}
	\begin{minipage}[t]{0.33\linewidth}
		\centering
		\includegraphics[trim = 8.5cm 6cm 7.9cm 5cm, width = \linewidth, clip]{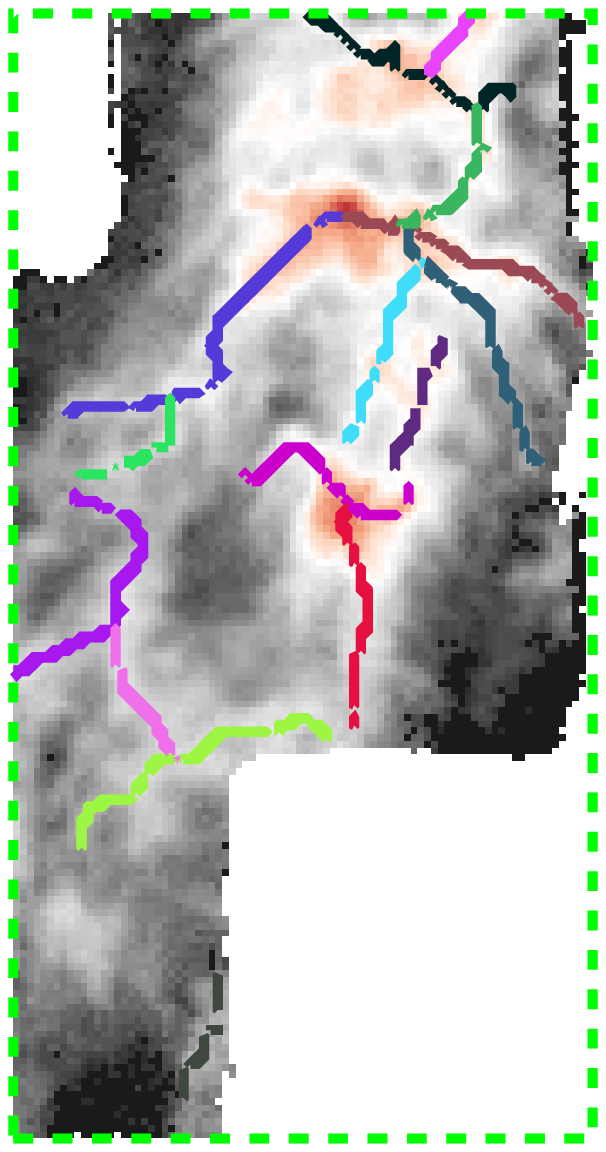}
		\\(c)
	\end{minipage}
	\begin{minipage}[t]{0.33\linewidth}
		\centering
		\includegraphics[trim = 8.5cm 6cm 7.9cm 5cm, width = \linewidth, clip]{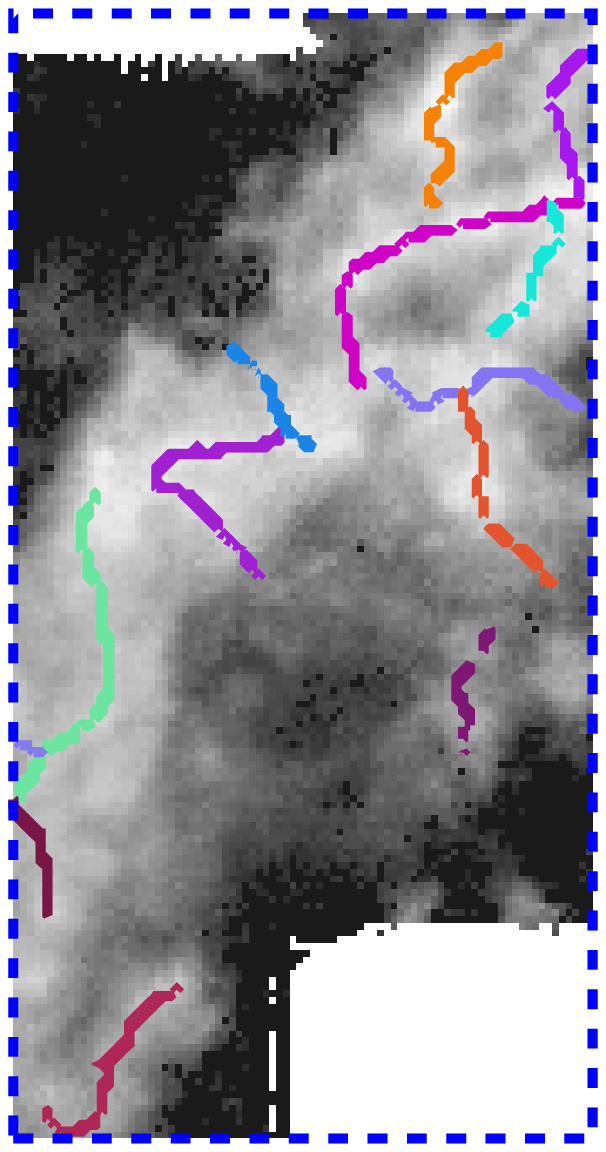}
		\\(d)
	\end{minipage}
	\begin{minipage}[t]{0.33\linewidth}
		\centering
		\includegraphics[trim = 8.5cm 6cm 7.9cm 5cm, width = \linewidth, clip]{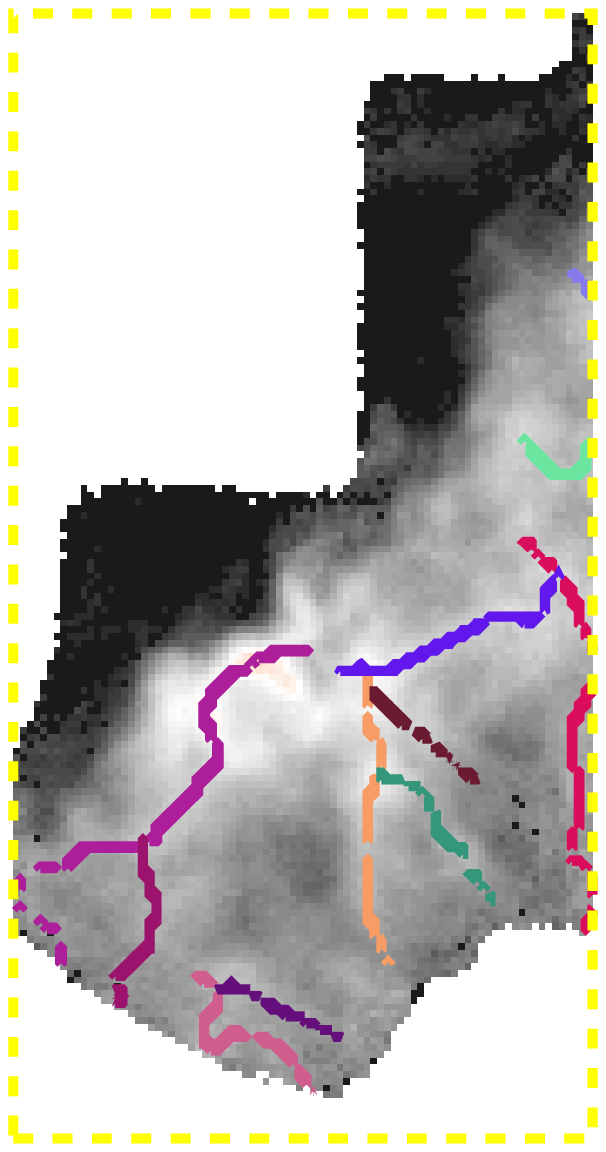}
		\\(e)
	\end{minipage}
	\caption{(a) Spatial distribution of 46 long filament skeletons. Different colors mark and distinguish different filament skeletons. The background is the spatial distribution of the $\mathrm{H_2}$ column density of Orion A GMC. (b)-(e) are enlarged views of the red, green, blue and yellow boxes in panel a, respectively.}
	\label{fig6}
\end{figure}

We identified 225 filaments in the $\mathrm{H_2}$ column density map of the Orion A GMC in total. Figure \ref{fig4} and \ref{fig5} show the spatial distribution and the histogram of the lengths of the 225 identified filaments, respectively. As shown in Figure \ref{fig4}, the identified skeletons of the filaments depict the overall internal structures of the Orion A GMC.  According to \citet{Andre2014}, filaments are defined as elongated overdense interstellar medium (ISM) structures with an aspect ratio larger than 5-10 . In molecular clouds such as Taurus and Polaris, the filaments have lengths of $\sim$1 pc or more \citep{Jackson2010, Beuther2011, Andre2014}. So, we selected filaments with lengths no less than 1.2 \mbox{pc} ($\sim$ 20 pixels) for further analysis to ensure no ``fake filaments'' in our sample. The spatial distribution of the selected 46 long filament skeletons is shown in Figure \ref{fig6}. For a selected filament, its local surrounding environment, and physical properties, such as width and central density, are not necessarily invariable.  Therefore, we divide these filaments into 397 segments for investigation of the properties of the column density profiles to keep the local characteristics of the filaments and to avoid the influence of the environment. Among the 397 segments, 356 segments have the length of five pixels ($\sim$ 0.30 \mbox{pc}, three times the spatial resolution of the observation) and 41 segments are at the ends of filaments and therefore have lengths less than five pixels. The 397 segments are the targets for detailed analysis of filamentary density profiles. 

\section{Column density profile analysis} \label{sec5}
To investigate the width of the identified filaments, we calculate the radial $\mathrm{H_2}$ column density profiles for each segment.

\subsection{Calculation of the column density profiles} \label{subsec5.1}
For a point in a given skeleton segment, the radial column density profile is extracted from the slice that is perpendicular to the line that connects the neighboring skeleton pixels on each side of the point. For each slice, because the skeleton point from by the DisPerSE algorithm may deviate from the local maxima of the column densities within two spatial pixels, 10 pixels along each side of the intensity peak of the slice including the peak are selected to calculate the column density profile. We apply this method to each point in the skeleton segments, which means that five, or fewer slices in some cases when the segment is at the end of a filament, are extracted for the calculation of the radial profile.

The $\mathrm{H_2}$ column density profile of each slice is normalized by the value of $N_{\mathrm{H_2}}$ at the skeleton point of the slice: 
\begin{equation}
	N_{r,i} = N_{r,i,\mathrm{raw}}/N_{0,i,\mathrm{raw}}
	\label{equation5}
\end{equation}
where $N_{r, i}$ and $N_{r,i,\mathrm{raw}}$ are the normalized value and the $\mathrm{H_2}$ column density at position $r$ in the $i$ th slice of the segment, respectively. $N_{0,i,\mathrm{raw}}$ is the value of $N_{\mathrm{H_2}}$ at the skeleton. The column density profile of each segment is the average of the profiles of the slices of the segment,
\begin{equation}
	\overline{N_r} = \sum\limits_{i}N_{r,i}/n,
	\label{equation6}
\end{equation}
where $n$ represents the number of the slices of the segment. $\overline{N}_r$ is the mean of the $\mathrm{H_2}$ column density at the position of distance $r$ from the central skeleton point. According to error propagation, the error of the normalized column density, $\sigma_{r, i}$,  at position $r$ in the $i$ th slice can be derived through \citep{Book2003},
\begin{equation}
	\sigma_{r,i} = N_{r,i}(\sigma_{r,i,\mathrm{raw}}^2/N_{r,i,\mathrm{raw}}^2 + \sigma_{0,i,\mathrm{raw}}^2/N_{0,i,\mathrm{raw}}^2)^{1/2},
	\label{equation7}
\end{equation}
where $\sigma_{r,i,\mathrm{raw}}$ is the measurement error at the position of distance $r$ from the skeleton point, and $\sigma_{0,i,\mathrm{raw}}$ is the measurement error of the skeleton point. The error of the averaged normalized column density, $\sigma_{r}$, at position $r$ can be expressed as \citep{Book2003},
\begin{equation}
	\sigma_r = (\sum\limits_{i}\sigma_{r,i}^2/n^2)^{1/2},
	\label{equation8}
\end{equation}
where $\sigma_r$ is the error of the normalized column density at position $r$.

Since the filament skeletons are identified in discrete and gridded data, the direction of the extracted slices may deviate from the real direction. To find out this influence on the filament profiles, we have done a smoothing test on the ISF skeleton. For each pixel in the ISF skeleton, we calculated its smoothed location using the locations of its nearest three pixels on each side and itself. In this way, we can obtain a smoothed skeleton. We find that there is no significant difference between the averaged column density profiles obtained from the smoothed skeleton and the original skeleton. Therefore, the influence from discrete and gridded data is negligible in the following analysis and results, and we use the original profiles in this work for simplicity.

\subsection{Symmetries of the column density profiles of the segments}\label{subsec5.2}
In practice, only the symmetrical column density profiles can be well fitted with a selected function, either Gaussian or Plummer-like function, therefore we check the observational symmetries of each H$_2$ column density profile and investigate the influence of the environments on the observed profile to get the intrinsic symmetry of the profile. We check whether the pixels in the slices of a selected segment are contaminated by any other filamentary segments. For this check, we restrict the slice length to 7 pixels on each side of the peak. If the slices of a selected segment intersect with other segment spines, we consider the profile of the slice is contaminated. Then we re-calculate the column density profile of the segment after removing all the slices that are considered to be contaminated, and check again the symmetry of the resulting profile.

\begin{figure}[!htb]
	\begin{minipage}[h]{0.5\linewidth}
		\centering
		\includegraphics[trim = 0cm 1cm 1cm 1cm, width = \linewidth, clip]{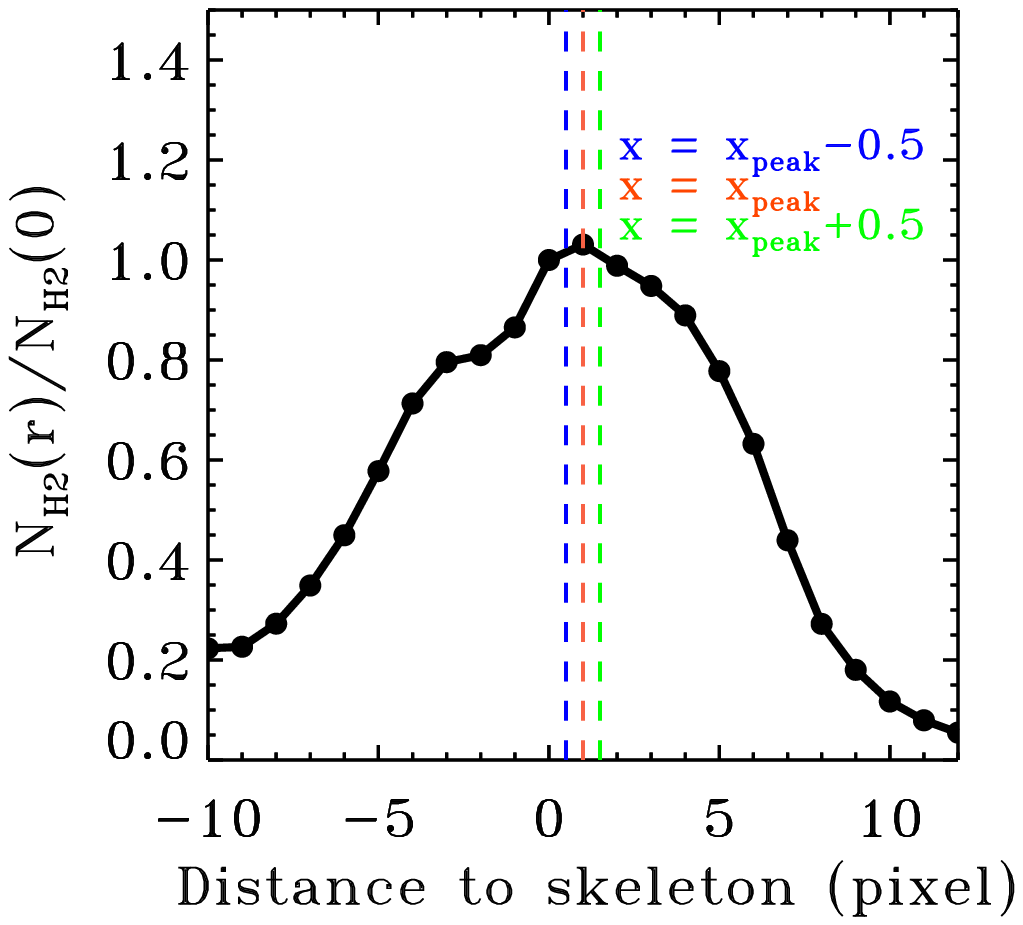}
		\\(a)
	\end{minipage}
	\begin{minipage}[h]{0.5\linewidth}
		\centering
		\includegraphics[trim = 0cm 1cm 1cm 1cm, width = \linewidth, clip]{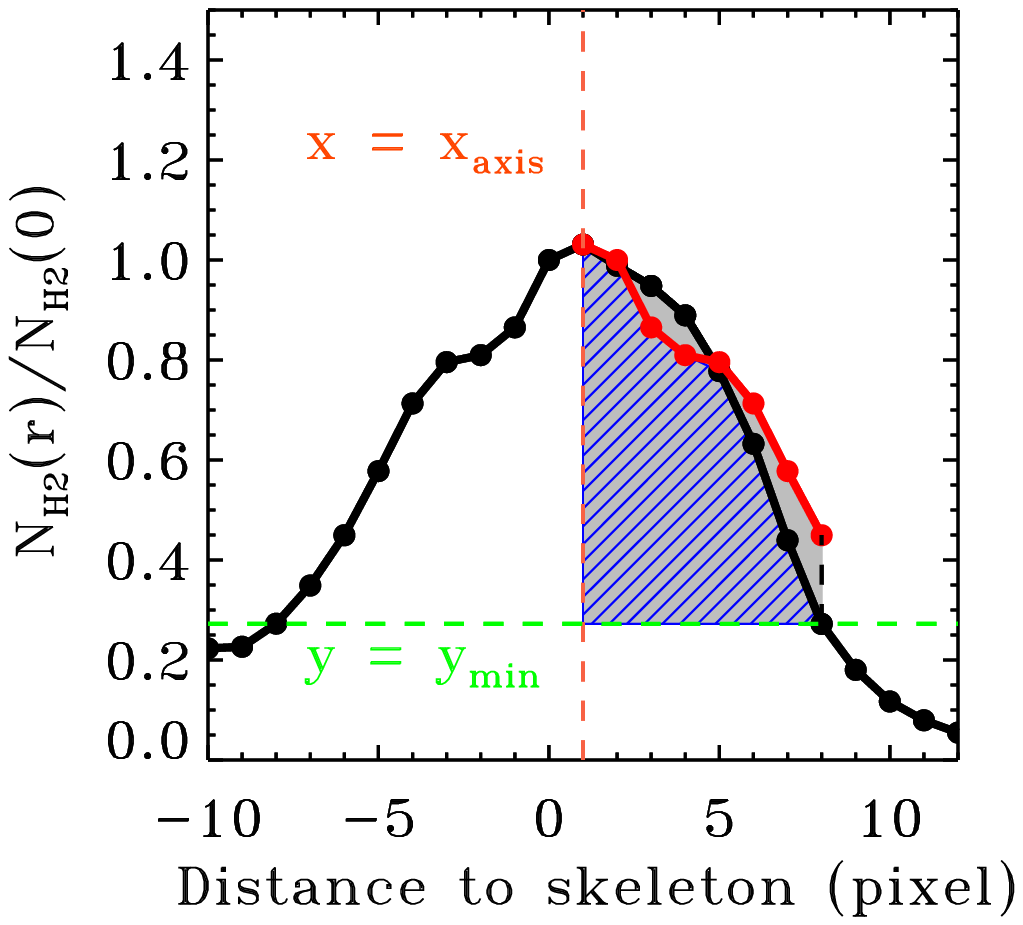}
		\\(b)
	\end{minipage}
	\caption{(a) Illustration for ``axis of symmetry'' candidates. The black line marks the column density profile. Three dash lines represent the ``axis of symmetry'' candidates. (b) Illustration for calculation of degree of symmetry. The red line marks the ``folded'' 7-pixel curves. The red dash line is the ``axis of symmetry''$x = x_{\mathrm{axis}}$. The zones in panel b filled by blue slashes and grey shadows are $S_{\mathrm{\cap}}$ and $S_{\mathrm{\cup}}$ sets, respectively.}
	\label{fig7} 
\end{figure}
\begin{figure}[!htb]
	\begin{minipage}[h]{0.5\linewidth}
		\centering
		\includegraphics[trim = 4.5cm 2.5cm 5cm 3cm, width = \linewidth, clip]{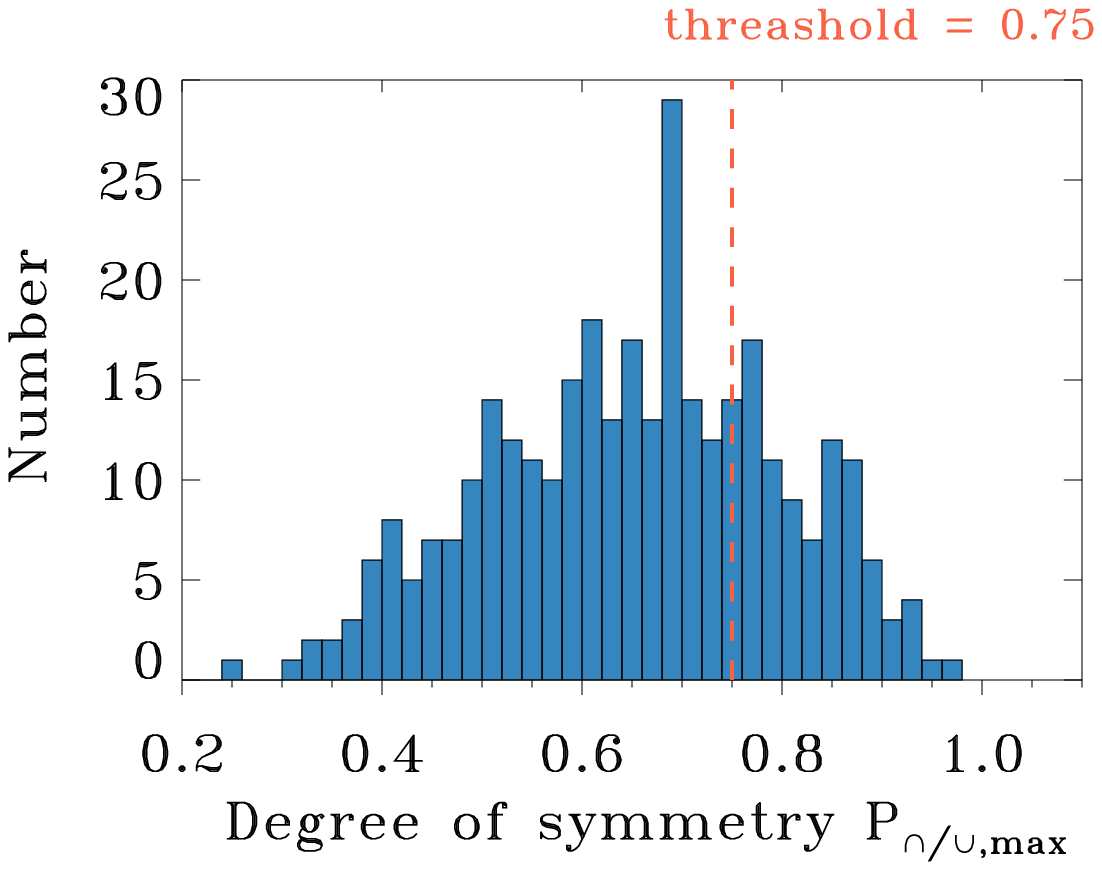}
		\\(a)
	\end{minipage}
	\begin{minipage}[h]{0.5\linewidth}
		\centering
		\includegraphics[trim = 4.5cm 2.5cm 5cm 3cm, width = \linewidth, clip]{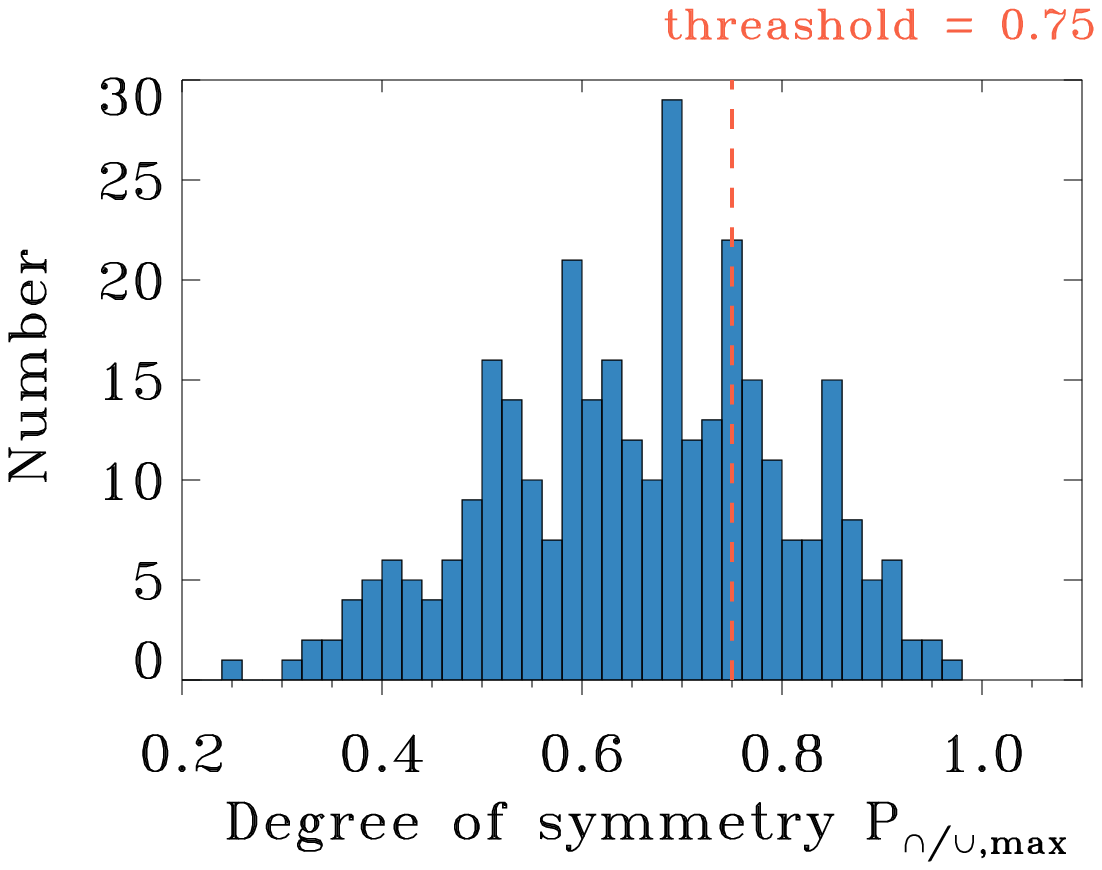}
		\\(b)
	\end{minipage}
	\caption{(a) Histogram of the degree of symmetry $P_{\cap/\cup,\mathrm{max}}$ of column density profiles before eliminating the contaminated slices. (b) Histogram of the degree of symmetry $P_{\cap/\cup,\mathrm{max}}$ of column density profiles after eliminating the contaminated slices. The red dashed lines mark the threshold to determine whether a profile is symmetrical or asymmetrical. The bin sizes in the two panels are both 0.02.}
	\label{fig8}
\end{figure}
We have applied a quantitative method to measure the symmetries of the averaged column density profiles. As shown in Figure \ref{fig7}(a), for an average column density profile, we have tried three positions for the ``axis of symmetry'' $x_{\mathrm{axis}}$:
	\begin{equation}
		x_{\mathrm{axis}} \in \{x_{\mathrm{peak}}-0.5, x_{\mathrm{peak}}, x_{\mathrm{peak}}+0.5\},
		\label{equation9}
	\end{equation}
	where $x_{\mathrm{peak}}$ is the radial coordinate corresponding to the peak of the column density profile, and $x_{\mathrm{peak}}$ and 0.5 are in unit of pixel. This is because that, for some profiles the line connecting the peak and the next highest point is very flat and it is reasonable to set the axis of symmetry at the middle of these two points rather than at the peak. Once $x_{\mathrm{axis}}$ is set, the left part, a 7-pixel ($\sim$ 0.42 pc) curve, of the column density profile is folded to the right side with respect to $x_{\mathrm{axis}}$, as illustrated in Figure \ref{fig7}(b). With the folded 7-pixel curve (red) and the original 7-pixel curve (black) of the right part of the profile, we construct two functions:
	\begin{equation}
		f_1(x) = \mathrm{Max}\{y_1(x), y_2(x)\},
		\label{equation10}
	\end{equation} 
	\begin{equation}
		f_2(x) = \mathrm{Min}\{y_1(x), y_2(x)\},
		\label{equation11}
	\end{equation}
	where $y_1$ and $y_2$ are the folded left-side 7-pixel curve and the original right-side 7-pixel curve of the profile, respectively. The functions $f_1$ and $f_2$ define two sets S$_{\cup}$ and S$_{\cap}$, which are the union and the intersection of the zones under $y_1$ and $y_2$, respectively. Sets S$_{\cup}$ and S$_{\cap}$ are illustrated in Figure \ref{fig7}(b) with grey shadows and blue slashes, respectively. The bottom line of the sets is the minimum $y_{\mathrm{min}}$ of the union of $y_1$ and $y_2$, as shown by the green dashed line in Figure \ref{fig7}(b). Then, the area ratio of the two zones, $P_{\cap/\cup} = S_{\mathrm{\cap}}/S_{\mathrm{\cup}}$, is used as quantitative measurement of the degree of symmetry of the profile within a selected range (seven pixels in this work). The value of $P_{\cap/\cup}$ changes when $x_{\mathrm{axis}}$ varies among the three tried positions. For each average column density profile, we choose the maximum of $P_{\cap/\cup}$ of the three tries, $P_{\cap/\cup,\mathrm{max}}$, to represent the degree of symmetry of the profile. Figure \ref{fig8} shows the histograms of the degree of the symmetry of the column density profiles of the 397 segments. The histogram without eliminating the contaminated slices is shown in Figure \ref{fig8}(a) and that after the elimination is shown in Figure \ref{fig8}(b). The criterion we set to determine whether a profile is symmetrical is that $P_{\cap/\cup,\mathrm{max}} \ge 0.75$.

\subsubsection{Categories of column density profiles}\label{subsubsec5.2.1}
The profiles of the 397 segments can be divided into eight categories according to their apparent symmetry and their environments. The number of profiles in each category and the percentage in the total number of segments are listed in Table \ref{table1}. 
\begin{table}[!htb]
	\bc
	\caption[]{Observed and intrinsic symmetry of segments}\label{table1}
	\begin{tabular}{c c c c c c}
		\hline\hline
		Category  & Observed symmetry & contamination & Intrinsic symmetry & Number & Fraction \\
		\hline
		1 & S & N & S & 44 & 11.1\% \\
		2 & S & P & S & 25 & 6.3\% \\
		3 & S & P & A & 13 & 3.3\% \\
		4 & A & N/W & A & 121 & 30.5\% \\
		5 & A & P & S & 16 & 4.0\% \\
		6 & A & P & A & 125 & 31.5\% \\
		7 & S/A & Y & not available & 49 & 12.3\% \\
		8 & not available & - & not available & 4 & 1.0\% \\
		\hline
	\end{tabular}
	\ec
	\tablecomments{\textwidth}{Column 1 gives the indexes of the categories. Column 2-4 are the observed properties, contamination conditions, and the intrinsic properties of the 397 segments, respectively. The total number and the percentage of each category in the whole segment sample are given in columns 5 and 6, respectively. In the table, S: Symmetrical profile; A: Asymmetrical profile; N: No contamination; Y: Strong contamination in all slices; W: Weak contamination in all slices; P: Contamination in some but not all slices.}
\end{table}

\begin{figure*}[!htb]
	\centering
	\begin{minipage}[t]{0.35\linewidth}
		\centering
		\includegraphics[trim = 1.9cm 2cm 6cm 4cm, width = \linewidth, clip]{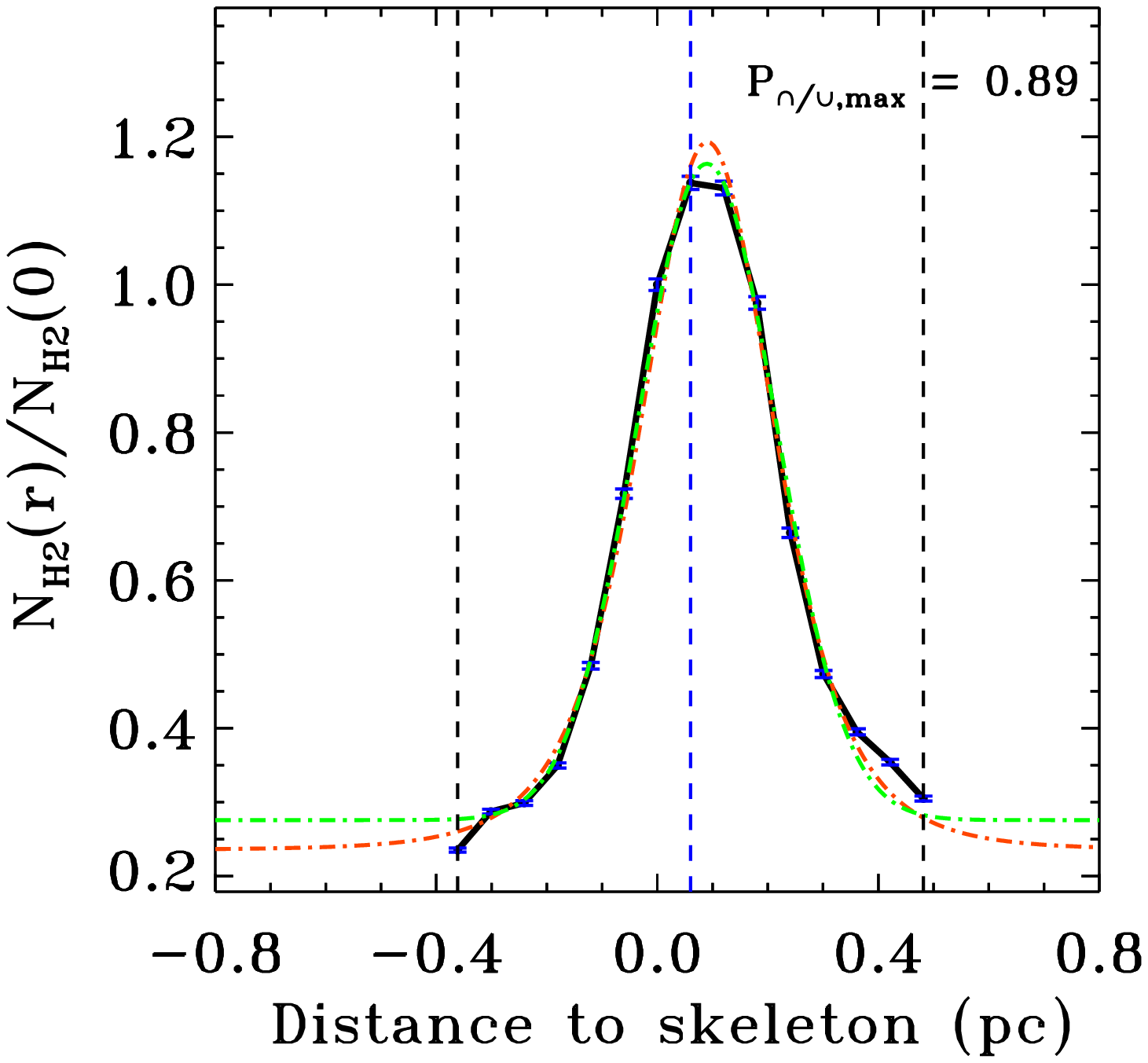}
		\\(a)
	\end{minipage}
	\begin{minipage}[t]{0.35\linewidth}
		\centering
		\includegraphics[trim = 3.5cm 2cm 3.7cm 3.5cm, width = \linewidth, clip]{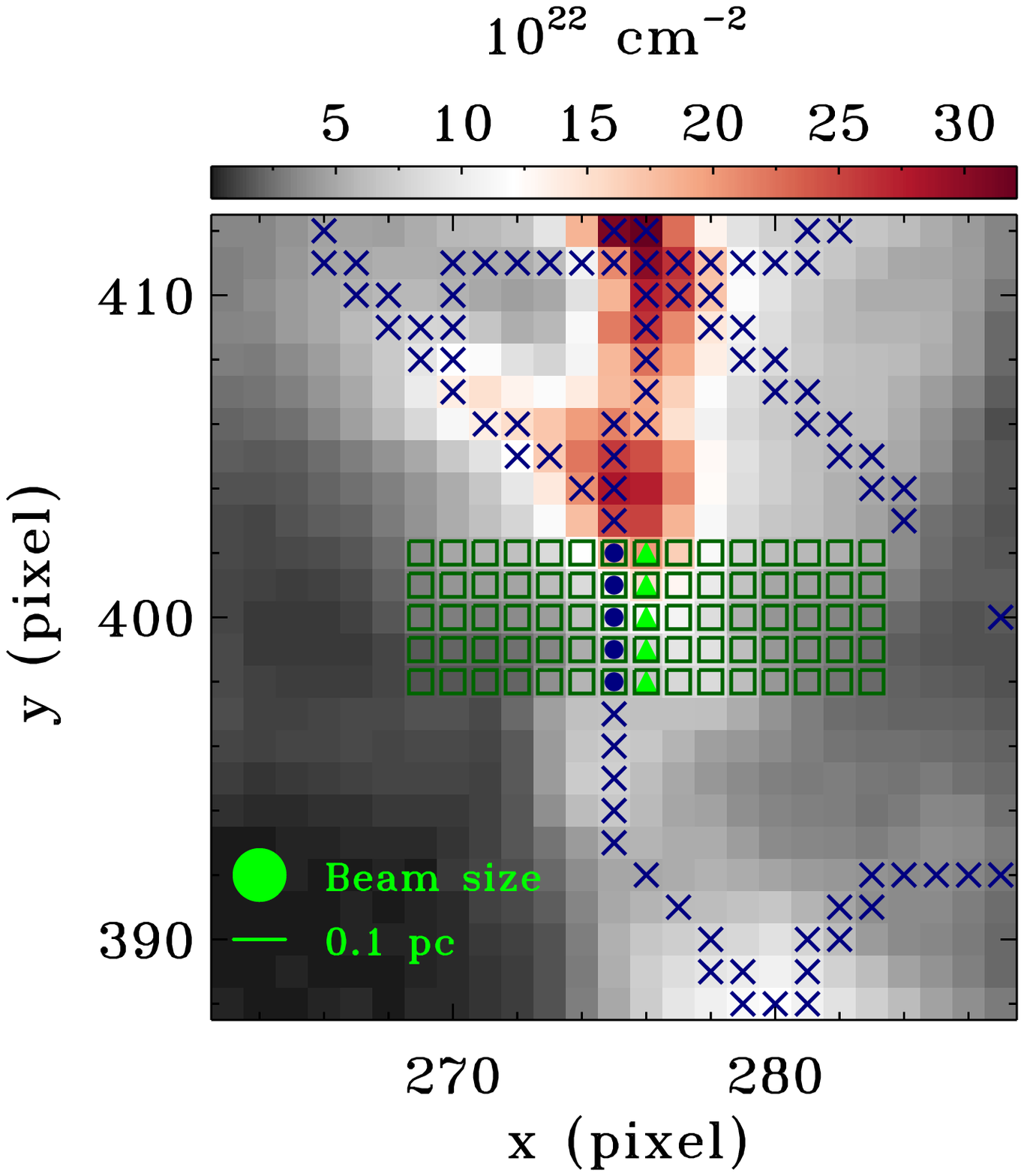}
		\\(b)
	\end{minipage}
	\caption{(a) An example of the uncontaminated column density profiles of category 1. The vertical black and blue dashed lines indicate the boundaries of the fitting range and the peak, respectively. The Red dot-dash line is the Plummer-like fitting curve, and the green line is the Gaussian fitting curve. The error bars are given according to equation \ref{equation8}. The degree of symmetry is indicated at the upper-right corner. (b) Spatial distribution and the environment of the segment. The background is the $\mathrm{H_2}$ column density map. Blue dots represent the skeleton of the segment in question, while blue crosses mark the skeleton of other segments. Green triangles mark the peak positions of the segment. Green boxes represent the pixels of uncontaminated slices of the segment in question. The beam size and the 0.1-pc scale bar are indicated at the lower-left corner.}
	\label{fig9}
\end{figure*}

The first category is the symmetrical column density profiles which are uncontaminated by other segment spines. Figure \ref{fig9} shows an example of this category. In Figure \ref{fig9}(a), the column density profile is symmetrical while the slices of the corresponding segment used to derive the profile do not overlap with any other segment spines, so we consider that there is no contamination in this profile. 

\begin{figure*}[!htb]
	\centering
	\begin{minipage}[t]{0.3\linewidth}
		\centering
		\includegraphics[trim = 1.9cm 2cm 6cm 4cm, width = \linewidth, clip]{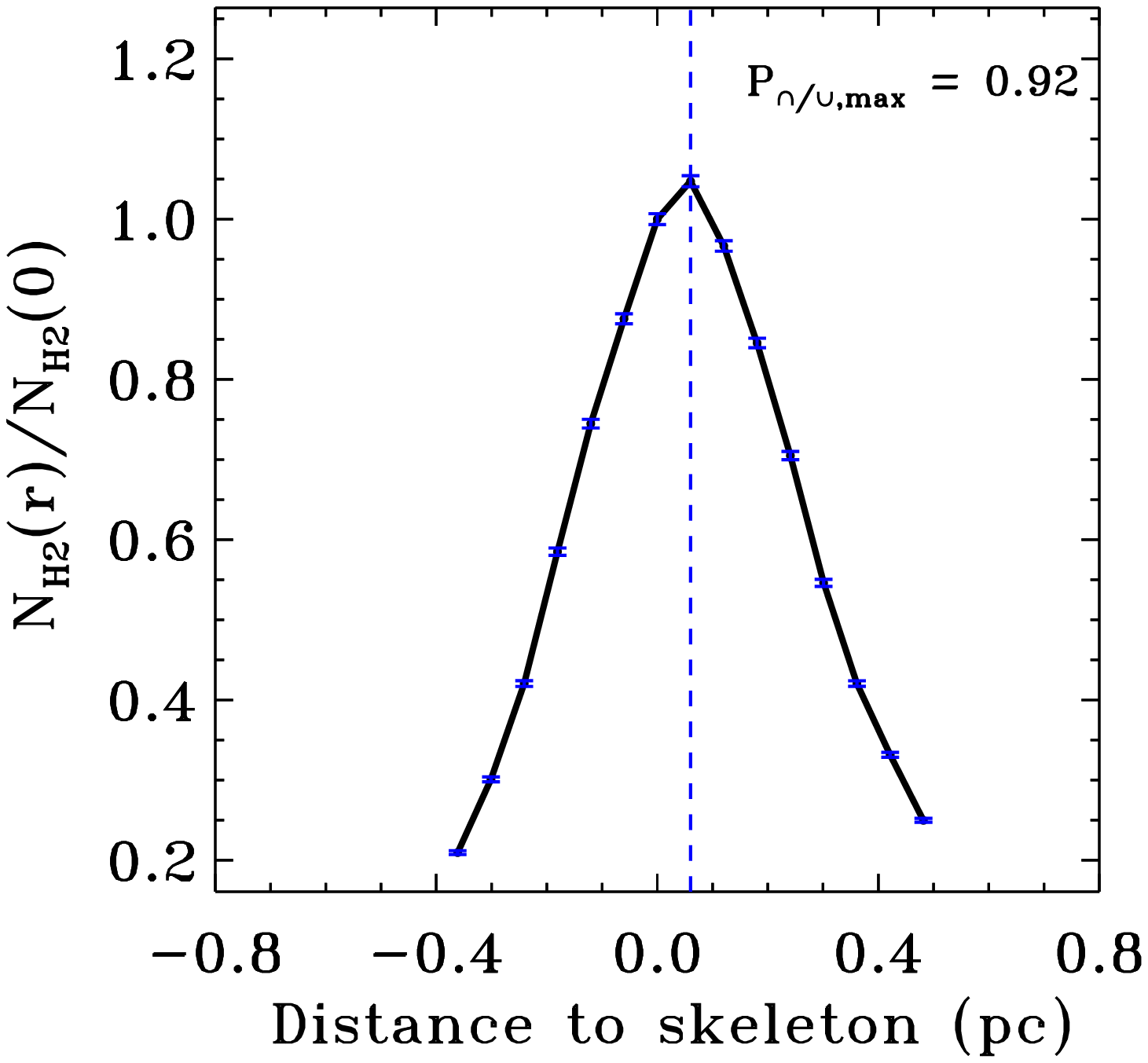}
		\\(a)
	\end{minipage}
	\begin{minipage}[t]{0.3\linewidth}
		\centering
		\includegraphics[trim = 1.9cm 2cm 6cm 4cm, width = \linewidth, clip]{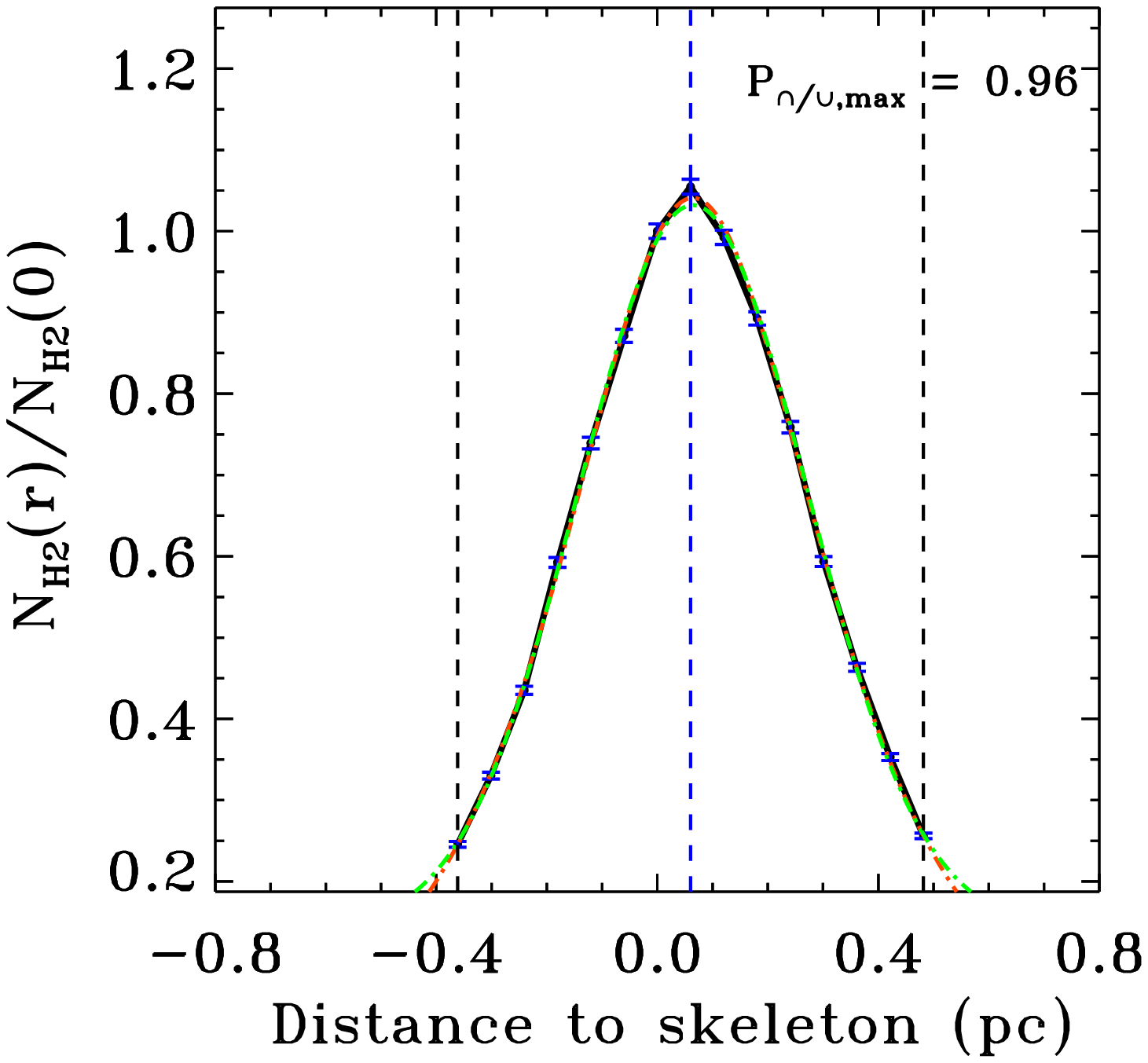}
		\\(b)
	\end{minipage}
	\begin{minipage}[t]{0.35\linewidth}
		\centering
		\includegraphics[trim = 3.5cm 2.3cm 3.7cm 3.5cm, width = \linewidth, clip]{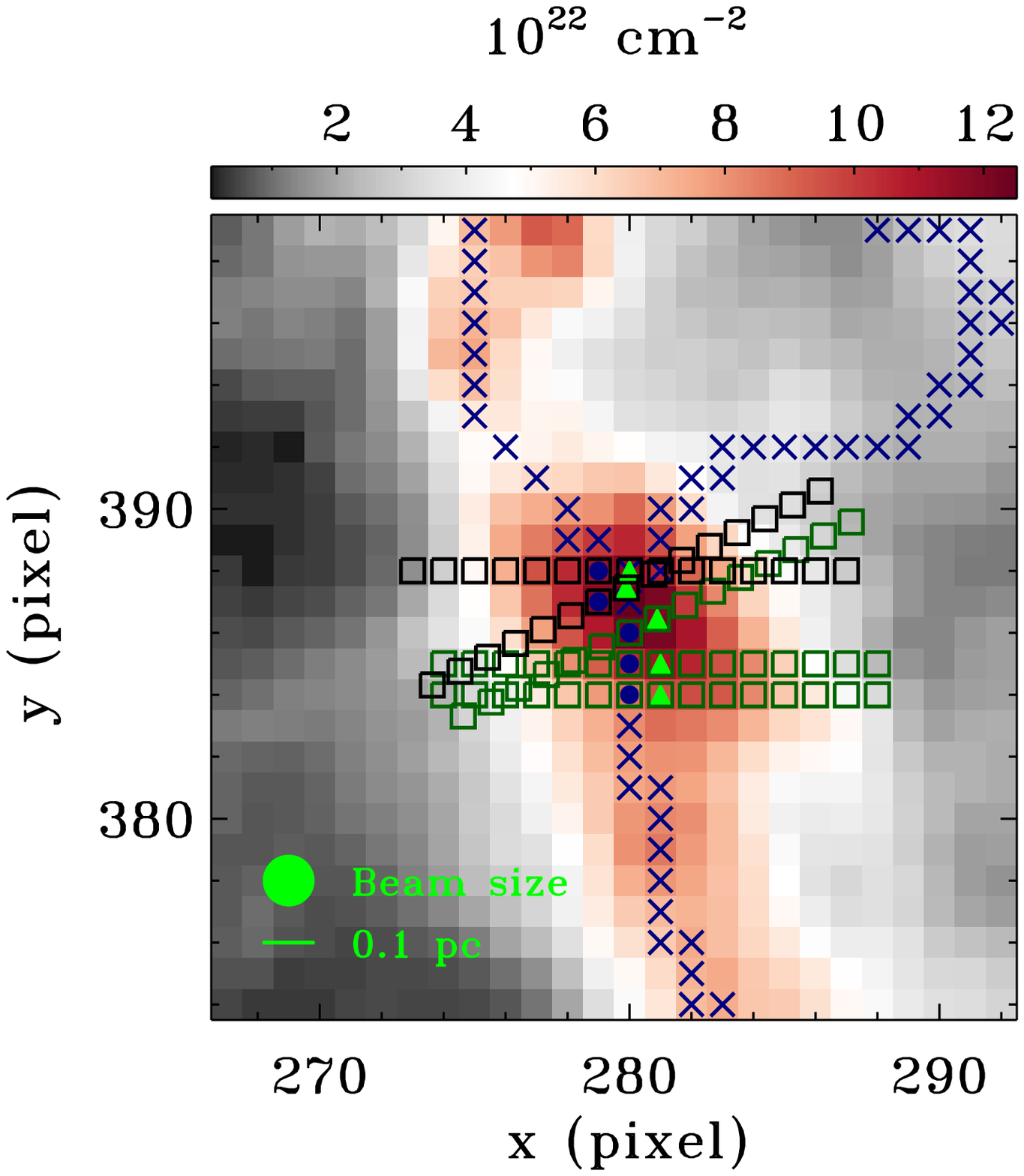}
		\\(c)
	\end{minipage}
	\caption{An example of the profiles of category 2. (a) Observed column density profile. (b) Intrinsic column density profile after removing the contaminated slices indicated with black boxes in panel c. The error bars in panels a and b are given according to equation \ref{equation8}. The vertical black and blue dashed lines indicate the boundaries of the fitting range and the peak, respectively. The Red dot-dash line is the Plummer-like fitting curve, and the green line is the Gaussian fitting curve. The degrees of symmetry are indicated at the upper-right corners of two panels. (c) Spatial distribution and the environment of the segment under analysis. The background is the $\mathrm{H_2}$ column density map. Blue dots represent the skeleton of the segment in question, while blue crosses mark the skeleton of other segments. Black boxes represent the pixels of contaminated slices of the segment in question. Green triangles mark the peak locations of the segment. Green boxes represent the pixels of uncontaminated slices of the segment in question. The beam size and the 0.1-pc scale are indicated at the lower-left corner.} 
	\label{fig10}
\end{figure*}

Figure \ref{fig10} shows an example of the profiles of Category 2. Figure \ref{fig10}(a) shows the observed symmetrical column density profile, while Figure \ref{fig10}(b) gives profile after removing the contaminated slices, showing that the intrinsic profile is symmetrical. The spatial distribution of the segment and its environments are shown in Figure \ref{fig10}(c), in which the black boxes mark the contaminated slices. The inner part of the slices indicated by the black boxes intersects with other filament spines, whereas the other three slices indicated by the green boxes do not intersect with any other filament spines. We consider this condition as partly contaminated.

\begin{figure*}[!htb]
	\centering
	\begin{minipage}[t]{0.3\linewidth}
		\centering
		\includegraphics[trim = 1.9cm 2cm 6cm 4cm, width = \linewidth, clip]{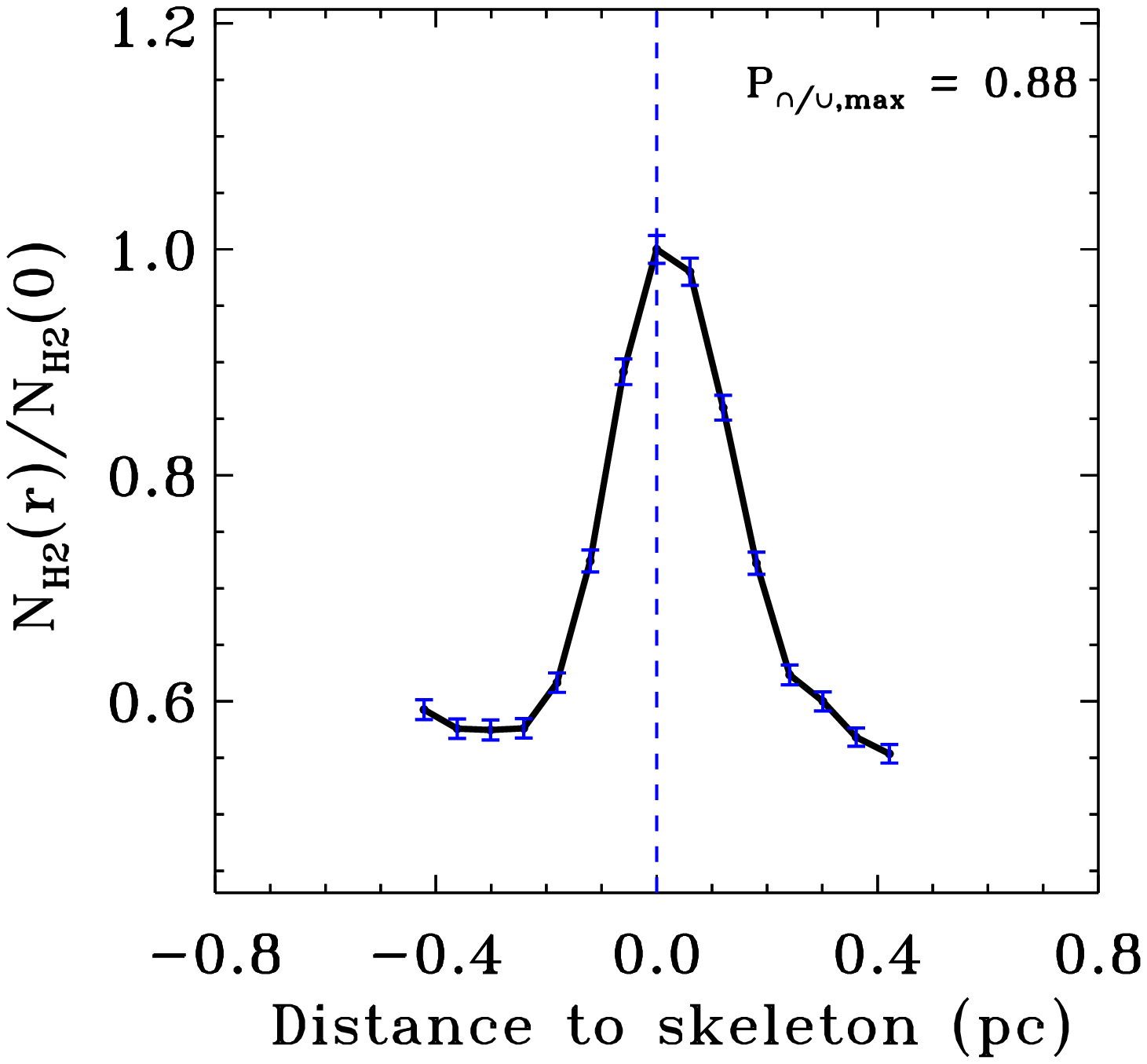}
		\\(a)
	\end{minipage}
	\begin{minipage}[t]{0.3\linewidth}
		\centering
		\includegraphics[trim = 1.9cm 2cm 6cm 4cm, width = \linewidth, clip]{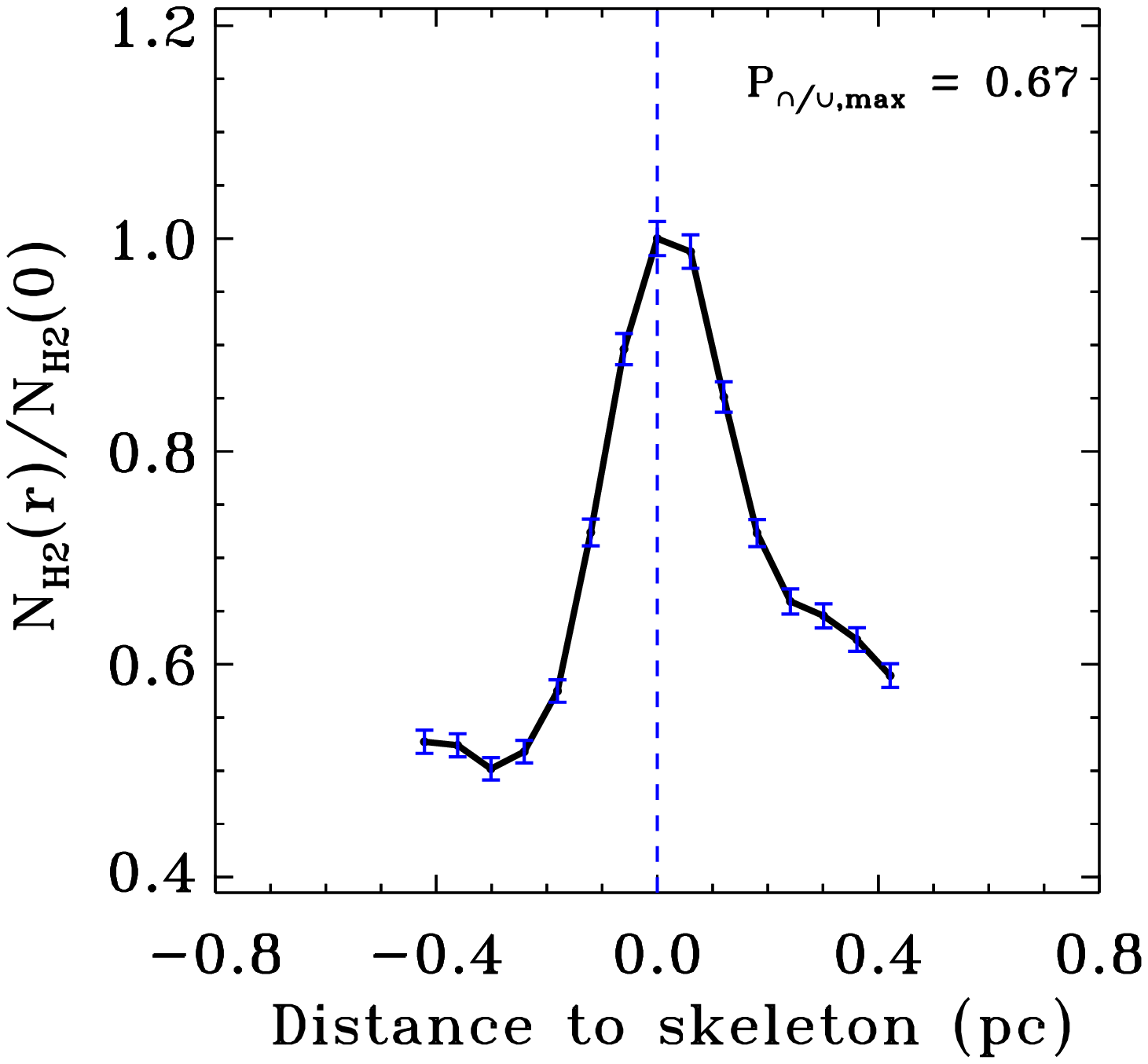}
		\\(b)
	\end{minipage}
	\begin{minipage}[t]{0.35\linewidth}
		\centering
		\includegraphics[trim = 3.5cm 2.3cm 3.7cm 3.5cm, width = \linewidth, clip]{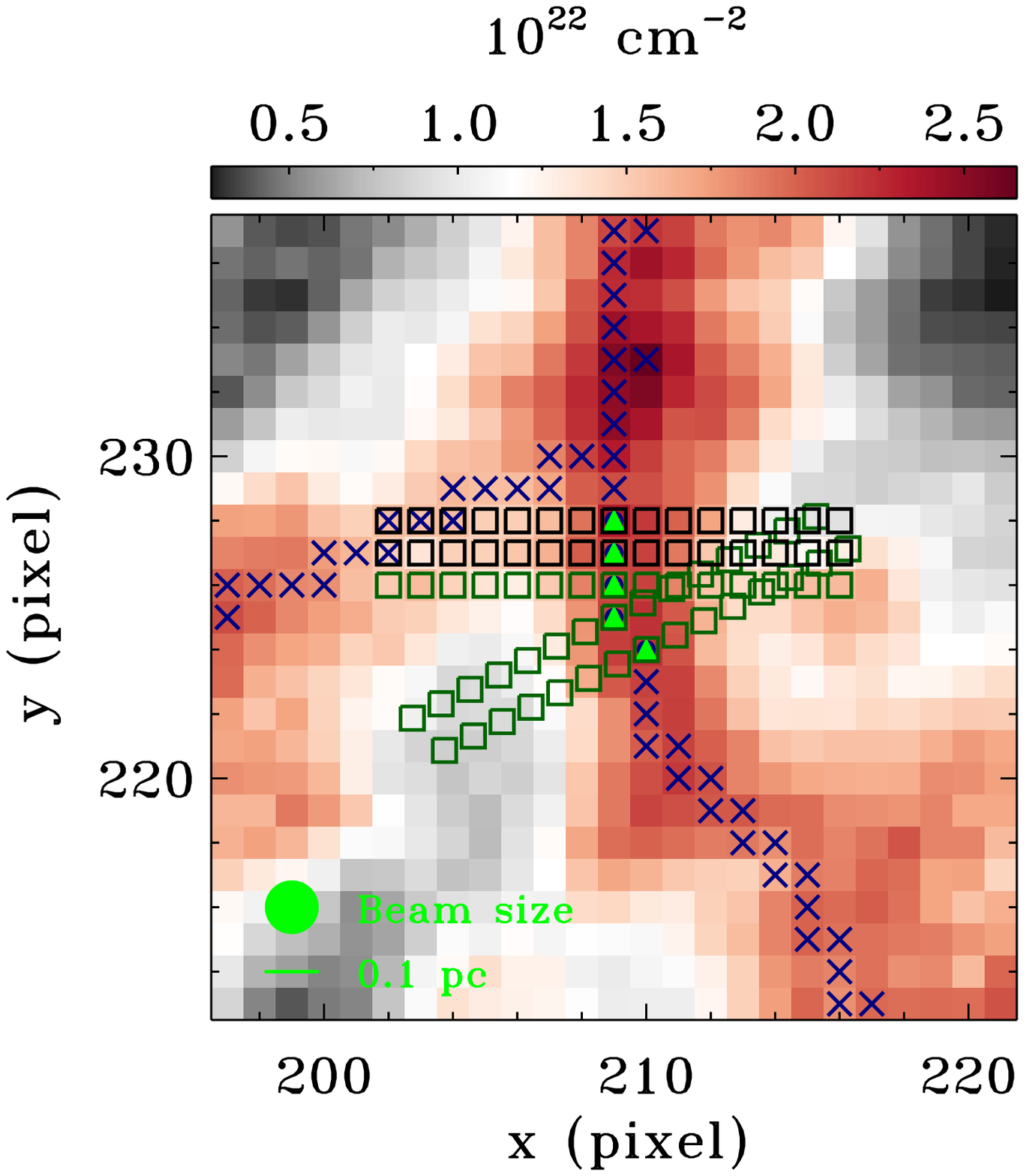}
		\\(c)
	\end{minipage}
	\caption{An example of the profiles of category 3. (a) Observed column density profile. (b) Intrinsic column density profile after removing the contaminated slices indicated with black boxes in panel c. The error bars are given according to equation \ref{equation8}. The blue dashed lines mark the peaks of the profiles. The degrees of symmetry are indicated at the upper-right corners of two panels. (c) Spatial distribution and the environment of the segment under analysis. The background is the $\mathrm{H_2}$ column density map. Blue dots represent the skeleton of the segment in question, while blue crosses mark the skeleton of other segments. Black boxes represent the pixels of contaminated slices of the segment in question. Green triangles mark the peak locations of the segment. Green boxes represent the pixels of uncontaminated slices of the segment in question. The beam size and the 0.1-pc scale are indicated at the lower-left corner.}
	\label{fig11}
\end{figure*}

Figure \ref{fig11} shows an example of the profiles of Category 3. We can see from Figure \ref{fig11}(a) that the observed column density profile is symmetrical. The slices of the corresponding segment in Figure \ref{fig11}(c) partly intersect with other filament spines. However, after removing the contaminated slices indicated with black boxes in Figure \ref{fig11}(c), the intrinsic column density profile of this segment is asymmetrical as shown in Figure \ref{fig11}(b). 

\begin{figure*}[!htb]
	\centering
	\begin{minipage}[t]{0.33\linewidth}
		\centering
		\includegraphics[trim = 1.9cm 2cm 6cm 4cm, width = \linewidth, clip]{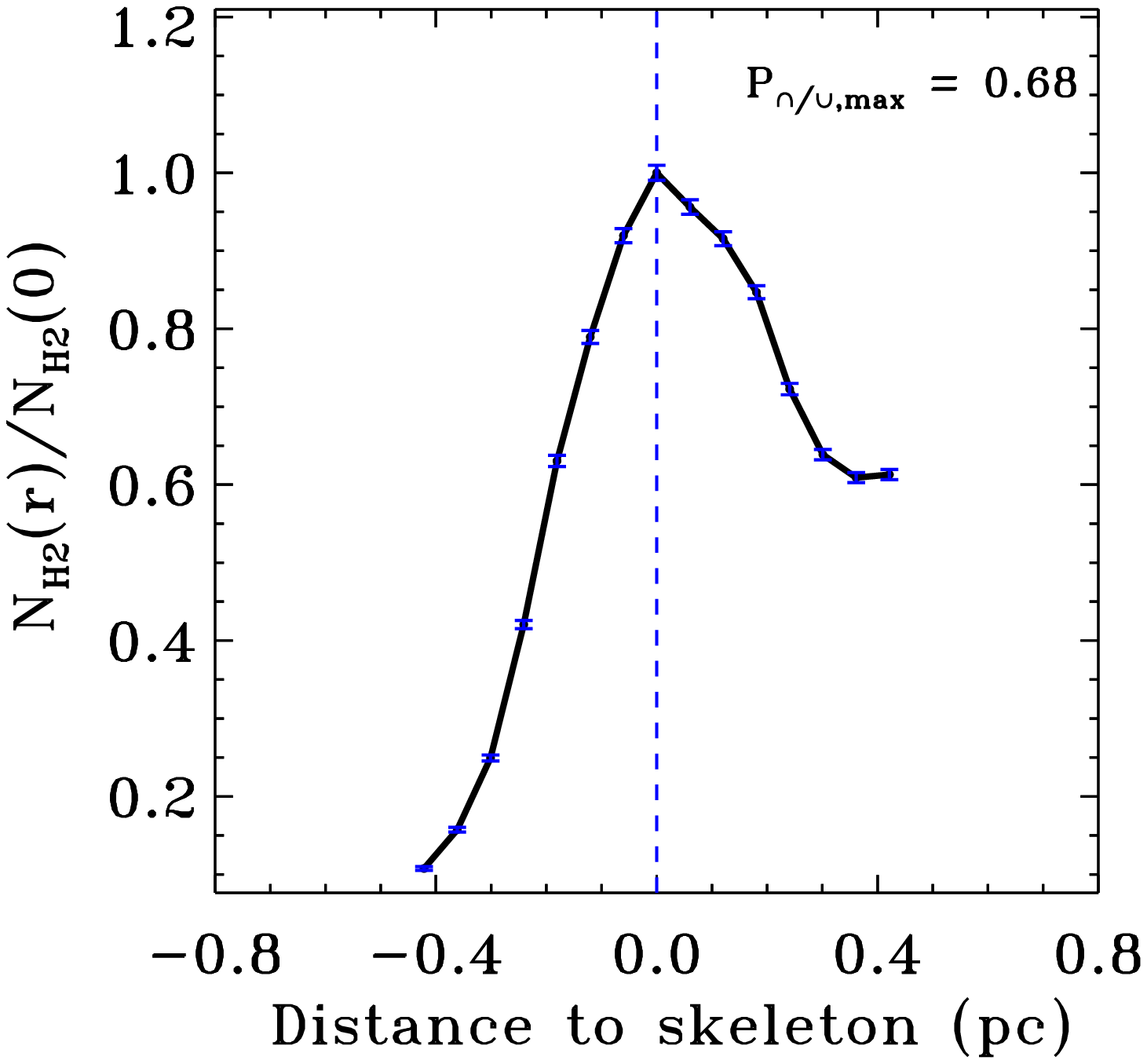}
		\\(a)
	\end{minipage}
	\begin{minipage}[t]{0.33\linewidth}
		\centering
		\includegraphics[trim = 3.5cm 2cm 3.7cm 3.5cm, width = \linewidth, clip]{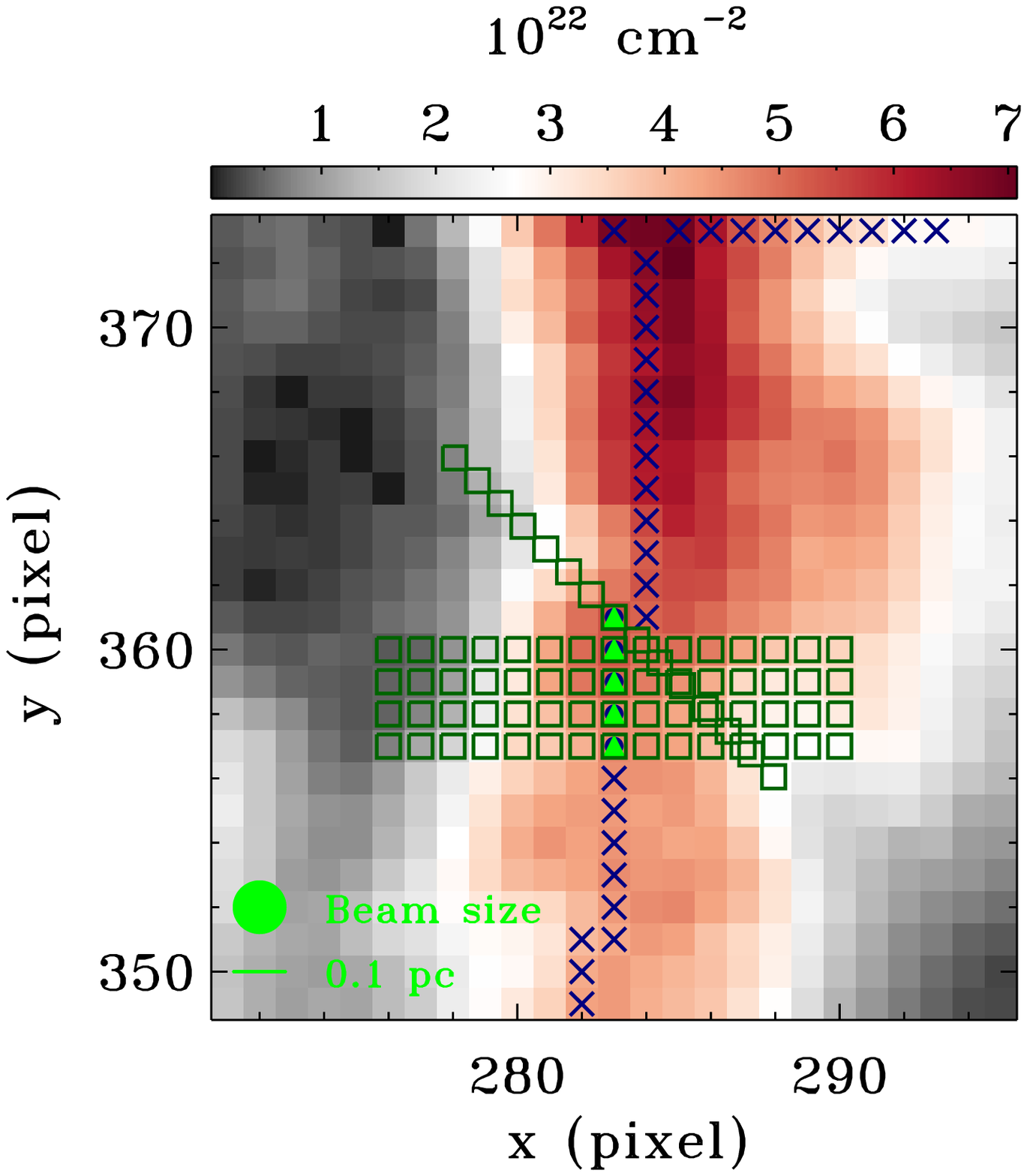}
		\\(b)
	\end{minipage}
	\begin{minipage}[t]{0.33\linewidth}
		\centering
		\includegraphics[trim = 1.9cm 2cm 6cm 4cm, width = \linewidth, clip]{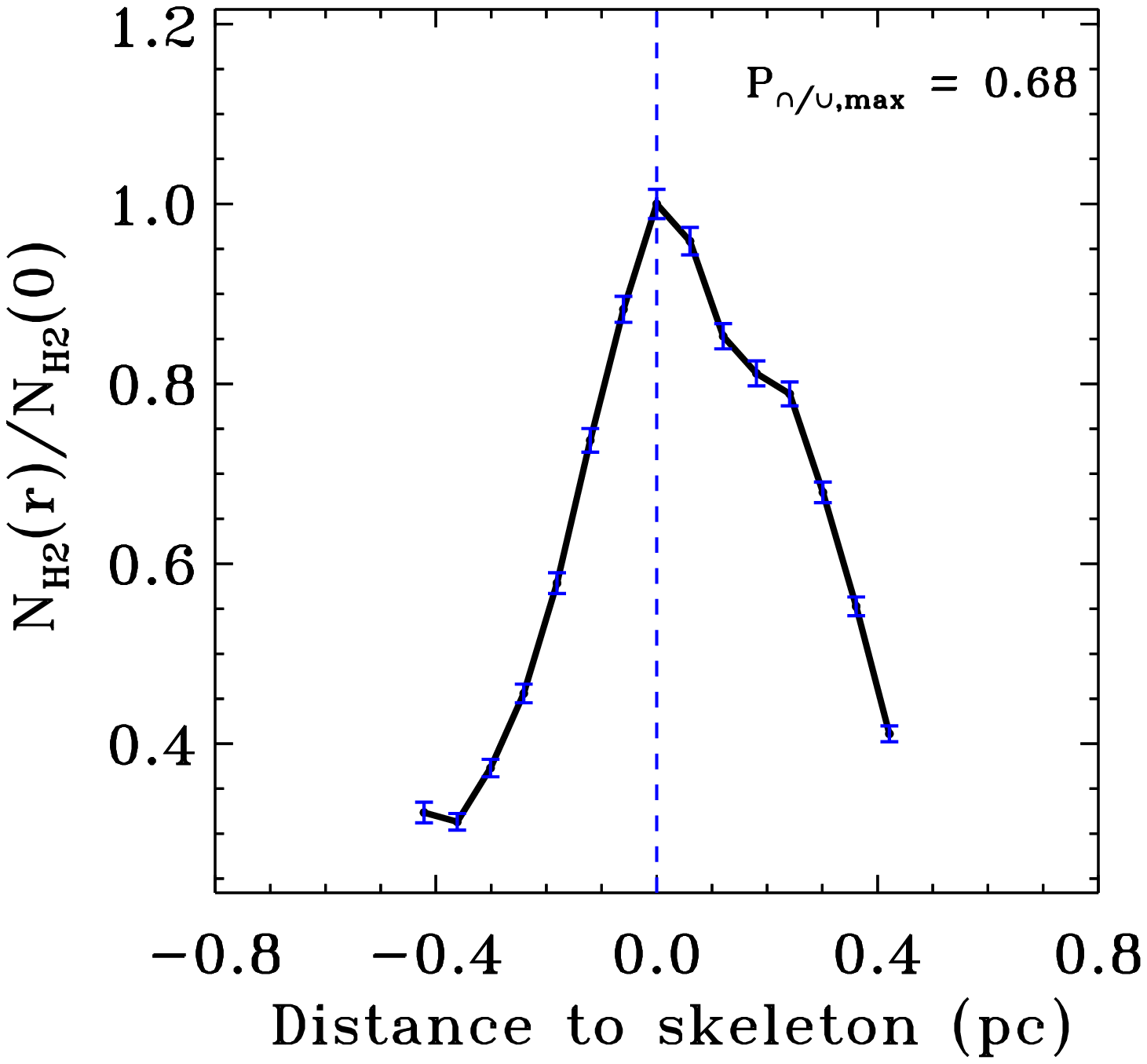}
		\\(c)
	\end{minipage}
	\begin{minipage}[t]{0.33\linewidth}
		\centering
		\includegraphics[trim = 3.5cm 2cm 3.7cm 3.5cm, width = \linewidth, clip]{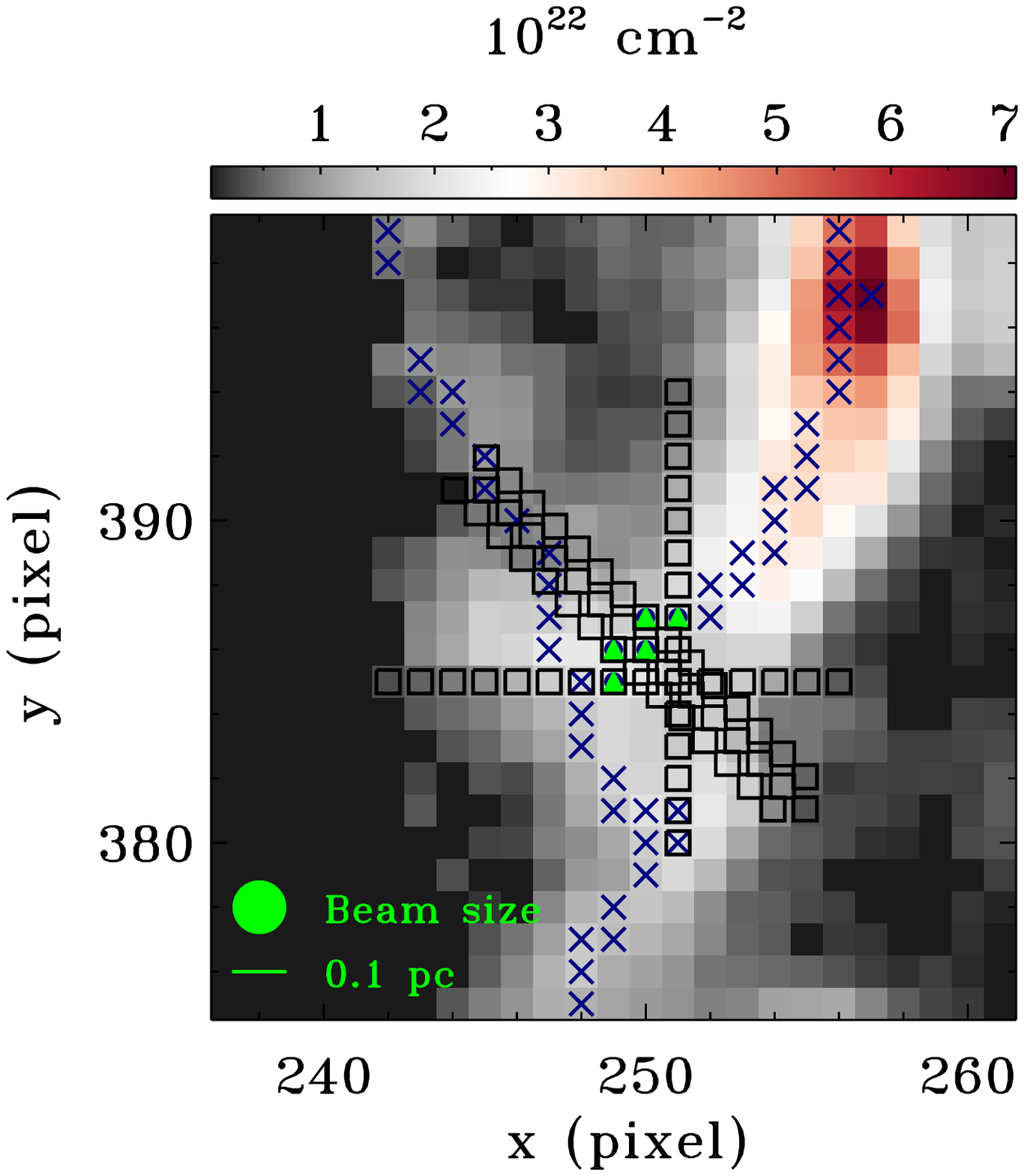}
		\\(d)
	\end{minipage}
	\caption{(a) and (c) are examples of the uncontaminated and mildly contaminated profiles of category 4, respectively. The error bars are given according to equation \ref{equation8}. The blue dashed lines mark the peaks of the profiles. The degrees of symmetry are indicated at the upper-right corners of two panels. (b) and (d) are spatial distribution and the environments of the two segments. The backgrounds are $\mathrm{H_2}$ column density maps. Blue dots represent the skeleton of the segment in question, while blue crosses mark the skeleton of other segments. Green triangles mark the peak locations of the segments. Green boxes represent the pixels of uncontaminated slices of the segment in question. The beam size and the 0.1-pc scale are indicated at the lower-left corners of two panels.}
	\label{fig12}
\end{figure*}

The profiles in Category 4 are asymmetrical, and they are uncontaminated or only weakly contaminated. Figure \ref{fig12} presents the corresponding examples of the two conditions, respectively. The column density profile in Figure \ref{fig12}(a) is asymmetrical, and the slices of the segment in Figure \ref{fig12}(b) do not intersect with other filament spines. The column density profile in Figure \ref{fig12}(c) is also asymmetrical. Although all slices of the segment intersect with another segment spine, the contamination is mild. The cases with weak contamination in Category 4 are rare and their number is only three.

\begin{figure*}[!htb]
	\centering
	\begin{minipage}[t]{0.3\linewidth}
		\centering
		\includegraphics[trim = 1.9cm 2cm 6cm 4cm, width = \linewidth, clip]{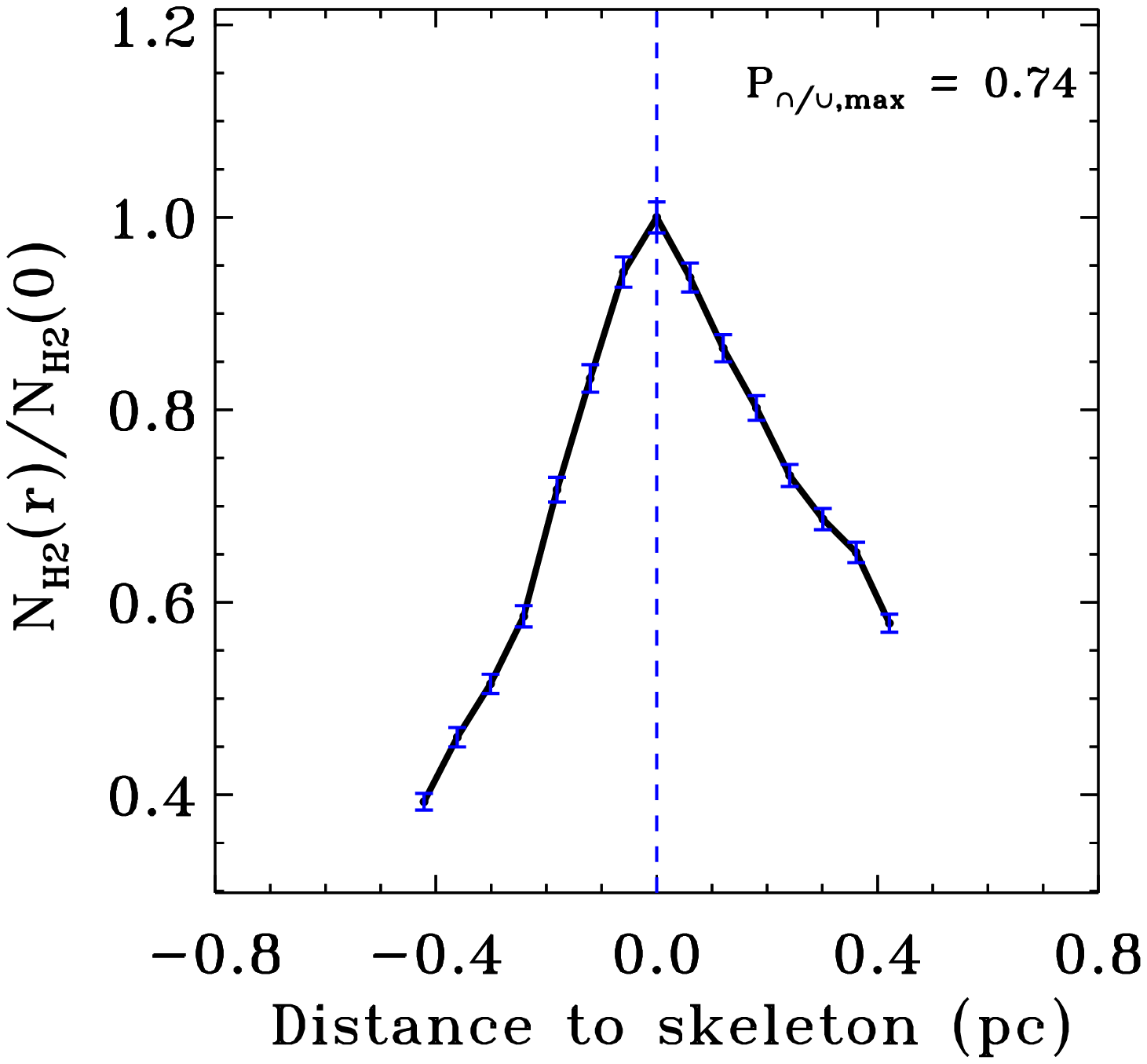}
		\\(a)
	\end{minipage}
	\begin{minipage}[t]{0.3\linewidth}
		\centering
		\includegraphics[trim = 1.9cm 2cm 6cm 4cm, width = \linewidth, clip]{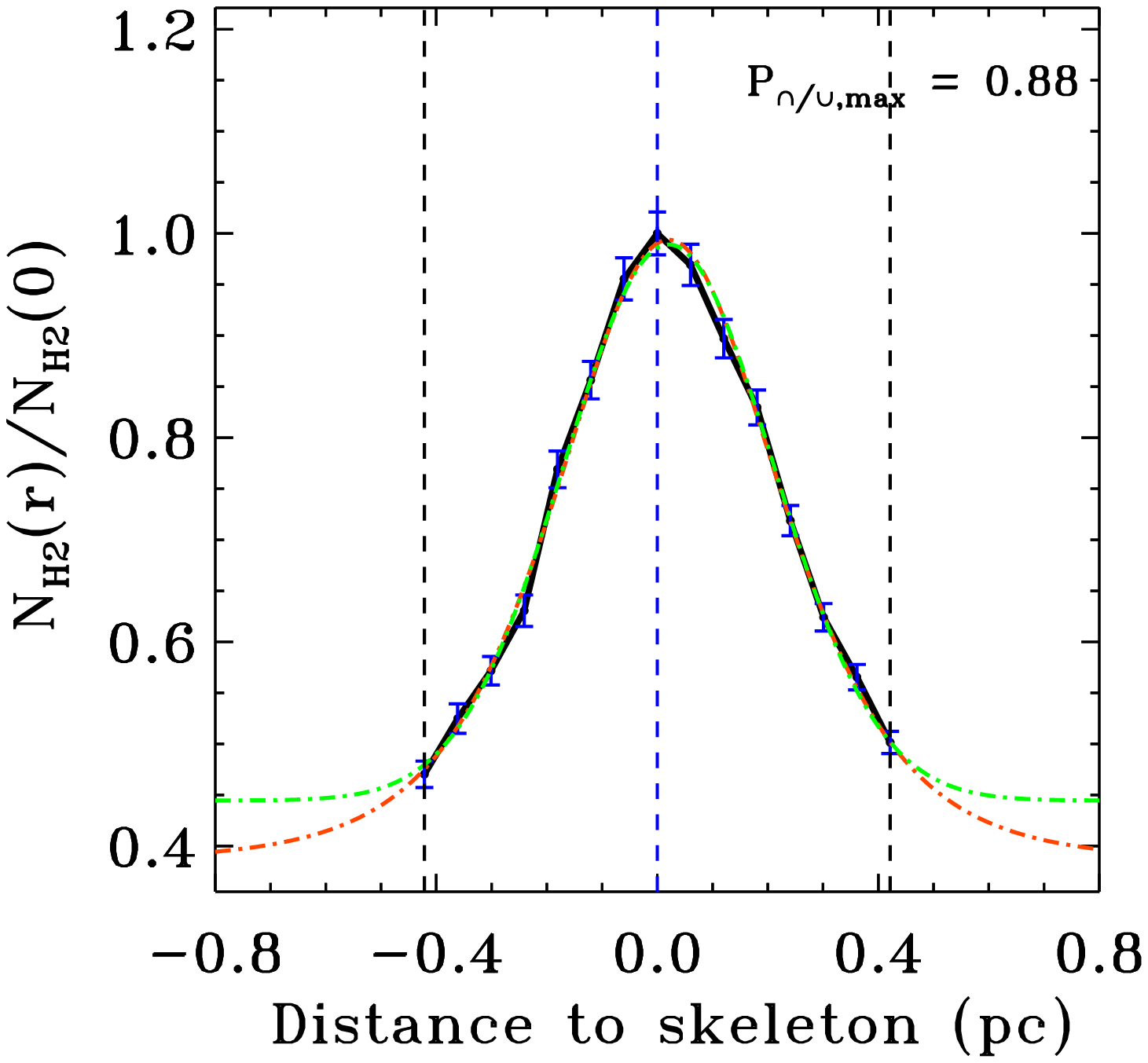}
		\\(b)
	\end{minipage}
	\begin{minipage}[t]{0.35\linewidth}
		\centering
		\includegraphics[trim = 3.5cm 2.3cm 3.7cm 3.5cm, width = \linewidth, clip]{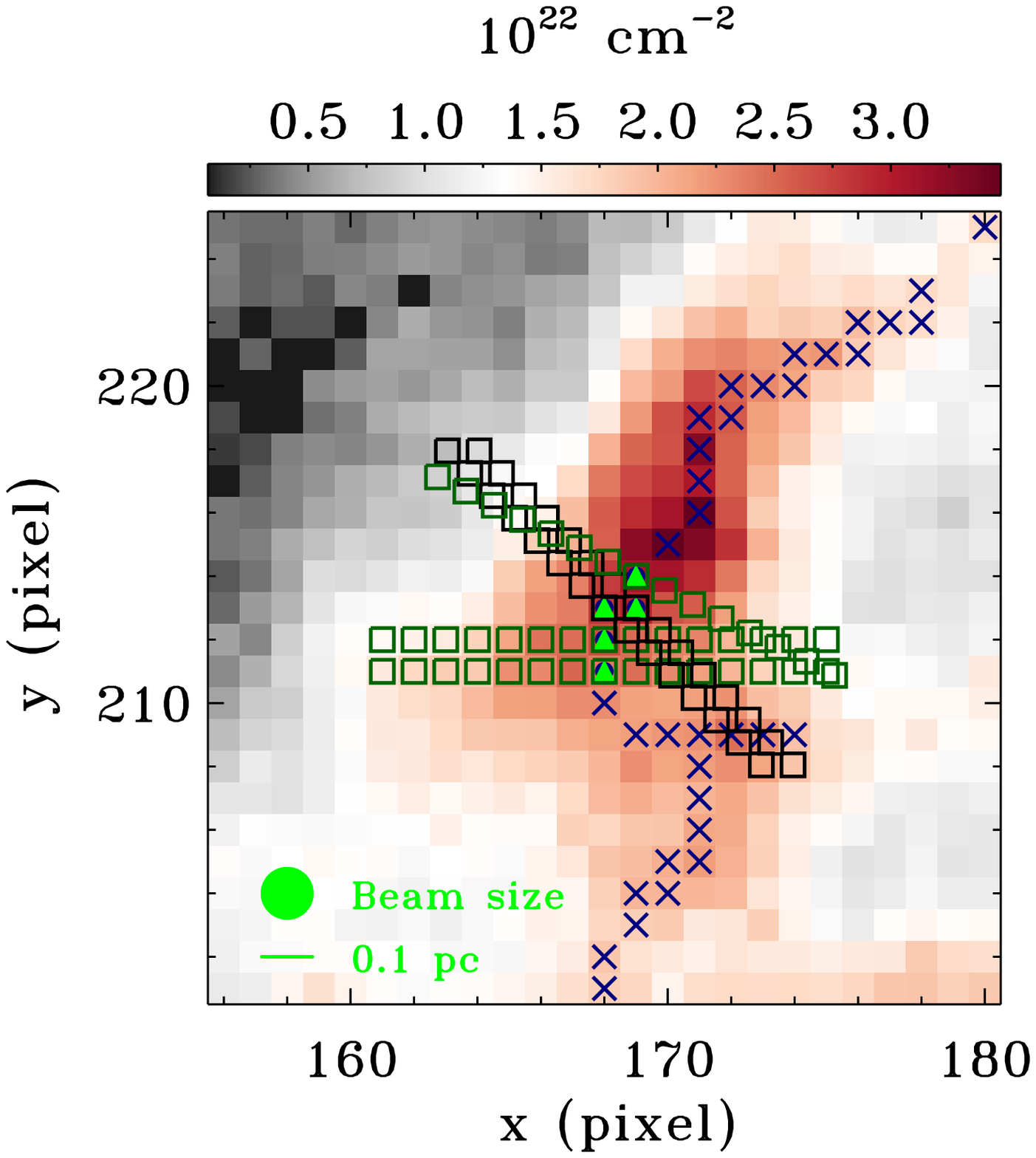}
		\\(c)
	\end{minipage}
	\caption{An example of the profiles of category 5. (a) Observed column density profile. (b) Intrinsic column density profile after removing the contaminated slices indicated with black boxes in panel c. The error bars in panels a and b are given according to equation \ref{equation8}. The vertical black and blue dashed lines indicate the boundaries of the fitting range and the peak, respectively. The Red dot-dash line is the Plummer-like fitting curve, and the green line is the Gaussian fitting curve. The degrees of symmetry are indicated at the upper-right corners of two panels. (c) Spatial distribution and the environment of the segment under analysis. The background is the $\mathrm{H_2}$ column density map. Blue dots represent the skeleton of the segment in question, while blue crosses mark the skeleton of other segments. Black boxes represent the pixels of contaminated slices of the segment in question. Green triangles mark the peak locations of the segment. Green boxes represent the pixels of uncontaminated slices of the segment in question. The beam size and the 0.1-pc scale are indicated at the lower-left corner.}
	\label{fig13}
\end{figure*}

Figure \ref{fig13} shows an example of the profiles of Category 5. The observed column density profile in Figure \ref{fig13}(a) is asymmetrical. However, similar to Category 3, the slices of the segment partly intersect with other filament spines, as indicated in Figure \ref{fig13}(c). After removing the contaminated slices, the intrinsic column density profile in Figure \ref{fig13}(b) is symmetrical.

\begin{figure*}[!htb]
	\centering
	\begin{minipage}[t]{0.3\linewidth}
		\centering
		\includegraphics[trim = 1.9cm 2cm 6cm 4cm, width = \linewidth, clip]{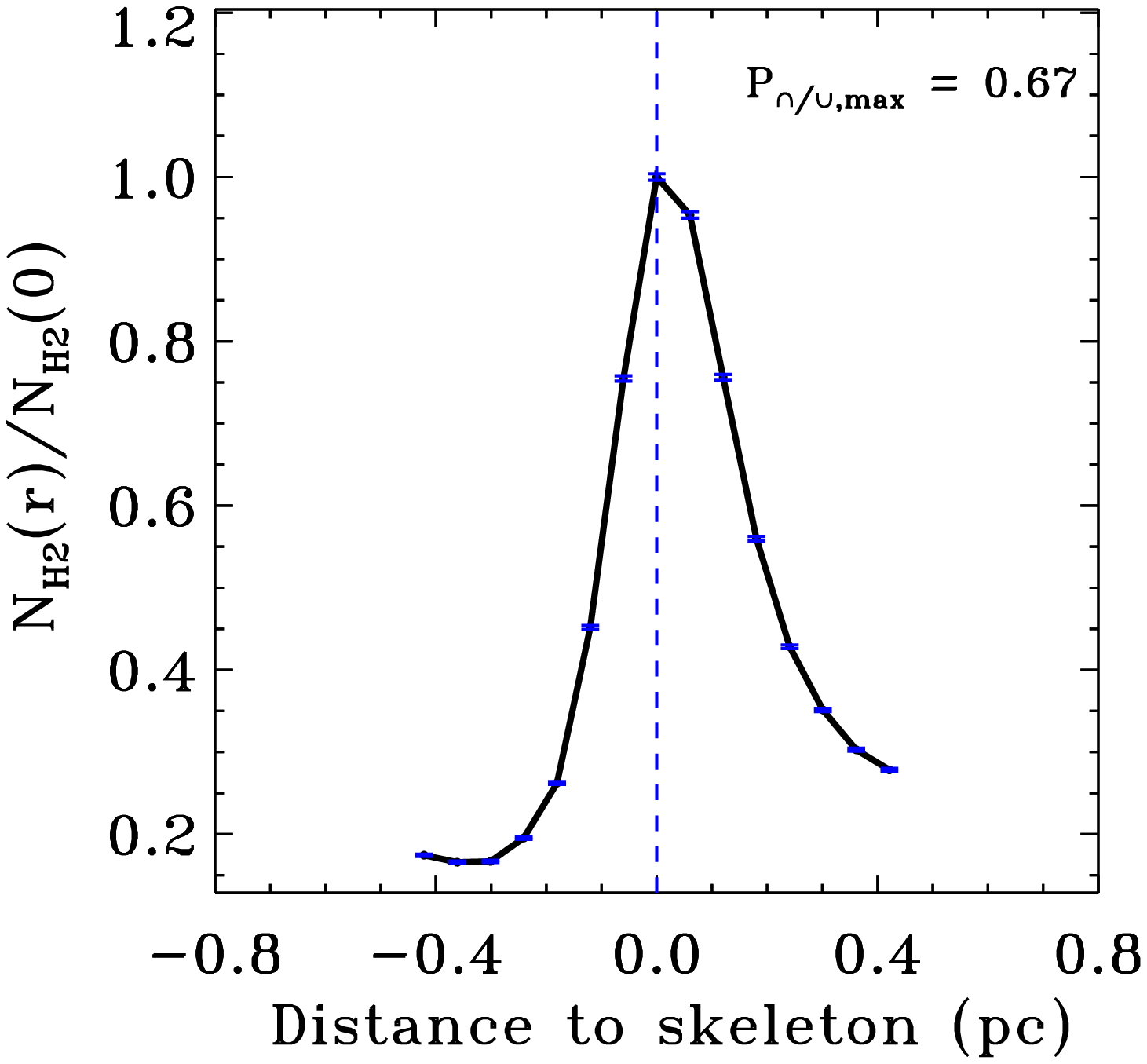}
		\\(a)
	\end{minipage}
	\begin{minipage}[t]{0.3\linewidth}
		\centering
		\includegraphics[trim = 1.9cm 2cm 6cm 4cm, width = \linewidth, clip]{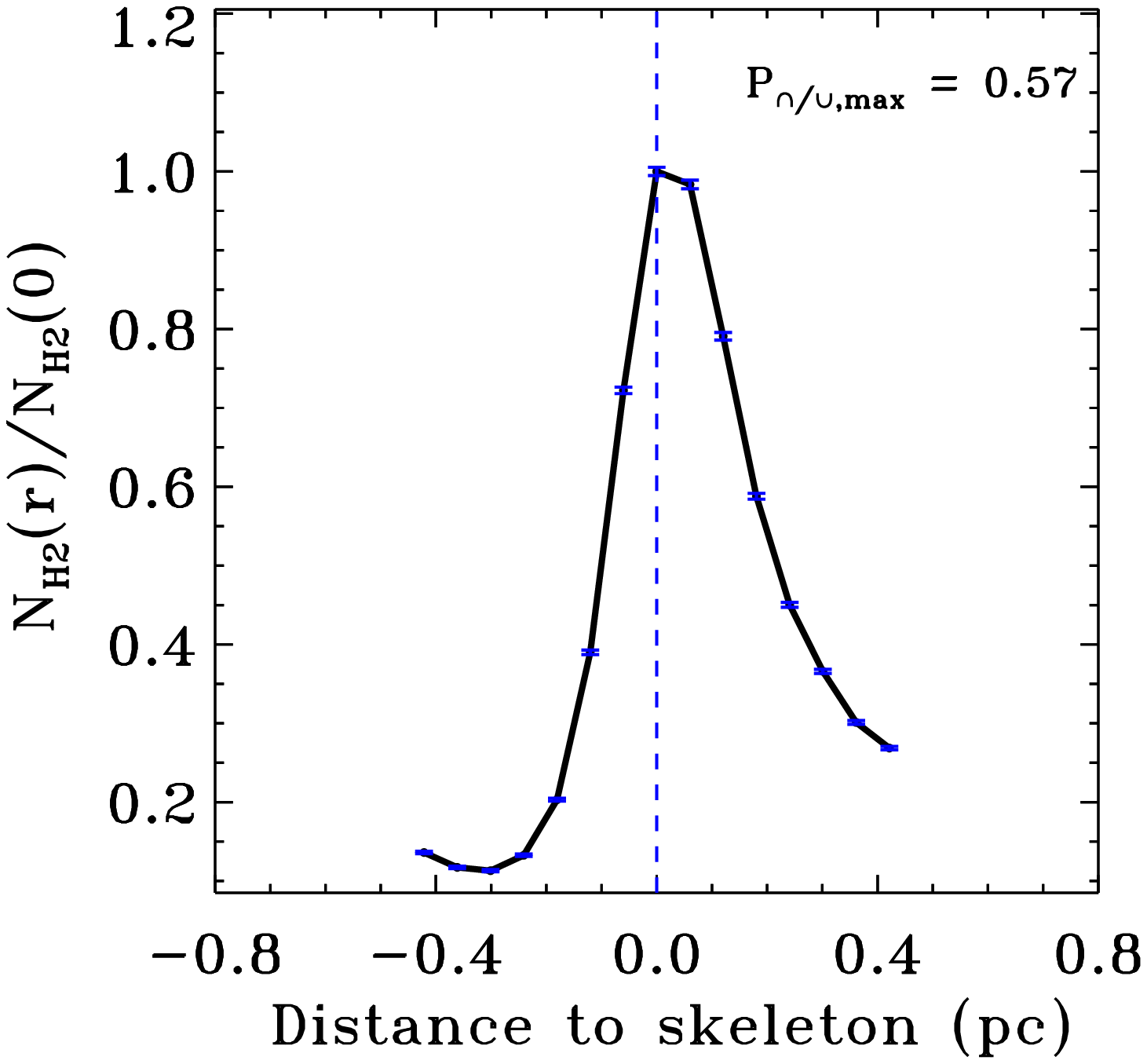}
		\\(b)
	\end{minipage}
	\begin{minipage}[t]{0.35\linewidth}
		\centering
		\includegraphics[trim = 3.5cm 2.3cm 3.7cm 3.5cm, width = \linewidth, clip]{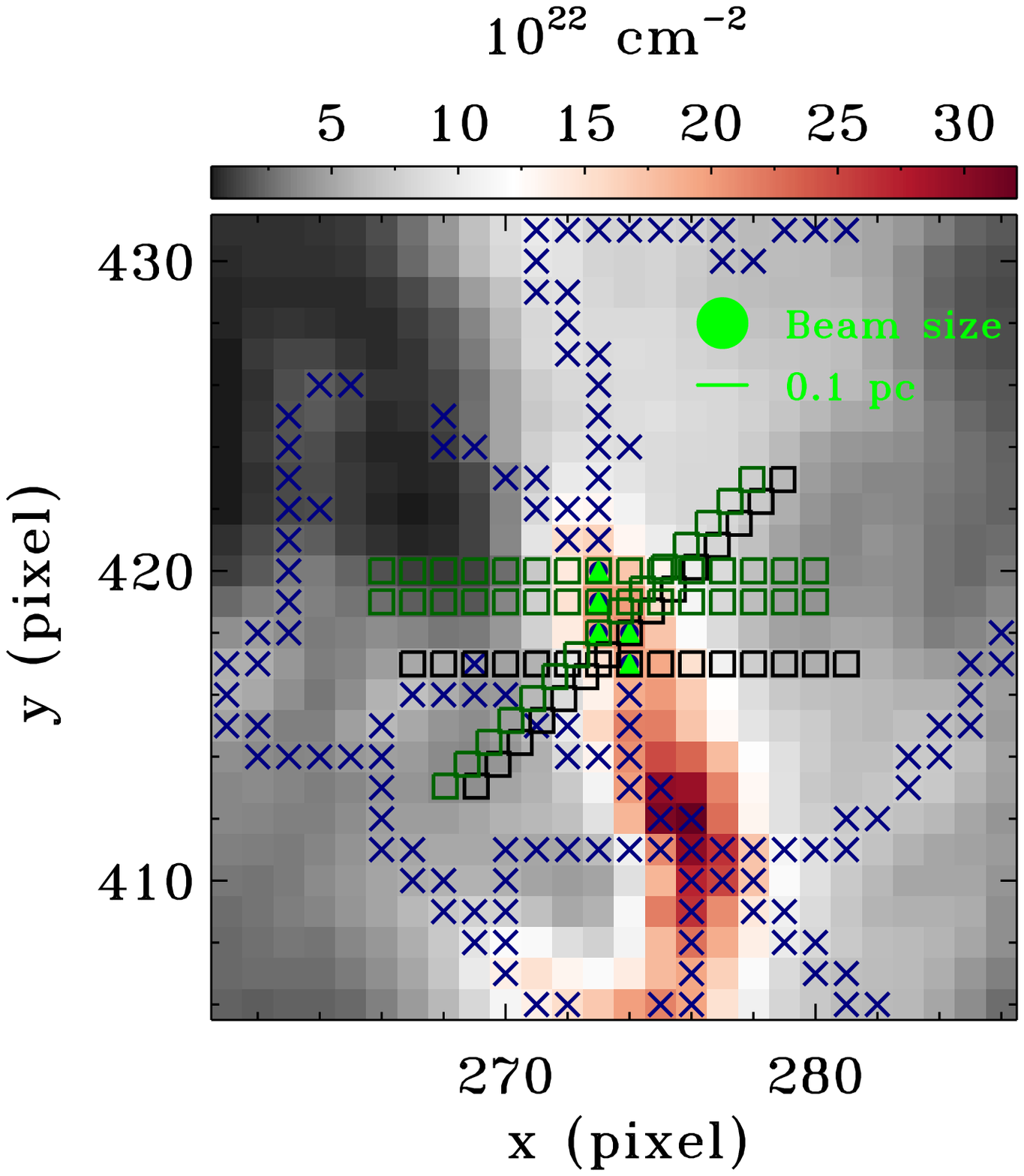}
		\\(c)
	\end{minipage}
	\caption{An example of the profiles of category 6. (a) Observed column density profile. (b) Intrinsic column density profile after removing the contaminated slices indicated with black boxes in panel c. The error bars are given according to equation \ref{equation8}. The blue dashed lines mark the peaks of the profiles. The degrees of symmetry are indicated at the upper-right corners of two panels. (c) Spatial distribution and the environment of the segment under analysis. The background is the $\mathrm{H_2}$ column density map. Blue dots represent the skeleton of the segment in question, while blue crosses mark the skeleton of other segments. Black boxes represent the pixels of contaminated slices of the segment in question. Green triangles mark the peak locations of the segment. Green boxes represent the pixels of uncontaminated slices of the segment in question. The beam size and the 0.1-pc scale are indicated at the upper-right corner.}
	\label{fig14}
\end{figure*}

Figure \ref{fig14} shows an example of the profiles of Category 6. We can see that the observed column density profile in Figure \ref{fig14}(a) is asymmetrical. The slices of the segment in Figure \ref{fig14}(c) partly intersect with other filament spines. After removing the contaminated slices, the column density profile of this segment in Figure \ref{fig14}(b) is still asymmetrical. 

\begin{figure*}[!htb]
	\centering
	\begin{minipage}[t]{0.35\linewidth}
		\centering
		\includegraphics[trim = 1.9cm 2cm 6cm 4cm, width = \linewidth, clip]{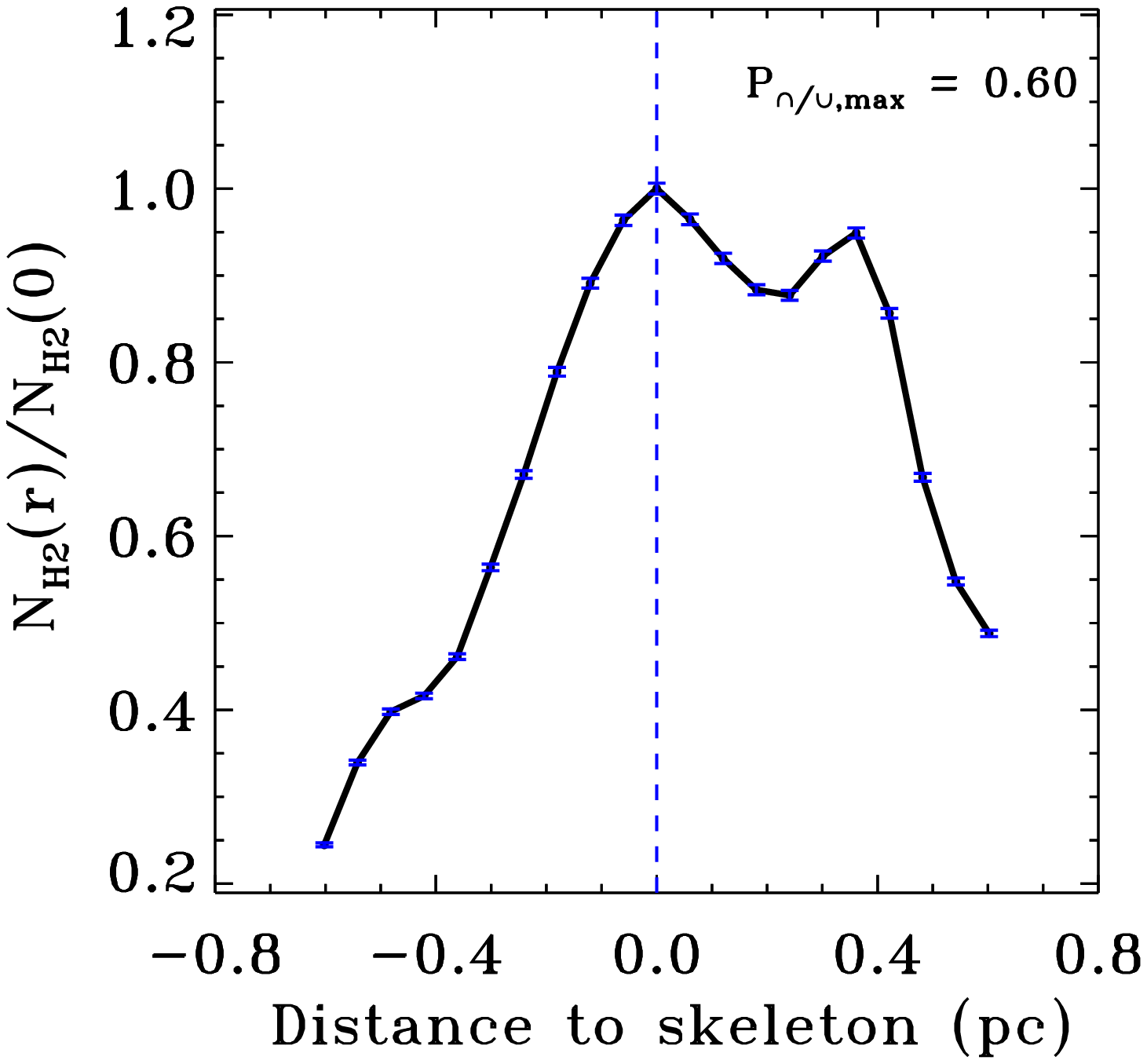}
		\\(a)
	\end{minipage}
	\begin{minipage}[t]{0.35\linewidth}
		\centering
		\includegraphics[trim = 3.5cm 2cm 3.7cm 3.5cm, width = \linewidth, clip]{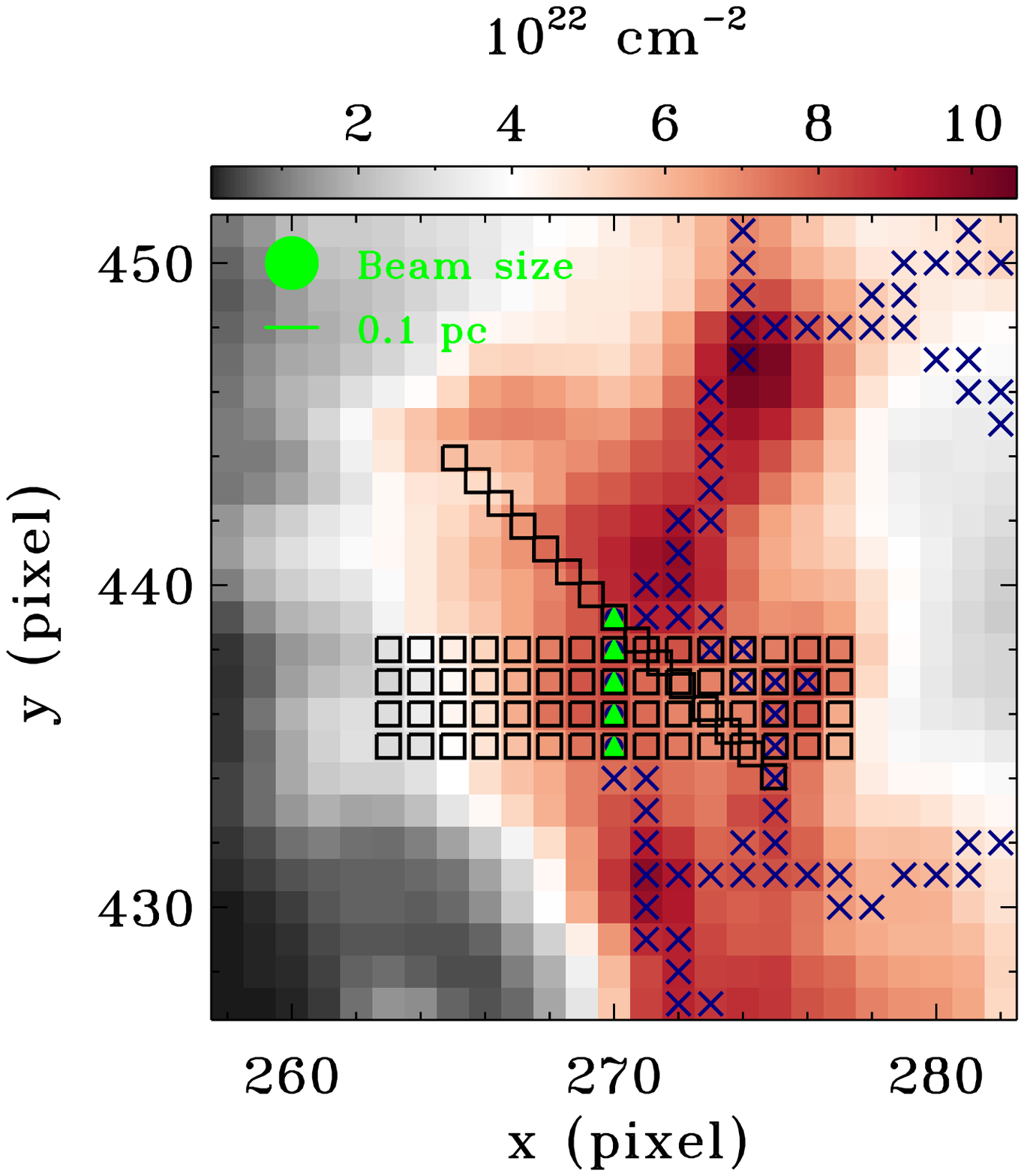}
		\\(b)
	\end{minipage}
	\caption{(a) An example of the profiles of category 7. The error bars are given according to equation \ref{equation8}. The blue dashed line marks the peak of the profile. The degree of symmetry is indicated at the upper-right corner of the panel. (b) Spatial distribution and the environment of the segment. The background is the $\mathrm{H_2}$ column density map. Blue dots represent the skeleton of the segment in question, while blue crosses mark the skeleton of other segments. Green triangles mark the peak locations of the segment. Black boxes represent the pixels of uncontaminated slices of the segment in question. The beam size and the 0.1-pc scale are indicated at the upper-left corner.}
	\label{fig15}
\end{figure*}

Figure \ref{fig15} shows an example of the profiles of Category 7. All the slices of the segment in Figure \ref{fig15}(b) intersect with other filament spines. The column densities at the overlapped regions are larger than 70\% of the peak column density of the profile, so all these slices are significantly contaminated. As all slices are strongly contaminated, the intrinsic symmetry of this segment is not available.

\begin{figure*}[!htb]
	\centering
	\begin{minipage}[t]{0.35\linewidth}
		\centering
		\includegraphics[trim = 1.9cm 2cm 6cm 4cm, width = \linewidth, clip]{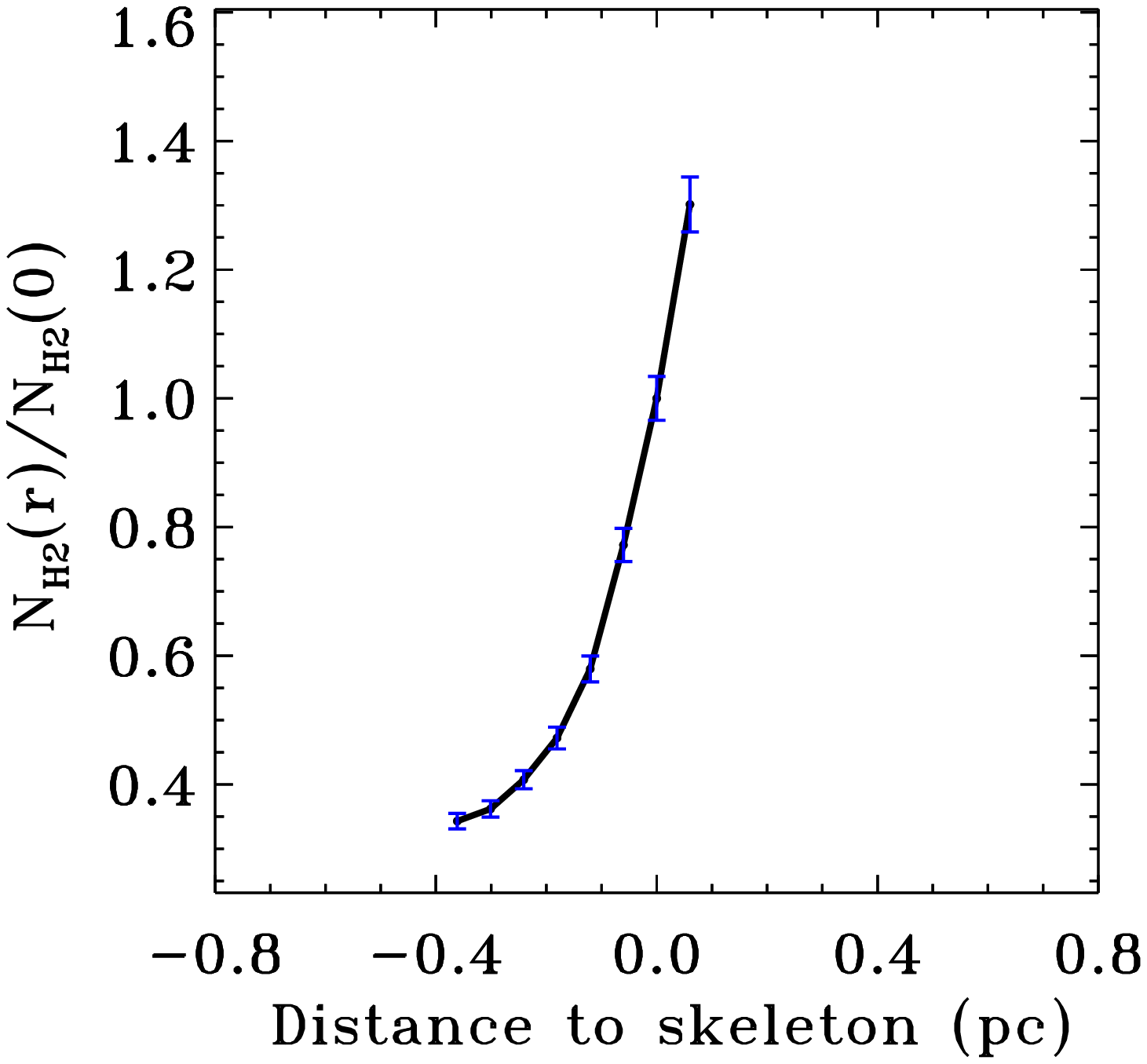}
		\\(a)
	\end{minipage}
	\begin{minipage}[t]{0.35\linewidth}
		\centering
		\includegraphics[trim = 3.5cm 2cm 3.7cm 3.5cm, width = \linewidth, clip]{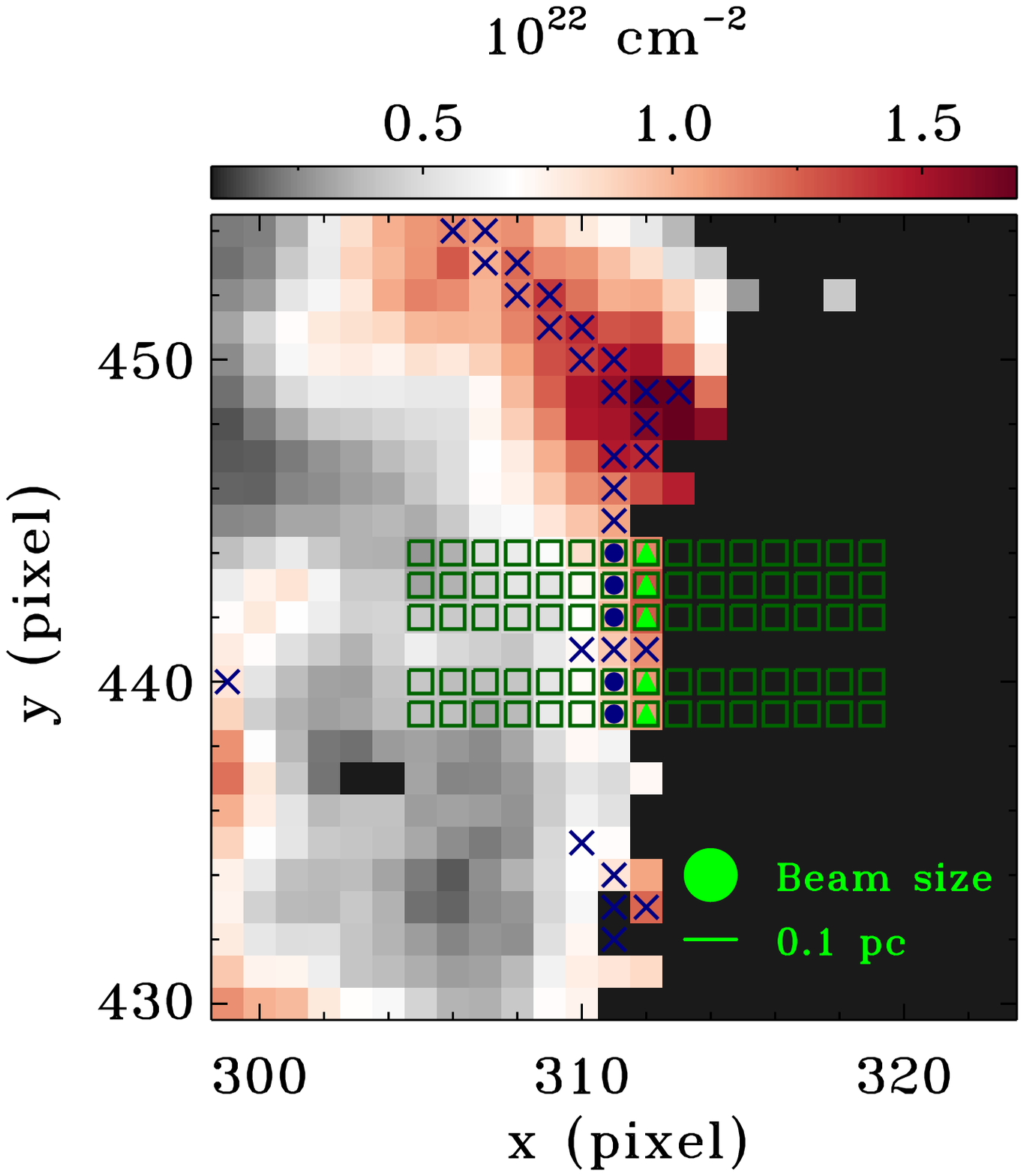}
		\\(b)
	\end{minipage}
	\caption{(a) An example of the profiles of category 8. The error bars are given according to equation \ref{equation8}. (b) Spatial distribution and the environment of the segment. The background is the $\mathrm{H_2}$ column density map. Blue dots represent the skeleton of the segment in question, while blue crosses mark the skeleton of other segments. Green triangles mark the peak locations of the segment. Green boxes represent the pixels of uncontaminated slices of the segment in question. The beam size and the 0.1-pc scale are indicated at the lower-right corner.}
	\label{fig16}
\end{figure*}

Figure \ref{fig16} shows an example of the profiles of Category 8. Some of the segments are located at the edge of the observational field, so the column density profiles of the segments as shown in Figure \ref{fig16} are incomplete. Therefore, the observed and intrinsic symmetry properties of these segments are not available.

\subsubsection{Statistics of the profiles of the eight categories}\label{subsubsec5.2.2}
\begin{figure}[!htb]
	\begin{minipage}[t]{0.56\linewidth}
		\centering
		\includegraphics[trim = 2cm 5cm 0cm 4cm, width = \linewidth, clip]{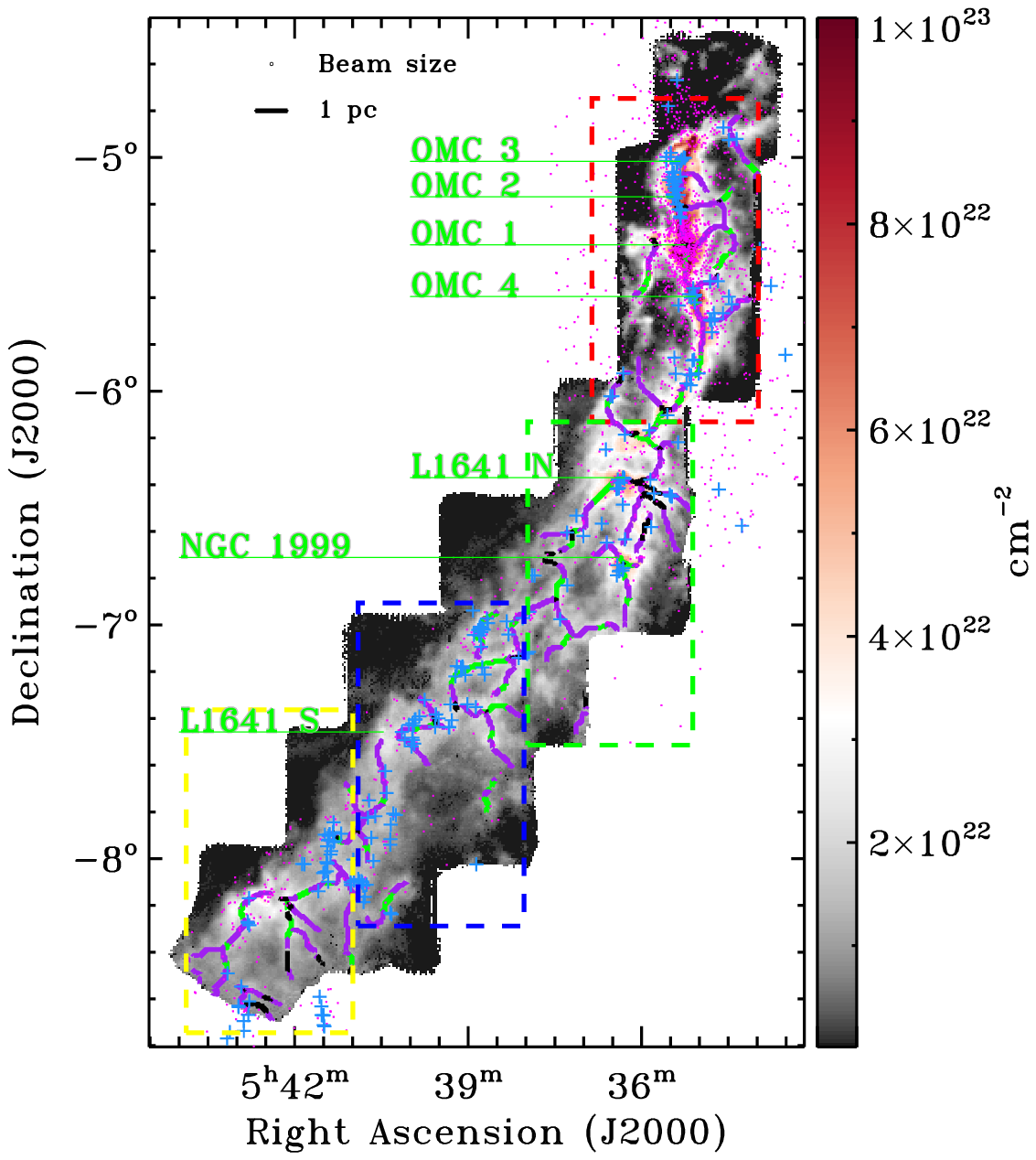}
		\\(a)
	\end{minipage}
	\begin{minipage}[t]{0.32\linewidth}
		\centering
		\includegraphics[trim = 8.5cm 5.5cm 7.9cm 5cm, width = \linewidth, clip]{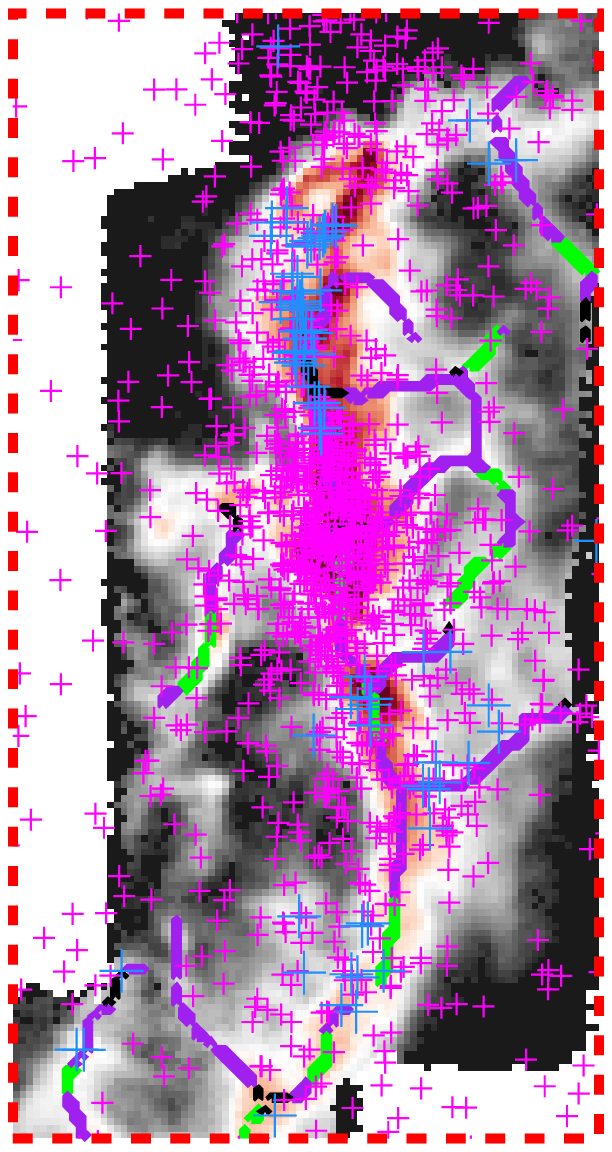}
		\\(b)
	\end{minipage}
	\begin{minipage}[t]{0.33\linewidth}
		\centering
		\includegraphics[trim = 8.5cm 6cm 7.9cm 5cm, width = \linewidth, clip]{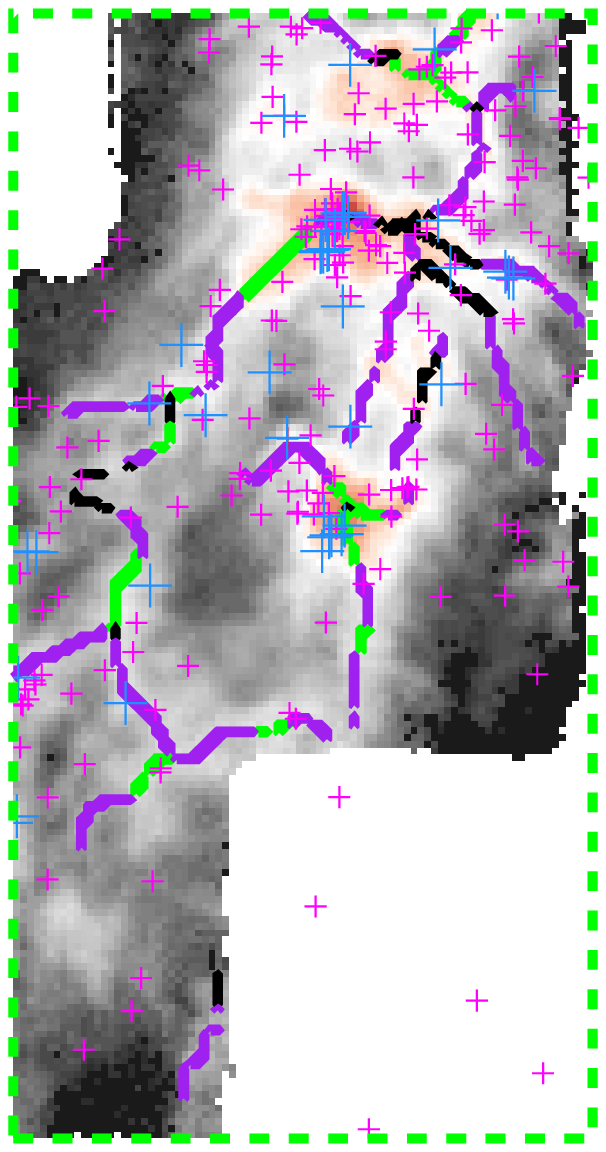}
		\\(c)
	\end{minipage}
	\begin{minipage}[t]{0.33\linewidth}
		\centering
		\includegraphics[trim = 8.5cm 6cm 7.9cm 5cm, width = \linewidth, clip]{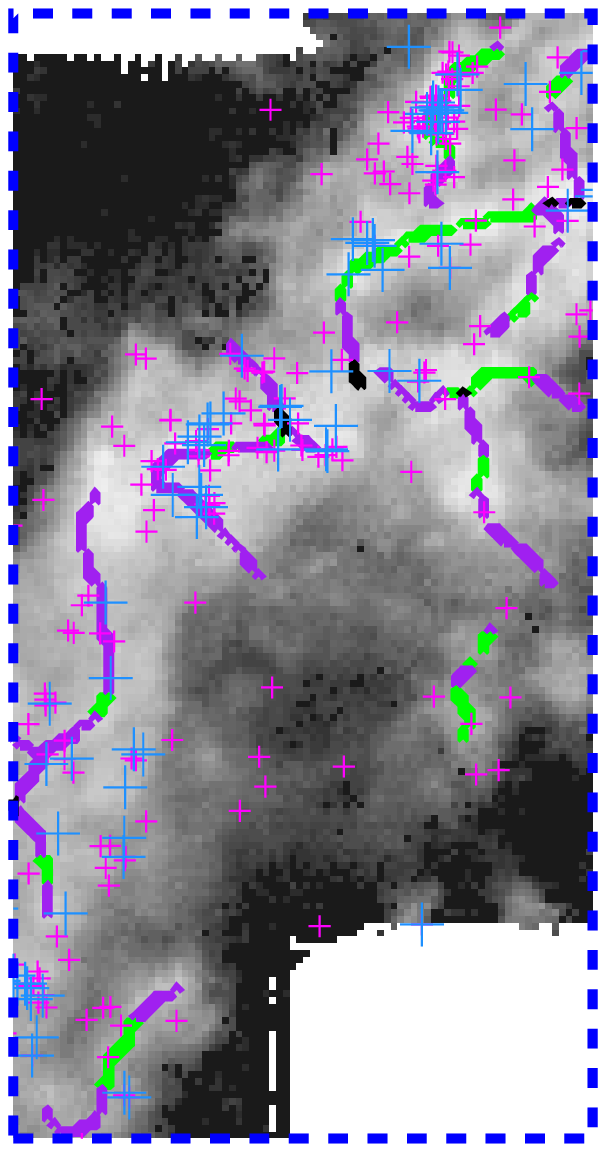}
		\\(d)
	\end{minipage}
	\begin{minipage}[t]{0.33\linewidth}
		\centering
		\includegraphics[trim = 8.5cm 6cm 7.9cm 5cm, width = \linewidth, clip]{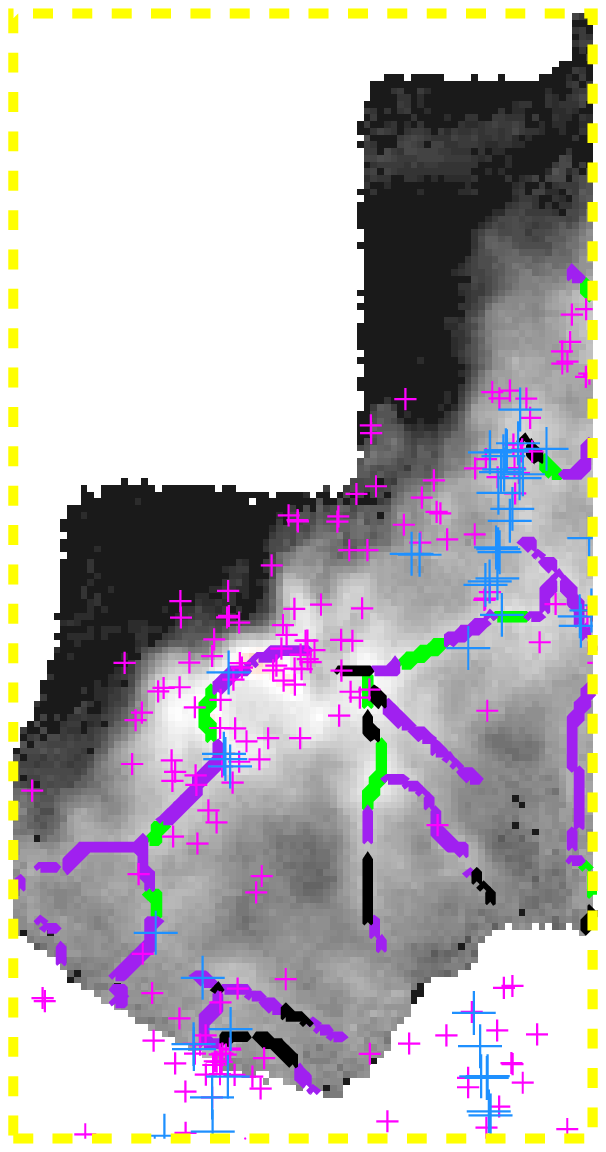}
		\\(e)
	\end{minipage}
	\caption{(a) Distribution of segments of different symmetries. The intrinsically symmetrical segments are marked in green. The intrinsically asymmetrical segments are marked in purple, while the segments with unknown intrinsic symmetries are marked in black. Active star-forming regions in the GMC are indicated with green letters. (b)-(e) are enlarged views of the red, green, blue and yellow boxes in panel a, respectively. The blue plus signs mark the protostars, and the magenta plus signs represent the disk dominated pre-main-sequence stars.}
	\label{fig17}
\end{figure}

The characteristics of the resulting categories of the 397 segment radial profiles are given in Table \ref{table1}. Figure \ref{fig17} shows the distributions of the segments of different intrinsic symmetry properties. From Figure \ref{fig17} and Table \ref{table1}, we can see that 65.3\% of the profiles are intrinsically asymmetrical, and about half of them belong to Category 6. The intrinsic symmetries of 13.3\% of the 397 profiles remain unknown. The fraction of the intrinsically symmetrical profiles, i.e., categories 1, 2, and 5, is 21.4\%, of which the FWHMs are derived by Plummer-like and Gaussian fitting methods in Section \ref{subsec5.3}.

\subsection{Widths of symmetrical segments} \label{subsec5.3}
\subsubsection{Morphology of the symmetrical radial profiles} \label{subsubsec5.3.1}
In previous studies \citep[e.g.,][]{Arzoumanian2011, Smith2014, Suri2019}, both Gaussian and Plummer-like functions are used to describe the morphology of the column density profiles of molecular filaments and to estimate the widths of filaments. In this work, we fitted the column density profiles of 85 intrinsically symmetrical profiles, which corresponds to category 1, 2, and 5, with the two forms of function. The procedures we used for the extraction of column density profiles and the fittings are written in the IDL programming language. We use the ``mpfitexpr'' routine in IDL to implement the fittings and the function forms, parameters, range and initial values of the parameters are all set a \textit{priori}.

\citet{Arzoumanian2011} and \citet{Clarke2019} suggest that the shape of column density profiles of filamentary structures can be well described by a Plummer-like function:
\begin{equation}
	\Sigma_{\mathrm{p}}(r) = A_{\mathrm{p}}\frac{\rho_{\mathrm{c}}R_{\mathrm{flat}}}{[1+(r/R_{\mathrm{flat}})^2]^{\frac{p-1}{2}}},
	\label{equation13}
\end{equation}
where $\rho_{\mathrm{c}}$ is the central density of the segment, $R_{\mathrm{flat}}$ is the inner flattening radius, and $p$ is the power-law exponent at large radii. $A_{\mathrm{p}}$ is a constant describing the effect of the inclination angle of the segment with respect to the plane of the sky. In this work, the column density profile of each segment has been normalized by the central column density at the skeleton points of the segment, and a zeroth-order baseline is also considered in the fitting processes \citep[e.g.,][]{Clarke2019, Suri2019}. However, we found that there exist some minor deviations between the skeleton positions derived from the DisPerSE algorithm and the actual local maxima of the column density distribution, which are within two spatial pixels. Therefore, the Plummer-like function we used to fit the column density profiles is written as: 
\begin{equation}
	\Sigma_{\mathrm{p}}(r) = \frac{A}{\{1+[(r-r_0)/R_{\mathrm{flat}}]^2\}^{\frac{p-1}{2}}}+B,
	\label{equation14}
\end{equation}
where constant $A$ is the normalized peak value of the column density profile, $B$ is the constant defining the baseline, and $r_0$ marks the deviation of the skeleton from the peak of the column density profile. The parameters in equation \ref{equation14} should be restrained during the fitting process, since too large values of the $p$ parameter in the Plummer-like function have no physical meaning in reality. Table \ref{table2} presents the ranges and the initial values of the parameters set a \textit{priori} for equation \ref{equation14}.
\begin{table}[h]
	\centering
	\caption{Ranges and initial values of parameters for Plummer-like fitting}
	\begin{tabular}{c c c}
		\hline\hline
		Parameter & Range & Initial value \\
		\hline
		$p$ & $2 \leq p \leq 4$ & 2 \\
		$R_{\mathrm{flat}}$ (\mbox{pc}) & $0<R_{\mathrm{flat}} \leq 1$ & 0.1 \\
		$r_0$ (\mbox{pc}) & $-0.18 \leq r_0 \leq 0.18$ & 0 \\
		$A$ & $0\leq A\leq 1.5$ & 1 \\
		$B$ & $B \geq 0$ & 0 \\
		\hline
	\end{tabular}
	\label{table2}
\end{table}

The Gaussian function we used to fit the column density profiles follows \citep[e.g.,][]{Arzoumanian2011, Andre2014, Panopoulou2014, Xiong2017, Clarke2019, Orkisz2019, Suri2019}:
\begin{equation}
	\Sigma_{\mathrm{p}}(r) = A\mathrm{e}^{-(r-r_0)^2/2\sigma^2}+B,
	\label{equation15}
\end{equation}
where $\sigma$ is the dispersion of the Gaussian function. Table \ref{table3} presents the ranges and the initial values of the parameter set for equation \ref{equation15}. 

\begin{table}[htb!]
	\centering
	\caption{Ranges and initial values of parameters for Gaussian fitting}
	\begin{tabular}{c c c}
		\hline\hline
		Parameter & Range & Initial value \\
		\hline
		$\sigma$ & $0 \leq \sigma \leq 1$ & 0.1 \\
		$r_0$ (\mbox{pc}) & $-0.18 \leq r_0 \leq 0.18$ & 0 \\
		$A$ & $0 \leq A \leq 1.5$ & 1 \\
		$B$ & $B \geq 0$ & 0 \\
		\hline
	\end{tabular}
	\label{table3}
\end{table}
In this work, the fitting range for the two methods above is determined through searching the column density profile for the lowest values on each side of the peak.  There are two circumstances in this process. The first is that the lowest value is found within the seven pixels from the peak, in other words, there is a ``valley'' in the profile on one side of the peak. In this case, the position of the ``valley'' is taken as one end of the fitting range. The second is that the profile keeps dropping, that is, no "valley" in the profile. In this case, the seventh pixel from the peak is taken as one end of the fitting range.

\subsubsection{``Widths'' of the segments} \label{subsubsec5.3.3}
Generally, the fitted results of the $R_{\mathrm{flat}}$ parameter from the Plummer-like fitting and the $\sigma$ parameter from the Gaussian fitting can not be compared directly. For consistency, we use the FWHMs of the well-fit Plummer-like and Gaussian functions to characterize the width of a segment. For the Plummer-like function, the width (FWHM) can be calculated through:
\begin{equation}
	w_{\mathrm{p}} = 2R_{\mathrm{flat}}\sqrt{2^{2/(p-1)}-1}.
	\label{equation17}
\end{equation}
For the Gaussian function, the width (FWHM) can be derived from:
\begin{equation}
	w_{\mathrm{g}} = 2\sigma\sqrt{2\ln2}.
	\label{equation18}
\end{equation}
In addition to fitting the intrinsically symmetrical profiles with two symmetrical functions, we also calculate the second moments of the profiles for both intrinsically symmetrical and asymmetrical segments (categories 1-6, 344 segments in total) to estimate the widths of the filaments in Orion A GMC. The second moment of a profile is derived after subtraction of a first-order baseline fitted using the next three pixels outside the inner seven pixels on each side of the profile.
The second moment is calculated according to the following formula, 
\begin{equation}
	m2 = \sqrt{\frac{\Sigma T_i(r_i-r_0)^2}{\Sigma T_i}}
	\label{equation19}
\end{equation}
where $T_i$ is the normalized column density at position $r_i$ in the profile after baseline subtraction, and $r_0$ is the mean of position $r_i$ with $T_i$ as the weight. The second moment is converted to FWHM through $w_{m2}=2\sqrt{2\ln2}\times m2$.

\begin{figure}[!htb]
	\centering
	\begin{minipage}[t]{0.55\linewidth}
		\centering
		\includegraphics[trim = 2cm 0cm 0cm 1cm, width = \linewidth, clip]{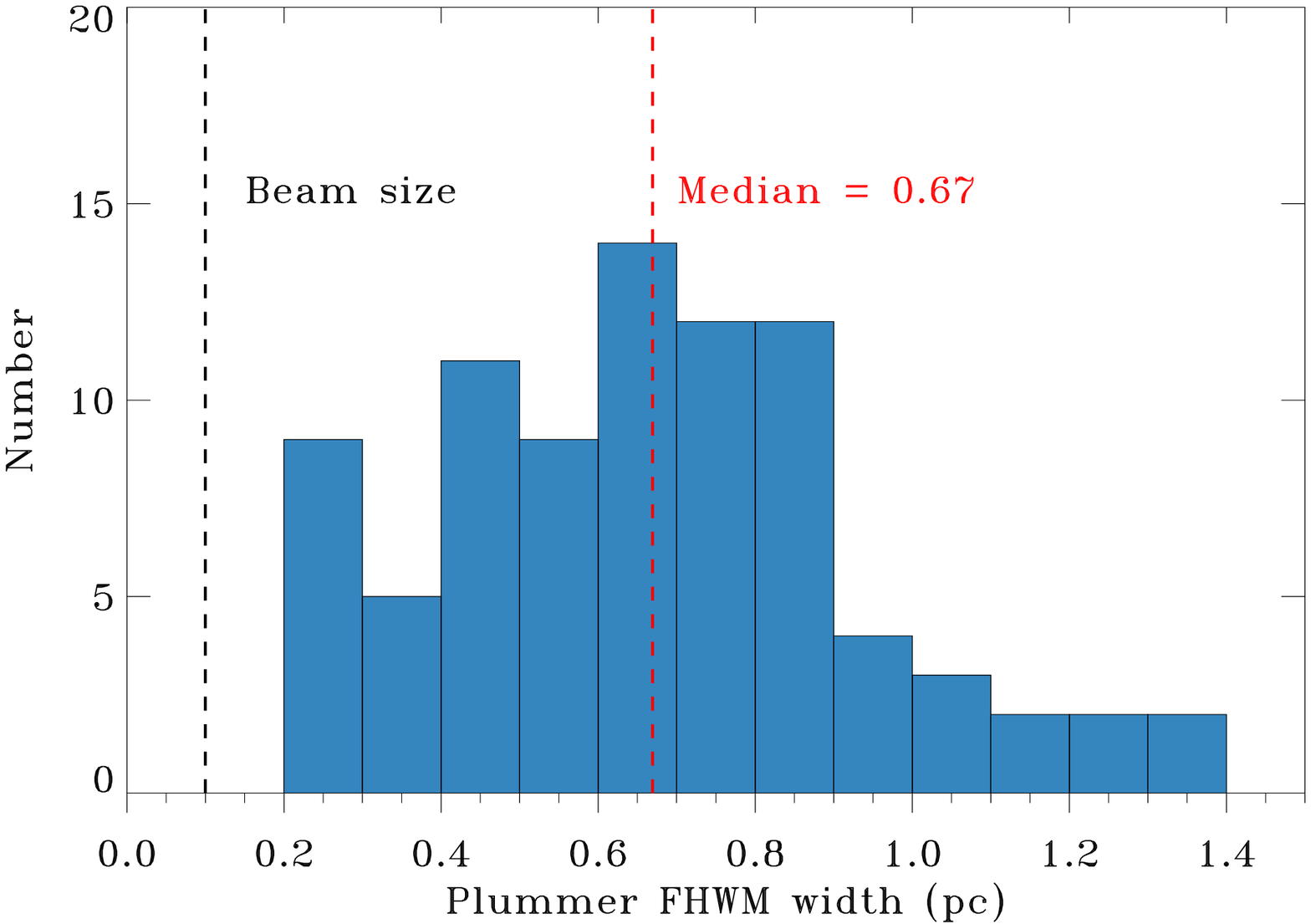}
		\\(a)
	\end{minipage}
	\begin{minipage}[t]{0.55\linewidth}
		\centering
		\includegraphics[trim = 2cm 0cm 0cm 1cm, width = \linewidth, clip]{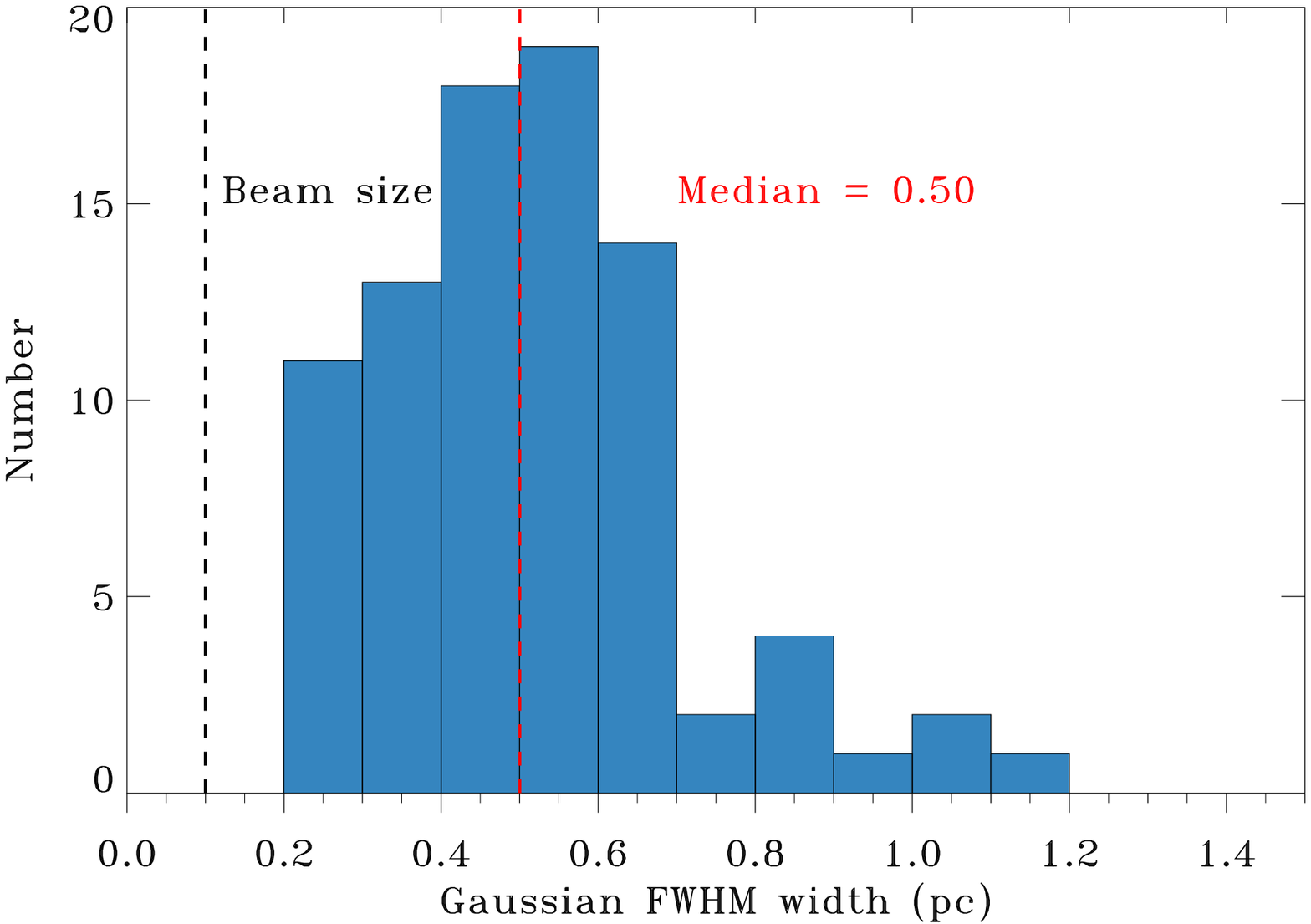}
		\\(b)
	\end{minipage}
	\begin{minipage}[t]{0.55\linewidth}
		\centering
		\includegraphics[trim = 2cm 0cm 0cm 1cm, width = \linewidth, clip]{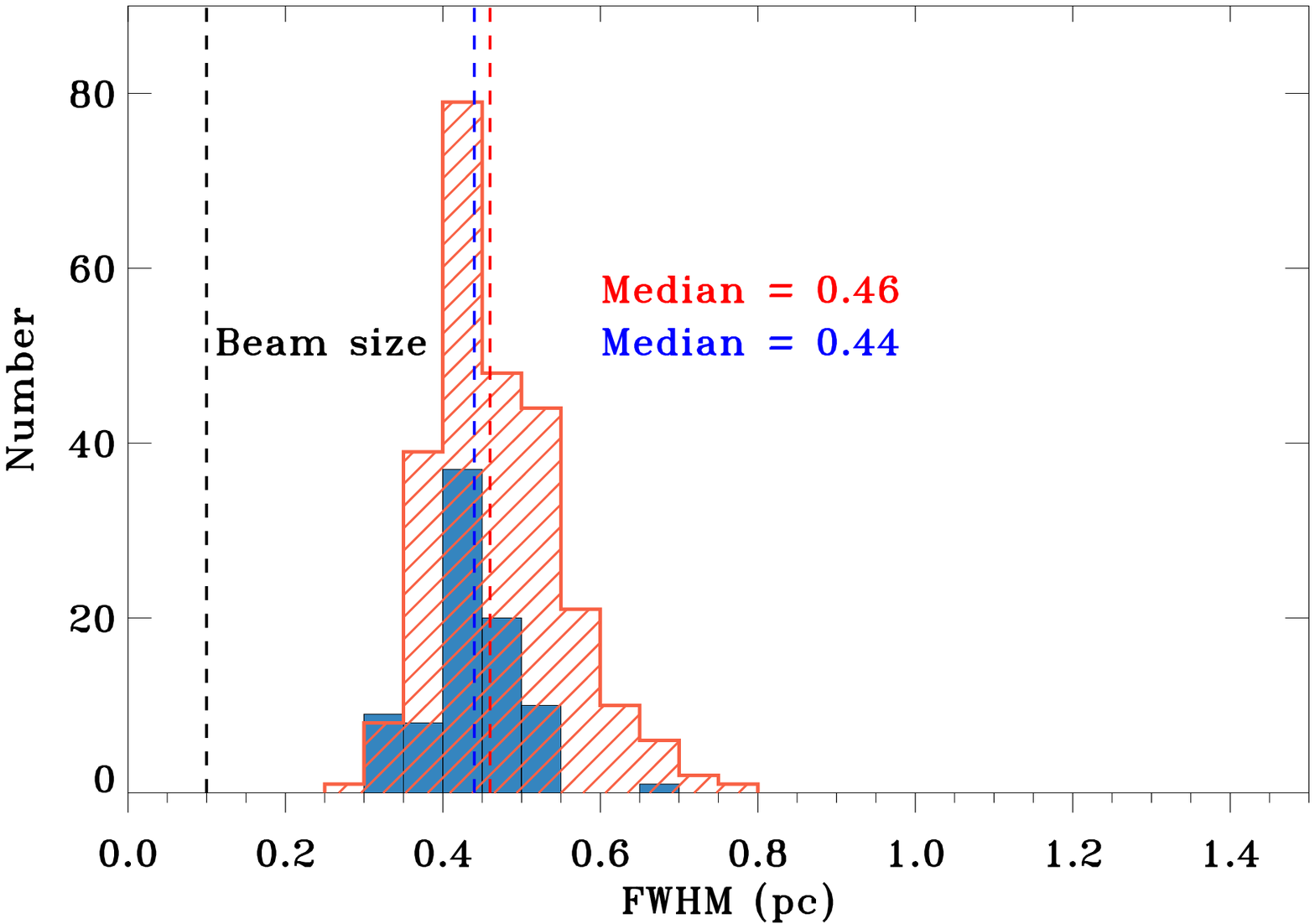}
		\\(c)
	\end{minipage}
	\caption{Histograms of the FWHMs of the (a) fitted Plummer-like function, (b) fitted Gaussian function, and (c) the calculated second moment. The bin sizes of panels a and b are 0.1 pc. The bin size of panel c is 0.05 pc. In panel c, the distributions of the widths derived from the second-moment method for intrinsically symmetrical and asymmetrical profiles are shown in blue columns and red slashes, respectively. The red dashed lines in panels a and b mark the median values of the distributions, and that in panel c indicate the median width of intrinsically asymmetrical profiles. The blue dashed line in panel c marks the median width of intrinsically symmetrical profiles. The black dashed line marks the beam size of the PMO 13.7m telescope in each panel.}
	\label{fig19}
\end{figure}

The distributions of the FWHMs derived using the above three methods are shown in Figure \ref{fig19}. For the two profile fitting methods, the median values of  $w_\mathrm{p}$ and $w_\mathrm{g}$ are 0.67 and 0.50 pc, respectively, while the mean values are 0.67 and 0.52 pc, respectively. For the second-moment method, the median values of $w_\mathrm{m2}$ for intrinsically symmetrical and asymmetrical profiles are 0.44 and 0.46 \mbox{pc}, and the mean values are 0.43 and 0.47 \mbox{pc}, respectively. For intrinsically symmetrical profiles, although the median values of $w_\mathrm{m2}$ and $w_\mathrm{g}$ are similar, the median $w_\mathrm{p}$ is $\sim$1.2 times larger than $w_\mathrm{m2}$ and $w_\mathrm{g}$. \citet{Nagahama1998} have studied the properties of filaments in the Orion A GMC using the Nagoya 4 m telescope with \element[][13]{CO} J = 1--0 emission. They obtained a typical width of $\sim$1.4 pc in the Orion A GMC, which is $\sim$3 times the widths obtained in this work. This difference may be caused by the lower spatial resolution (2.7\arcmin) of their observation on one hand and the larger distance (480 pc) they adopted for the Orion A GMC on the other hand \citep{Nagahama1998}. The derived median widths from the three methods are much broader than the universal width (0.1 pc) of filaments obtained from Herschel observations \citep{Arzoumanian2011}. \citet{Suri2019} also conducted a detailed study on the profiles of the filaments in the Orion A GMC using data of quite higher angular resolution (8\arcsec, corresponding to 0.015 pc at d = 388 pc used in their work). The typical width derived in their work is $\sim$0.13 pc, which is consistent with the results from Herschel observations. Whether the characteristic width of the filamentary structures obtained by Herschel observations is universal is still under debate. At the distance of the Orion GMC, 414 pc, the angular resolution of our data (50\arcsec) corresponds to a spatial resolution of 0.1 pc. The median filament width derived in this work is around 0.5-0.7 pc. After deconvolution from beam size used in this work, the median filament width is 0.49-0.69 pc. Therefore, we are convinced that the filaments identified in this work are resolved structures. The measured widths of filaments are found to be influenced either by the maps used for identification (velocity channel maps or column density maps) or the adopted fitting method and ranges \citep{Smith2014}. For example, \citet{Panopoulou2014} used the FCRAO $^{13}$CO emission line data to analyze the widths of the filamentary structures in the Taurus molecular cloud. They have identified filaments in both the velocity-integrated map and the velocity channel maps, and their results did not show any typical filament width of $\sim$0.1 pc. However, their results show that the distributions of the filament widths derived from the velocity-integrated map and the velocity channel maps are peaked at $\sim$0.4 and $\sim$0.25 pc, respectively, which is inconsistent with the universal width obtained from Herschel observations. We identified filaments in the column density map derived from the integrated intensity of the $^{13}$CO emission, i.e., covering the full velocity range of the Orion A GMC, while \citet{Suri2019} identified filaments in velocity slices. The different maps used for identification in the two work can result in different filament widths, as suggested in \citet{Panopoulou2014}. Molecular clouds have hierarchical structures on various spatial scales. With higher resolution and denser gas tracers, narrower widths of filaments can be found. However, the structures on large scale, such as the filaments identified in this work, should still exist. A possible explanation for the difference of filament widths between \citet{Suri2019} and this work is that the filaments identified by \citet{Suri2019} may be sub-scale denser structures embedded in the filaments identified in this work.

\section{Discussion} \label{sec6}
\subsection{Influence of the fitting method and fitting range on the fitted filament widths}\label{subsec6.1}
\begin{figure*}[!htb]
	\centering
	\begin{minipage}[t]{0.45\linewidth}
		\centering
		\includegraphics[trim = 1.5cm 2cm 4cm 3cm, width = \linewidth, clip]{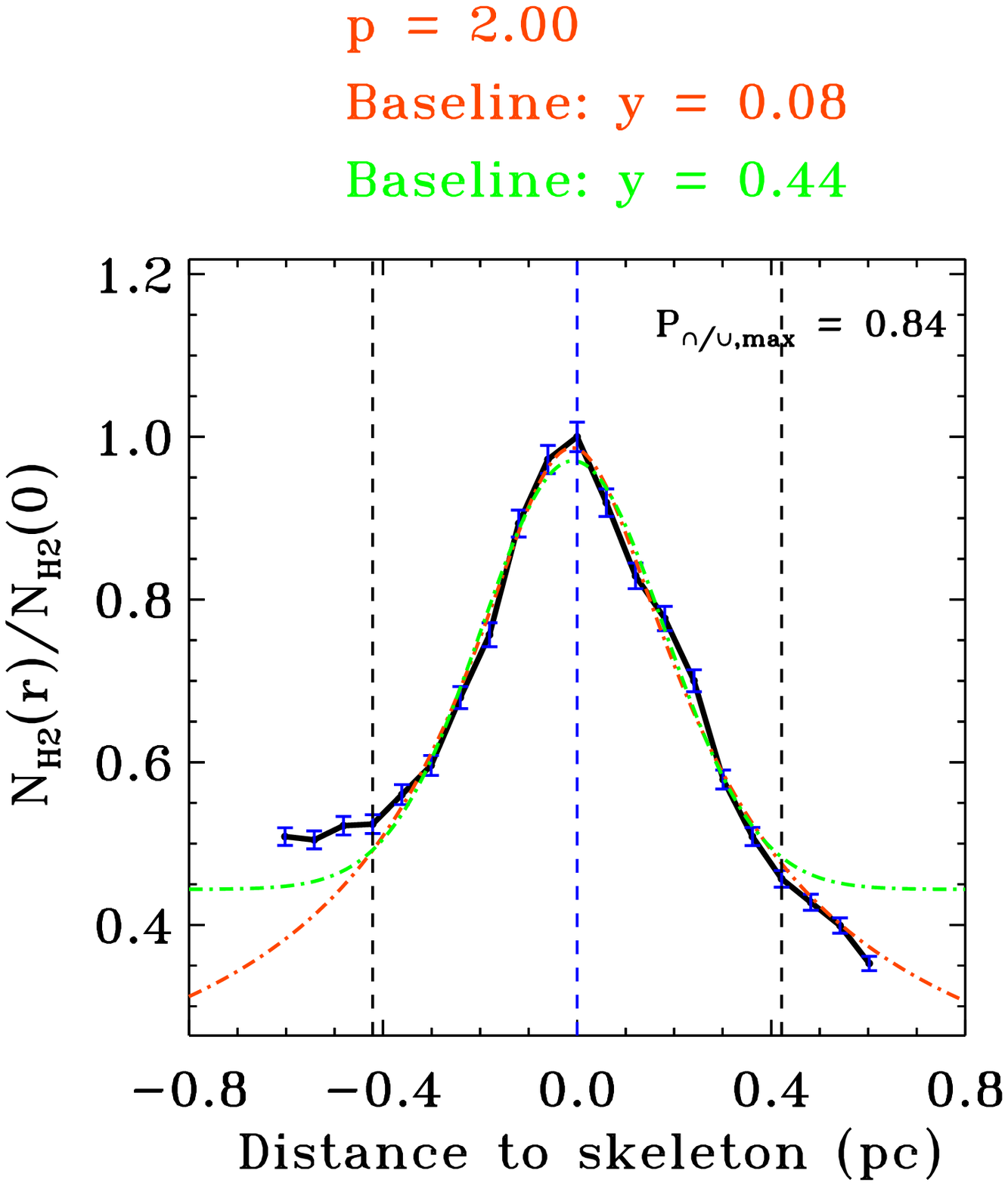}
		\\(a)
	\end{minipage}
	\begin{minipage}[t]{0.45\linewidth}
		\centering
		\includegraphics[trim = 1.5cm 2cm 4cm 3cm, width = \linewidth, clip]{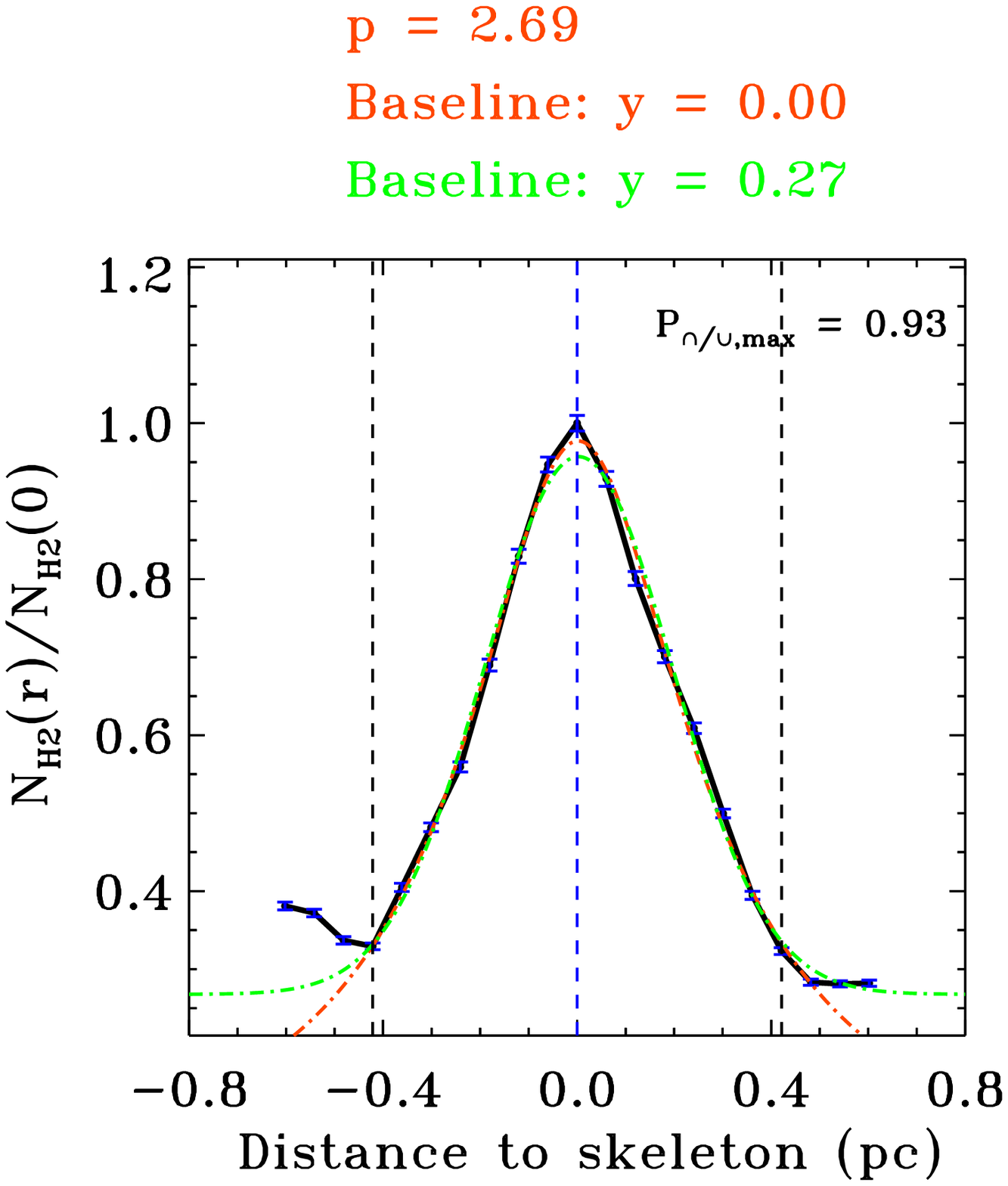}
		\\(b)
	\end{minipage}
	\begin{minipage}[t]{0.45\linewidth}
		\centering
		\includegraphics[trim = 1.5cm 2cm 4cm 3cm, width = \linewidth, clip]{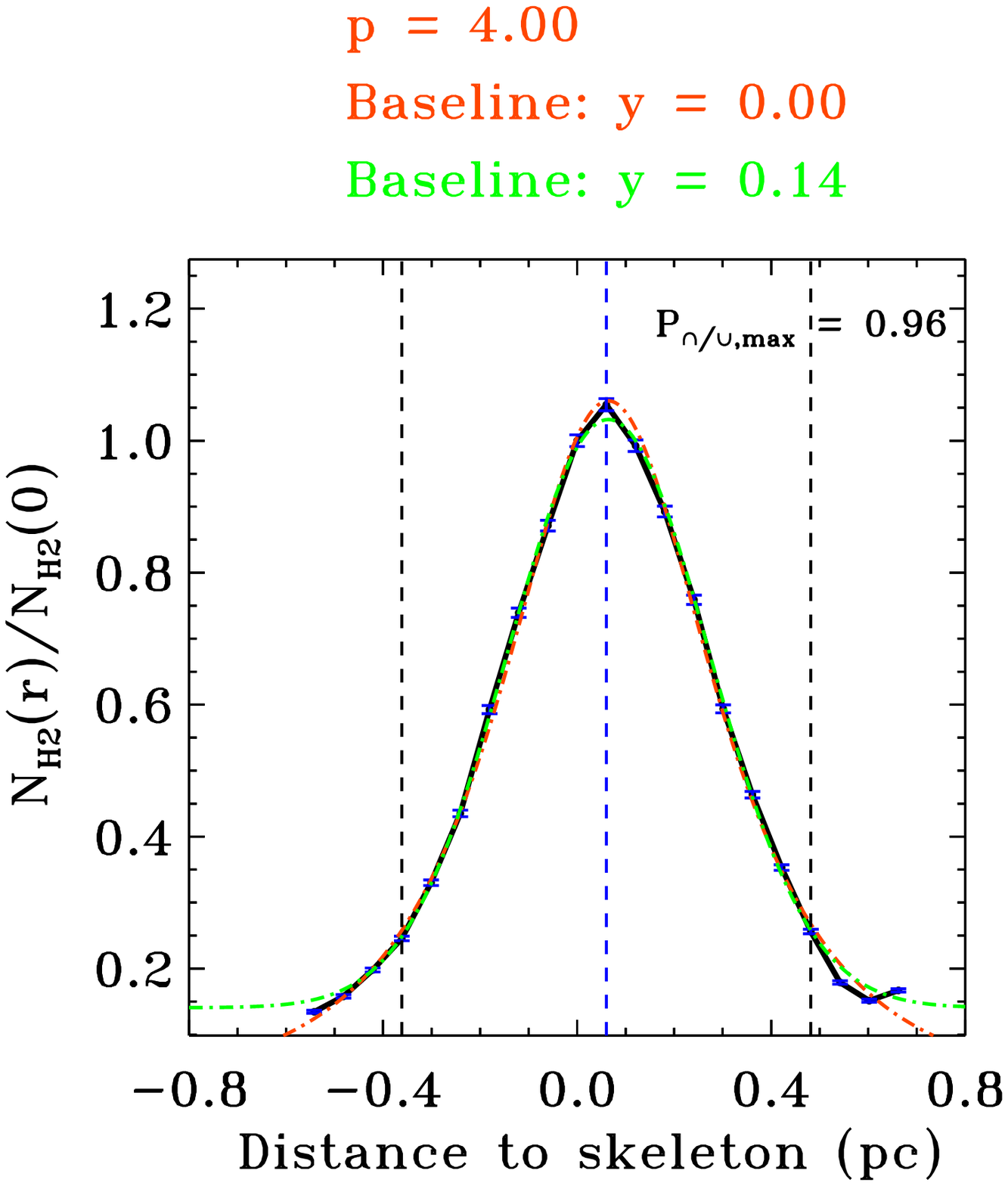}
		\\(c)
	\end{minipage}
	\begin{minipage}[t]{0.45\linewidth}
		\centering
		\includegraphics[trim = 1.5cm 2cm 4cm 3cm, width = \linewidth, clip]{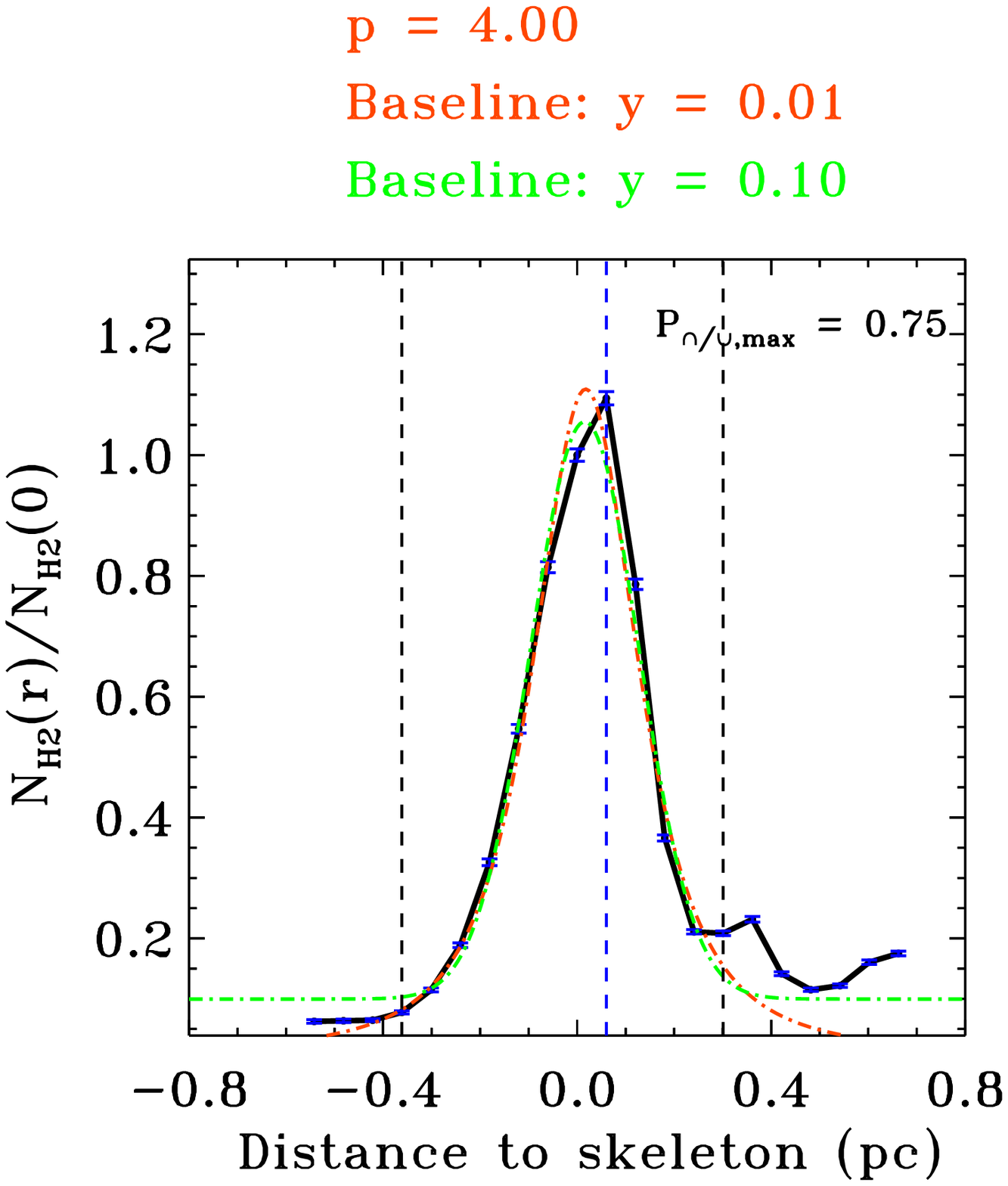}
		\\(d)
	\end{minipage}
	\caption{(a) and (b) are examples of the profiles in the southern region of Orion A GMC, while (c) and (d) are examples of the profiles in the northern region. The error bars are given according to equation \ref{equation8}. The black and blue vertical dashed lines indicate the boundaries of the fitting range and the peak. The Red dot-dash line is the Plummer-like fitting curve, and the green line is the Gaussian fitting curve. The fitted $p$ parameter for the Plummer-like fitting and the baselines for the two fitting functions are given at the top of each panel, with the red for the Plummer-like fitting and the green for the Gaussian fitting. The degree of symmetry of the profile is indicated at the upper-right corner.}
	\label{figA}
\end{figure*}

\begin{figure}[htb!]
	\centering
	\includegraphics[trim = 0cm 0cm 0cm 0cm, width = 0.7\linewidth, clip]{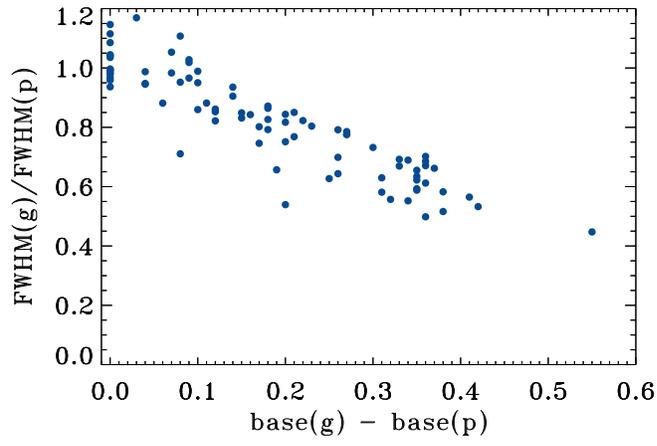}
	\caption{Relation between the difference of the baselines derived from the Gaussian fitting and the Plummer-like fitting and the ratio of $w_\mathrm{g}$ to $w_\mathrm{p}$.}
	\label{fig20}
\end{figure}

In the fitting processes, we find that there exist some situations like those presented in Figure \ref{figA}(a) and \ref{figA}(b). The two fitting methods give significantly different widths. The FWHMs given by Plummer fitting for the two profiles in Figures \ref{figA}(a) and \ref{figA}(b) are 0.78 and 0.59 pc, respectively, while those given by Gaussian fitting are 0.44 and 0.46 pc. We note that the two profiles all have non-zero baselines in the Gaussian fittings whereas nearly zero baselines in the Plummer-like fittings. The $w_\mathrm{p}$ derived from the Plummer-like fitting may overestimate the actual width of the profile in Figures \ref{figA}(a) and \ref{figA}(b). The profiles with a relatively homogeneous background found, for example, Figure \ref{figA}(a) and \ref{figA}(b), are mostly located in the southern part of the Orion A GMC, whereas profiles in the northern part of the GMC, for example Figure \ref{figA}(c) and \ref{figA}(d), show relatively small baseline difference between the Gaussian and Plummer-like fittings. The $w_\mathrm{p}$ and $w_\mathrm{g}$ are consistent with each other in Figure \ref{figA}(c) ($\sim$0.50 pc) and \ref{figA}(d) ($\sim$0.26 pc). To demonstrate the difference in the influence of the ``baseline'' on the width between the two fitting methods, we plot in Figure \ref{fig20} the relationship of the ratio between the FWHMs obtained from the Gaussian function and the Plummer-like function with the difference between the fitted baselines. The FWHM ratios decrease when the difference between the baselines increases. We can see that when the Gaussian fitting returns a higher baseline than the Plummer-like fitting, it also returns a narrower width.

\begin{figure}[htb!]
	\centering
	\includegraphics[trim = 2cm 2cm 6cm 3cm, width = 0.6\linewidth, clip]{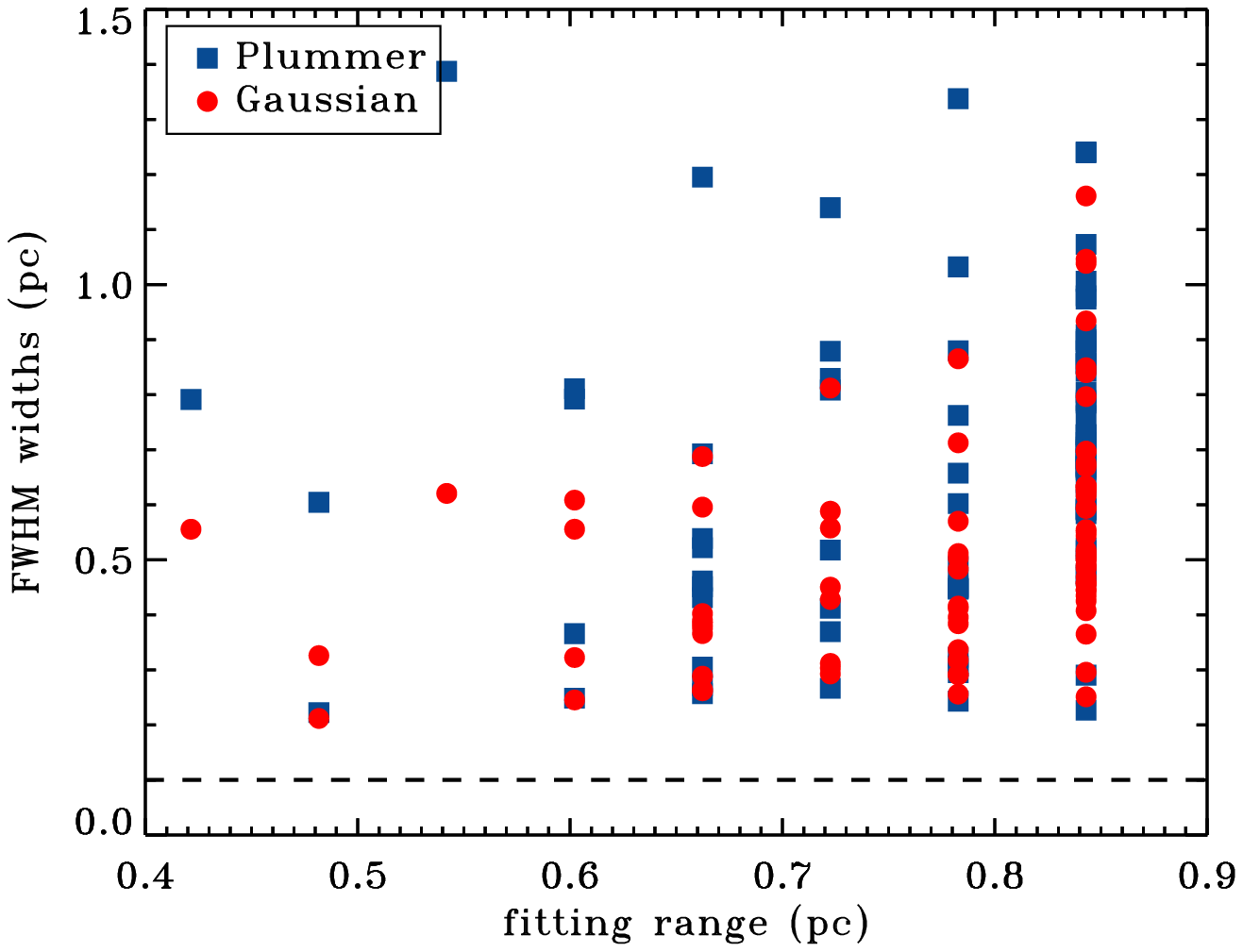}
	\caption{Relation between the fitted widths of the symmetrical segments and the applied fitting ranges. The beam size of our observations is indicated by the horizontal dashed line.}
	\label{fig21}
\end{figure}
The above results indicate that the filaments can not be treated as isolated structures from the environments, which is consistent with the results of \citet{Smith2014}. \citet{Smith2014} studied the column density profiles of simulated molecular filaments and suggested that the widths of the filaments are sensitive to the environment of filaments. 

Figure \ref{fig21} presents the relation between the applied fitting ranges and the derived widths for the two profile fitting methods. The Pearson correlation coefficients between the FHWMs and the fitting ranges for the Plummer-like and Gaussian fitting are 0.18 and 0.33, respectively, which shows weak correlation of fitted widths with the applied fitting ranges. 

The $p$ parameter in the Plummer-like function is often considered as an indicator of the presence of an isothermal cylinder in hydrostatic equilibrium ($p = 4$) \citep[e.g.,][]{Ostriker1964} or a filament supported by magnetic field ($p = 2$) \citep[e.g.,][]{Fiege2000}. In our Plummer-like fitting, we have found no typical value for the $p$ parameter. 

\subsection{Variation of the segment widths with spatial locations}\label{sec6.2}
\begin{figure}[h]
	\centering
	\includegraphics[trim = 0cm 0cm 0cm 0cm, width = 0.7\linewidth, clip]{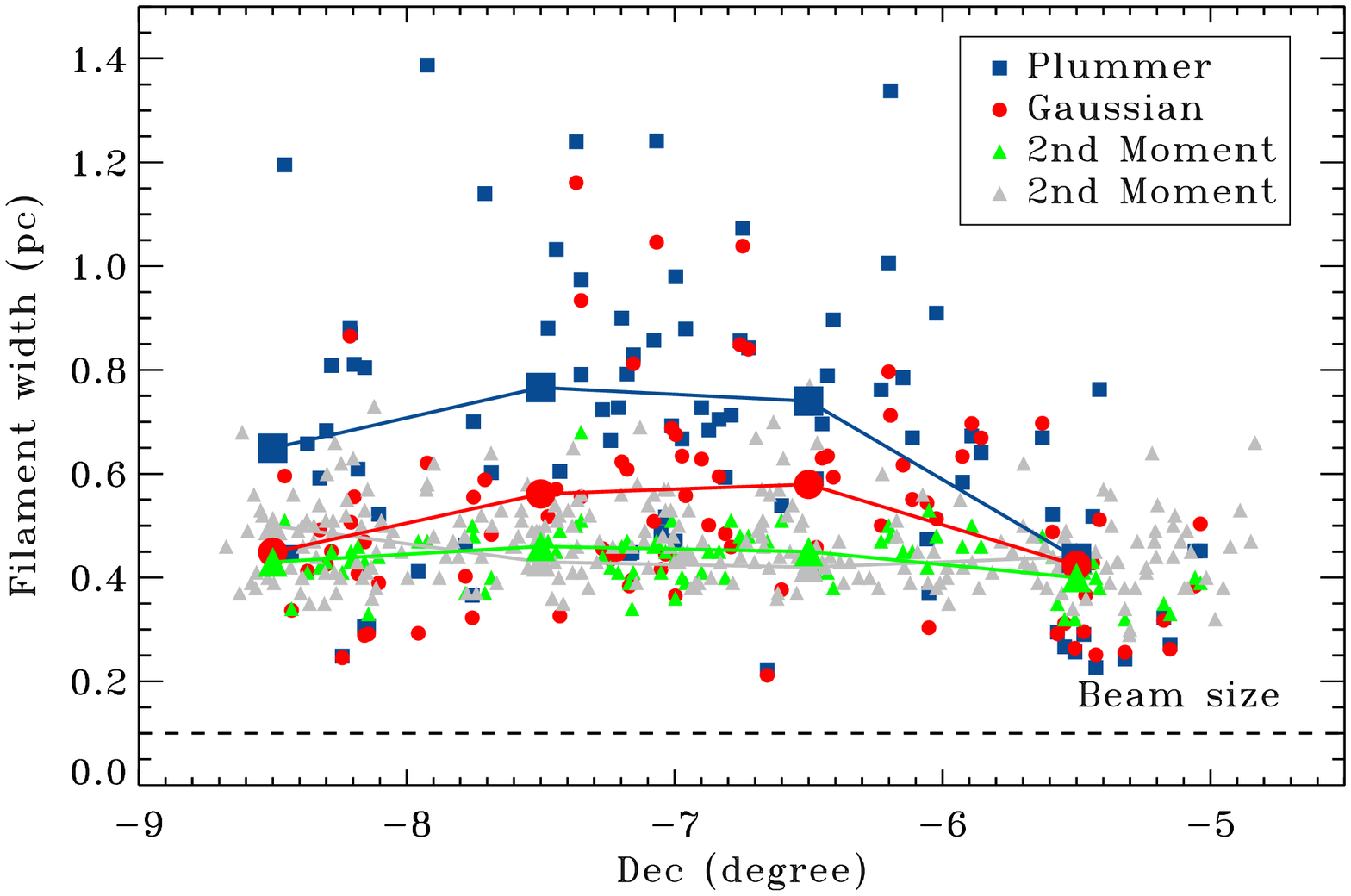}
	\caption{Relation between declination and FWHM widths. Blue blocks represent the Plummer FWHM widths. Red dots represent the Gaussian FWHM widths. The widths derived from the second moment for intrinsically symmetrical profiles are shown in green triangles, and those for intrinsically asymmetrical profiles are shown in grey triangles. The same symbols with larger sizes are the median widths in every one-degree interval from $-9^{\circ}$ to $-5^{\circ}$.}
	\label{fig22}
\end{figure}

It has long been realized that the Orion A GMC is composed of two parts that have different properties in many aspects, such as temperature, column density, chemistry, and star formation activity \citep[e.g.,][]{Nagahama1998, Ripple2013, Stutz2015, Kong2018, Ma2020}. We present the variation of the filament width along the direction of declination from south to north in Figure \ref{fig22}. The median widths from the three methods in every one-degree interval from $\delta = -9^{\circ}$ to $-5^{\circ}$ are also given in Figure \ref{fig22}. Across the whole extent of the declination, the widths derived from Plummer-like fitting are the largest, followed by the widths from the Gaussian fitting, and then those from the second-moment method. There is also a trend that the widths of the segments derived from Plummer-like fitting moderately increase from $-5^{\circ}$ to $-7.5^{\circ}$ and then decrease from $-7.5^{\circ}$ to $-9^{\circ}$. The widths derived from Gaussian fitting increase from $-5^{\circ}$ to $-6.5^{\circ}$, and then decrease from $-6.5^{\circ}$ to $-9^{\circ}$. The median $w_\mathrm{p}$ and $w_\mathrm{g}$ are consistent in the northern ISF region above $\delta \sim -6^{\circ}$, lying around 0.4 pc. The $w_\mathrm{p}$ and $w_\mathrm{g}$ are $\sim0.2 - 0.3$ \mbox{pc} in the declination range from $-5.1^{\circ}$ to $-5.5^{\circ}$ . From the second-moment method, the difference in the median width between intrinsically symmetrical and asymmetrical profiles in a given declination is negligible, and both values are distributed in a narrow range around $\sim$0.4 \mbox{pc} from north to south. In addition to the median width, the scatters of the widths at a given declination for the two fitting methods also slightly increase from the northern to the southern part of the GMC, while the scatters from the second-moment method almost remain the same from north to south. The distance of the southern part to the Sun is larger than that of the northern part \citep{Grossschedl2018}. Considering the distance difference, the widths of the filaments in the southern part should be larger than what we calculated. \citet{Grossschedl2018} suggest that the Orion A GMC consists of a denser and enhanced star-forming northern Head and a lower density and star formation quieter southern Tail. Our results suggest that different physical properties, such as column densities and the intensities of star formations, may lead to different widths of filaments in the Orion A GMC.

\begin{figure}[h]
	\centering
	\includegraphics[trim = 1cm 2cm 3cm 3cm, width = 0.8\linewidth, clip]{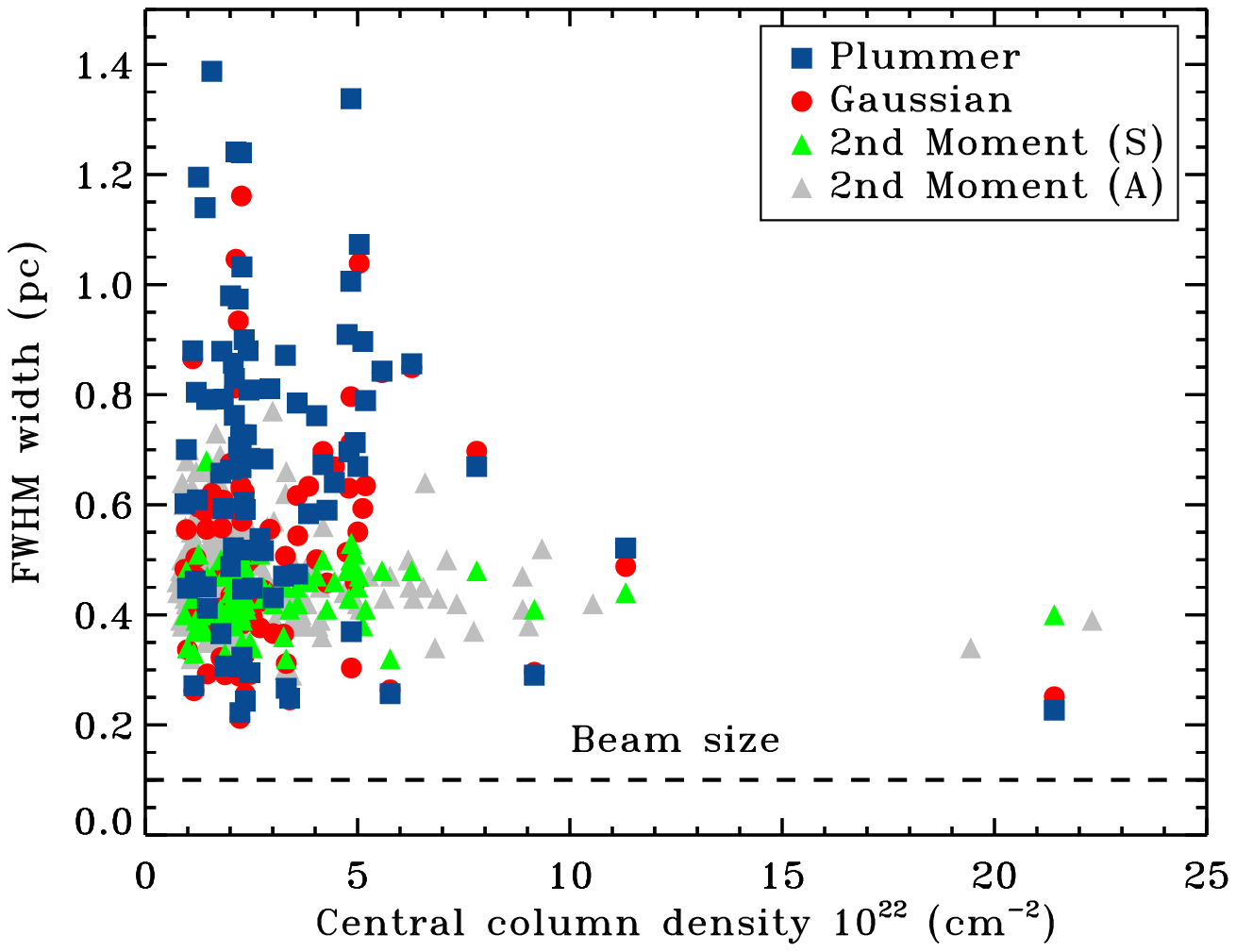}
	\caption{Relation between the FWHM widths and the central column densities of the selected segments. Blue blocks, red dots, green triangles correspond to the intrinsically symmetrical profiles with widths obtained from the Plummer-like fitting, Gaussian fitting, and the second-moment method, respectively, while grey triangles to the asymmetrical profiles with widths obtained from the second-moment method.}
	\label{fig23}
\end{figure}

The relationship between the widths and the central column densities of the fitted profiles is presented in Figure \ref{fig23}. The central column density of each segment is the median of the central column densities of all the slices (five at most) in that segment. We can see from Figure \ref{fig23} that the widths of the segments are independent on the central column densities of the segments up to $N_\mathrm{H_2}\sim7\times10^{22}$ cm$^{-2}$. This independence is consistent with the results of \citet{Arzoumanian2011}, \citet{Panopoulou2014}, and \citet{Suri2019}. However, both the widths and the scatter of the widths slightly decease when $N_\mathrm{H_2}$ is above $\sim7\times10^{22}$ cm$^{-2}$. We have checked the locations of the segments with $N_\mathrm{H_2}>7\times10^{22}$ cm$^{-2}$ and found that all these segments are located in the northern ISF region and the segments with the highest $N_\mathrm{H_2}$ correspond to the Orion KL region. The widths of the filaments are supposed to decrease with the increasing central column density, $\Sigma_0$, of the filaments, since the Jeans length $\lambda_{\mathrm{J}} = c_{\mathrm{s}}^2/(G\mu_{\mathrm{H}}\Sigma_{0})$ is inversely proportional to $\Sigma_0$ \citep[e.g.,][]{Arzoumanian2011}, where $c_\mathrm{s}$ is the sound speed and $\mu_\mathrm{H}$ is the molecular weight of the atomic hydrogen. When the central column density of a filament is high enough, its width will exceed the Jeans length and the filament will collapse and fragment under the influence of self-gravity. One possible explanation for the independence between the widths and the central column densities of the segments below $N_\mathrm{H_2}\sim7\times10^{22}$ cm$^{-2}$ is that these segments of the filaments are accreting materials from the environments, as proposed by \citet{Arzoumanian2011}. 

\subsection{Spatial  association with YSOs}\label{sec6.3}
Previous studies suggest that filaments and filamentary ``hubs'' are the places for the formation of stars and stellar clusters \citep[e.g.,][]{Myers2009}. \citet{Li2018} found that the majority of the young stellar objects (YSOs) in the regions of filaments in the Rosette Molecular Cloud are distributed along molecular filaments. The YSOs in the Orion A GMC have been surveyed by a lot of near-far infrared observations \citep[e.g.,][]{Stutz2013, Furlan2016}. \citet{Stutz2015} and \citet{Ma2020} both used a joint catalog of YSOs from Herschel and Spitzer surveys toward the Orion A GMC to study the relationship between the star formation activity and the shape of the N-PDF of molecular clouds. In this work, we use the same catalog to investigate the association between the YSOs and the filaments. Figure \ref{fig17} shows the spatial distribution of YSOs in the Orion A GMC. We can see that the YSOs are strongly associated with the GMC. The majority of the protostars (Class 0, Class I and Flat-spectrum sources) are distributed along the filaments, whereas the association between the disk dominated pre-main-sequence stars (Class II sources) and the filaments is relatively weaker. The protostars are clustered in the conjunctions of filaments, the so-called ``hubs'' of filaments, such as the locations of OMC-2, OMC-4, L1641 N, and NGC 1999. Besides, the protostars also show concentration at the positions where the filaments turn over, such as the location of the L 1641 S dark cloud. In Section \ref{subsubsec5.3.3}, we have obtained the median width of the filaments from Gaussian fitting, 0.50 pc. Here, we use this characteristic width to determine whether a YSO is located in a filament. If a YSO is located within 0.25 pc ($\sim$4 spatial pixels) of a filament skeleton pixel, it is recorded as ``in'' filament. According to this criterion, we find that 72.1\% of the protostars are associated with the filaments marked in Figure \ref{fig17}. For the Class II sources, a total of 52.5\% of them are located in the filaments. These results indicate that the filamentary structures are strongly coupled with recent and on-going star formation in the Orion A GMC.

\section{Conclusions} \label{sec7}
We present a study on the filamentary structures in Orion A GMC with focus on the symmetry and width of the filaments. The filaments are identified in the H$_2$ column density map using the DisPerSE algorithm. We use the Monte Carlo method to estimate the measurement errors of the H$_2$ column densities. We  divide the prominent filaments into segments with similar lengths and calculated their column density profiles. The symmetry properties of the column density profiles and the influence of environments on the symmetries of segments are studied. All the profiles of the segments are divided into eight categories according to their observed and intrinsic symmetry. The symmetrical column density profiles are fitted by Plummer-like and Gaussian functions for the extraction of the widths of the segments. We also used the second moments to calculate the widths of intrinsically symmetrical and asymmetrical profiles. We note that the $^{13}$CO emission used in this work may be unable to trace the densest gas in the Orion A GMC very well, and therefore the widths of filaments derived from $^{13}$CO data may be somewhat overestimated. The key points of the results are listed as follows. 

We have identified 225 filaments in the Orion A GMC, which include 46 filaments with lengths larger than 1.2 \mbox{pc}. These 46 filaments are divided into 397 segments, among which 65.3\% have intrinsically asymmetrical profiles, while 21.4\% have intrinsically symmetrical profiles, and the symmetry properties of 13.3\% of the segments are not available.

The median width derived from the Plummer-like fitting is 0.67 pc, and is 0.50 \mbox{pc} from Gaussian fitting. Derived from the second moment, the median width of the intrinsically symmetrical profiles is 0.44 pc, and that of the intrinsically asymmetrical profiles is 0.46 pc.

The ratio $w_{\mathrm{g}}/w_{\mathrm{p}}$ decreases as the difference between the Gaussian and Plummer-like fitted baselines increases. Both $w_{\mathrm{g}}$ and $w_{\mathrm{p}}$ exhibit weak correlations with the fitting range.

The widths of the segments derived from Plummer-like fitting moderately increase from $-5^{\circ}$ to $-7.5^{\circ}$ and decrease from $-7.5^{\circ}$ to $-9^{\circ}$. The widths derived from Gaussian fitting increase from $-5^{\circ}$ to $-6.5^{\circ}$ and decrease from $-6.5^{\circ}$ to $-9^{\circ}$, while median widths at a given declination derived from the second-moment method are all distributed in a narrow range around 0.4 pc. Above $\delta \sim -6^{\circ}$, the median widths from the three methods lie around $\sim$0.4 pc. The widths of the segments are independent on the central column densities when $N_\mathrm{H_2}$ is below $\sim7\times10^{22}$ cm$^{-2}$.

The YSOs are coincident with the filaments in the Orion A GMC. The protostars show concentration at the filament ``hubs'' and the places where the filaments turn over. We find that 72.1\% of the protostars are located within the median width of the filaments of 0.50 pc. For the Class II sources, 52.5\% of them are located in the filaments.

\begin{acknowledgements}		
We thank the anonymous referee for constructive suggestions. We thank the PMO-13.7 m telescope staffs for their supports during the observation. MWISP project is supported by National Key R\&D Program of China under grant 2017YFA0402701 and Key Research Program of Frontier Sciences of CAS under grant QYZDJ-SSW-SLH047. We acknowledge supports by NSFC under grant 11973091. Y. Zheng acknowledges supports by NSFC grants 11503086, 11503087.
\end{acknowledgements}
\clearpage
\bibliographystyle{raa}
\bibliography{ref_final}
\end{document}